\documentclass[1pt]{article}

\usepackage{setspace}

\usepackage{times}

\usepackage{amsmath} 
\usepackage{amssymb}
\usepackage{bm}
\usepackage[ruled,vlined]{algorithm2e} 
\usepackage{algorithmic}
\usepackage{graphicx}
\usepackage{float}
\usepackage{appendix}
\usepackage{hyperref}
\usepackage{cleveref}
\usepackage{geometry}
\usepackage{nameref}
\usepackage{xcolor} 
\usepackage{caption}
\usepackage{subcaption}
\usepackage{comment}
\usepackage[normalem]{ulem}
\usepackage{authblk} 

\newcommand{\argmax}{\mathop{\mathrm{argmax}}} 

\renewcommand*{\mathbf}[1]{\ifmmode\bm{#1}\else\textbf{#1}\fi} 

\newcommand{\fmten}[1]{\emph{#1}}

\title{A 2-stage elastic net algorithm for estimation of sparse networks with heavy tailed data}

\author[a]{Davide Bernardini}
\author[a]{Sandra Paterlini}
\author[a]{Emanuele Taufer}

\affil[a]{\small{Department of Economics and Management, University of Trento}}

\date{\today}

\begin{document} 


\baselineskip24pt


\maketitle


\begin{abstract}
We propose a new 2-stage procedure that relies on the elastic net penalty to estimate a network based on partial correlations when data are heavy-tailed. The new estimator allows to consider the lasso penalty as a special case. Using Monte Carlo simulations, we test the performance on several underlying network structures and four different multivariate distributions: Gaussian, t-Student with 3 and 20 degrees of freedom and contaminated Gaussian. Simulation analysis shows that the 2-stage estimator performs best for heavy-tailed data and it is also robust to distribution misspecification, both in terms of identification of the sparsity patterns and numerical accuracy. Empirical results on real-world data focus on the estimation of the European banking network during the Covid-19 pandemic. We show that the new estimator can provide interesting insights both for the development of network indicators, such as network strength, to identify crisis periods and for the detection of banking network properties, such as centrality and level of interconnectedness, that might play a relevant role in setting up adequate risk management and mitigation tools.
\end{abstract}




\newpage
\section{Introduction} \label{sec:intro}

An undirected graphical model is a set of two elements: a joint probability distribution $f$ and a graph $\mathcal{G}$, which encodes the conditional dependence structure of a set of random variables (see Lauritzen \cite{GM1996}, Koller and Friedman \cite{GM2009}). So far, most of the existing literature has focused on Gaussian graphical models, where the joint distribution is multivariate Gaussian $\mathcal{N}_p(\mathbf{\mu},\mathbf{\Sigma})$. In such a set-up, the conditional dependence structure, and thus the graph of the Gaussian graphical model, can be retrieved by looking, for example, at the inverse matrix $\mathbf{\Theta}$ of the covariance matrix $\mathbf{\Sigma}$. In particular, $\theta_{ij}=0$  implies conditional independence between variable $i$ and $j$ (see Lauritzen \cite{GM1996}).

Several methods have been proposed to estimate $\mathbf{\Theta}$ and thus the conditional dependence graph. Typically, a penalized estimator is used to recover the sparsity pattern and then reconstruct the graph of conditional dependencies (see, for example, Banerjee et al. \cite{Bane08}, Cai et al. \cite{Cai11}, Friedman et al. \cite{glasso08}, Lee et al. \cite{gslope18}, Mazumder and Hastie \cite{glasso12}, Meinshausen and B{\"u}hlmann \cite{Mein06}, Yuan \cite{Yuan10}, Zhou et al. \cite{gelato}). A widely used penalty is the Least Absolute Shrinkage and Selection Operator (LASSO) penalty proposed by Tibshirani \cite{Tib96} and based on $\ell_1$-norm. For example, Meinshausen and B{\"u}hlmann \cite{Mein06} introduced a conditional LASSO-penalized regression approach. Friedman et al. \cite{glasso08} proposed the graphical LASSO (\emph{glasso}) algorithm. It adds an element-wise $\ell_1$-norm penalty to the multivariate Gaussian log-likelihood function in order to estimate a sparse $\mathbf{\Theta}$. More recently, Bernardini et al. \cite{gelnet_ber} and Kovács et al. \cite{gelnet_kov} introduced independently penalized estimators of the precision matrix based on the \fmten{elastic net} penalty (see Zou and Hastie \cite{ElNet05}). Bernardini et al. \cite{gelnet_ber} suggested three different alternative approaches and investigate their performance through simulations, while Kovács et al. \cite{gelnet_kov} focused on a direct modification of the \emph{glasso} algorithm augmented with target matrices. The \fmten{elastic net} penalty extends the LASSO penalty by combining together $\ell_1$-norm (LASSO) and $\ell_2$-norm. Thus, LASSO is a special case of the \fmten{elastic net} penalty when the weight given to $\ell_2$-norm is zero.

When the distribution of data is not Gaussian, these estimators may lead to poor estimates because the assumption about the underlying distribution is violated. Thus, extensions to other multivariate distributions can be useful to address such situations. This paper extends the 2-stage graphical \fmten{elastic net} estimator (\emph{2Sgelnet}) in Bernardini et al. \cite{gelnet_ber} to the case where the joint distribution is a multivariate t-Student. The goal is to estimate the sparse precision matrix from which then retrieve the sparse partial correlation matrix. In order to achieve it, we propose a modification of the \emph{tlasso} algorithm by Finegold and Drton \cite{tlassoFD11}, introducing then the 2-stage t-Student \fmten{elastic net} (\emph{2Stelnet}) estimator, which relies on the Expectation-Maximization (EM) algorithm (see Dempster \cite{Demp72}), by exploiting the scale-mixture representation of the multivariate t-Student. As for \emph{2Sgelnet}, this new proposed estimator includes both LASSO and \fmten{elastic net} cases since the former is a special case of the latter.

Note that for t-Student distribution, a zero element in $\mathbf{\Theta}$ does not necessarily imply conditional independence, but only a zero partial correlation. Thus a partial correlation matrix is estimated. Monte Carlo simulations show that the proposed extension (\emph{2Stelnet}) to the t-Student distribution of \emph{2Sgelnet} of Bernardini et al. \cite{gelnet_ber} leads, in general, to better estimates in a wide range of situations where the data exhibit heavier tails than in the normal case and in presence of model misspecification. Fat tails are, for example, a well-known stylized fact of financial time series (see Cont \cite{Cont01}). Furthermore, economics and finance have recently begun to pay more attention to the importance of several network structures (see Jackson\cite{AREnet09}, Carvalho and Tahbaz-Salehi\cite{AREprod19}, Diebold and Yilmaz\cite{finnet15}, Jackson and Pernoud \cite{FinNetSurv20}, Bardoscia et al. \cite{FinNetSurvPhy}). There is, consequently, a growing interest in broadening the set of tools available to estimate network structures. For example, financial networks have received a renewed attention in the aftermath of the recent global financial crisis. Several empirical studies have focused their attention on the estimation and analysis of these networks; see, for example, Barigozzi and Brownlees \cite{NETS19}, Bilio et al. \cite{GrangerNet12}, Dermier et al. \cite{globalnet18} and Torri et al. \cite{Torri18}. In this paper, we utilize \emph{2Stelnet} to estimate the relationships among a large set of important European banks, for the period 2018-2020. In order to improve our understanding of systemic risk of European banking network and assess the impact of Covid-19 pandemic, we track some common network statistics and the evolution of the average intensity of the connections. We also identify the most central banks by looking at different centrality measures. Finally, we carry out an exploratory simulation study to map the effects of shock in the estimated network.

This paper is organized as follows. Section \ref{sec:methods} describes the \emph{2Stelnet} estimator. Section \ref{sec:simsetup} describes the set-up of our simulation analysis and Section \ref{sec:simresults} reports the results obtained. Finally, in Section \ref{sec:appl}, we estimate the European banking network during the period of 2018-2020 and analyze its characteristics.
\section{t-Student graphical models} \label{sec:methods}
Let $\mathbf{X}=[X_1,...,X_p]^\top$ be a $p$-dimensional random vector with joint multivariate t-Student distribution $t_p$($\mathbf{\mu},\mathbf{\Psi}^{-1},\nu$) where $\mathbf{\mu}$ is the mean vector, $\mathbf{\Psi^{-1}}$ is the positive definite scatter, or dispersion, matrix and $\nu$ are the degrees of freedom. The covariance matrix $\mathbf{\Sigma}$ of $\mathbf{X}$ and its inverse, the precision matrix $\mathbf{\Theta}$, are then equal to $\mathbf{\Sigma}=\frac{\nu}{\nu-2}\mathbf{\Psi}^{-1}$  and $\mathbf{\Theta}=\frac{\nu-2}{\nu}\mathbf{\Psi}$, with $\nu>2$.
Our goal is to estimate a sparse precision matrix $\mathbf{\Theta}$, from which to retrieve a sparse graph whose weights are partial correlations between couples of variables. Partial correlations are obtained by properly scaling the off-diagonal elements in $\mathbf{\Theta}$. Let $\theta_{jk}$ be the element in the $j$-th row and $k$-th column of $\mathbf{\Theta}$, the partial correlation p$_{jk}$ between components $X_j$ and $X_k$ is then equal to:
\begin{equation}\label{formula:prec2pcorr}
    \text{p}_{jk}=-\frac{\theta_{jk}}{\sqrt{\theta_{jj}\theta_{kk}}}.
\end{equation}
Thus, we build a graph $\mathcal{G}(\text{\textbf{V}},\text{\textbf{E}})$ with the set of nodes \textbf{V}=$\{1,...,p\}$ representing the elements in $\mathbf{X}$ and the set of edges \textbf{E} $\subseteq$ \textbf{V}$\times$\textbf{V} and ($j,k$) $\in$ \textbf{E} if p$_{jk}\neq0$, where ($j,k$) represents the edge between elements $X_j$ and $X_k$. Note that, differently from Gaussian graphical models where the distribution of $\mathbf{X}$ is multivariate Gaussian, here p$_{jk}=0$ does not necessarily imply that elements $X_j$ and $X_k$ are conditionally independent (see Baba et al. \cite{Baba04}). Nonetheless, it can be shown that, in this case, if nodes $j$ and $k$ are separated by a subset of nodes \textbf{C} $\subseteq$ $\{h\mid h \in \text{\textbf{V}}\wedge h\neq j,k\}$ in the graph $\mathcal{G}$, then the elements $X_j$ and $X_k$ are conditionally not correlated given the subset \textbf{C} (see Finegold and Drton \cite{tlassoFD11}). For elliptical distributions, conditional and partial correlations are equivalent. (see Baba et al. \cite{Baba04}).

\noindent
Following the scale-mixture representation of a multivariate t-Student distribution as in Finegold and Drton \cite{tlassoFD11}, we have that:
\begin{equation} \label{tmixture}
    \mathbf{X} = \mathbf{\mu} + \frac{\mathbf{Y}}{\sqrt{\tau}} \sim t_p(\mathbf{\mu},\mathbf{\Psi}^{-1},\nu).
\end{equation}
where $\mathbf{Y} \sim \mathcal{N}_p(\mathbf{0},\mathbf{\Psi}^{-1})$ and $\tau \sim \Gamma\big{(}\frac{\nu}{2},\frac{\nu}{2}\big{)}$.
Thus, the multivariate $p$-dimensional t-Student distribution can be seen as a mixture of a multivariate $p$-dimensional Gaussian distribution with an independent univariate gamma distribution. By properly exploiting this representation, it is possible to rely on the EM algorithm (Dempster et al. \cite{EMart77}) to estimate the parameters of the multivariate t-Student distribution. Following closely the \emph{tlasso} procedure proposed by Finegold and Drton \cite{tlassoFD11}, we suggest a similar EM algorithm to produce a sparse estimate of $\mathbf{\Theta}$. Differently from the \emph{tlasso} that uses LASSO, or $\ell_1$-norm, penalty (see Tibshirani \cite{Tib96}) to induce sparsity, we propose an approach that utilizes the \fmten{elastic net} penalty, a linear combination of $\ell_1$-norm and $\ell_2$-norm \cite{ElNet05}, to do a penalized estimation of $\mathbf{\Theta}$. In fact, we use the \emph{2Sgelnet} by Bernardini et al. \cite{gelnet_ber}, instead of relying on \emph{glasso} by Friedman et al. \cite{glasso08}. The core idea behind is to estimate a sparse $\mathbf{\Psi}$, the precision matrix of the multivariate Gaussian in the mixture, since its elements are proportional to the elements of $\mathbf{\Theta}$.

\subsection{The EM algorithm} \label{sec:EM}
Let $\text{\textbf{x}}_1,...,\text{\textbf{x}}_n$ be $n$ $p$-vectors of observations drawn from $t_p(\mathbf{\mu},\mathbf{\Psi}^{-1},\nu)$ distribution, realizations of $\mathbf{X}$. The random variable $\tau$ in the mixture (\ref{tmixture}) is considered the hidden, or latent, variable whose value is updated given the current estimate of the parameters and the observed data. Let also $\tau_i$ be the value of the latent variable $\tau$, associated with observation $\text{\textbf{x}}_i$. As in the \emph{tlasso} of Finegold and Drton \cite{tlassoFD11}, we also assume that the degrees of freedom $\nu$ are known in advance. This simplifies the procedure, but $\nu$ can be treated also as an unknown parameter and thus estimated (see Liu and Rubin \cite{EMtdof}). The EM algorithm proceeds as follows \cite{tlassoFD11}. At time step $t+1$:

\begin{itemize}
    \item \emph{Expectation step} (\emph{E-step})
    \begin{itemize}
        \item The expected value of $\tau$, given a generic vector of observations \textbf{x} and parameters $\mathbf{\mu}$, $\mathbf{\Psi}$ and $\nu$, is:
        \begin{equation}
            \text{\textbf{E}}(\tau \mid \mathbf{X}=\text{\textbf{x}}) = \frac{\nu + p}{\nu + (\text{\textbf{x}}-\mathbf{\mu})^\top\mathbf{\Psi}(\text{\textbf{x}}-\mathbf{\mu})}.
        \end{equation}
        \item Using the current estimates $\hat{\mathbf{\mu}}^{(t)}$ and $\hat{\mathbf{\Psi}}^{(t)}$ of the parameters, we update the estimated value $\hat{\tau}_i^{(t+1)}$ of $\tau_i$:
        \begin{equation} \label{updatetau}
           \hat{\tau}_i^{(t+1)} = \frac{\nu + p}{\nu + (\text{\textbf{x}}_i-\hat{\mathbf{\mu}}^{(t)})^\top\hat{\mathbf{\Psi}}^{(t)}(\text{\textbf{x}}_i-\hat{\mathbf{\mu}}^{(t)})}.
        \end{equation}
        \item[] for each $i=1,...,n$.
    \end{itemize}
    \item \emph{Maximization step} (\emph{M-step})
    \begin{itemize}
        \item Compute the updates of parameters given the data and $\hat{\tau}_i^{(t+1)}$:
        \begin{equation} \label{updatemu}
            \hat{\mathbf{\mu}}^{(t+1)} = \frac{\sum_{i=1}^n\hat{\tau}_i^{(t+1)}\text{\textbf{x}}_i}{\sum_{i=1}^n\hat{\tau}_i^{(t+1)}}. 
        \end{equation}
        \begin{equation} \label{penMstep}
            \hat{\mathbf{\Psi}}^{(t+1)} = \hat{\mathbf{\Psi}}_{\alpha,\lambda}(\hat{\tau}_i^{(t+1)},\text{\textbf{x}}_1,...,\text{\textbf{x}}_n).
        \end{equation}
        \item[] where $\hat{\mathbf{\Psi}}_{\alpha,\lambda}(.)$ is a penalized estimator of $\mathbf{\Psi}$, whose penalty is controlled by hyperparameters $\alpha$ and $\lambda$
    \end{itemize}
\end{itemize}

The EM algorithm cycles sequentially through the E and M steps until a convergence criterion is satisfied. Let $\hat{\psi}_{jk}$ be the element in $j$-th row and in $k$-th column of $\hat{\mathbf{\Psi}}$, we stopped the iterations in the algorithm when max$_{j,k}(|\hat{\psi}^{(t+1)}_{jk} - \hat{\psi}^{(t)}_{jk}|)$ is smaller than a given threshold value $\delta$.

In the following, we discuss the estimator $\hat{\mathbf{\Psi}}_{\alpha,\lambda}(.)$ of $\mathbf{\Psi}$ for (\ref{penMstep}) based on \emph{2Sgelnet} of Bernardini et al. \cite{gelnet_ber}.

\subsection{Two-stage t-Student elastic net - [\emph{2Stelnet}]} \label{sec:2Stelnet}

This estimator consists in a 2-step procedure. At first, it estimates the sparsity structure of $\hat{\mathbf{\Psi}}^{(t+1)}$ by using conditional regressions with \fmten{elastic net} penalty. This is inspired by the neighborhood selection approach of Meinshausen and B\"uhlmann \cite{Mein06} where the LASSO penalty is used. At each iteration, we first transform the observed data as follows (from (\ref{tmixture})):
\begin{equation}\label{datatransformed}
    \widetilde{\text{\textbf{x}}}_i = (\text{\textbf{x}}_i-\hat{\mathbf{\mu}}^{(t+1)})\sqrt{\hat{\tau}_i^{(t+1)}}.
\end{equation}
Let $\widetilde{\text{\textbf{X}}}$ be the $n$ by $p$ matrix of transformed observations, such that the $i$-th row is equal to $\widetilde{\text{\textbf{x}}}_i^\top$. Let also $\widetilde{\text{\textbf{X}}}_k$ be the $k$-th column of $\widetilde{\text{\textbf{X}}}$ and $\widetilde{\text{\textbf{X}}}_{-k}$ be $\widetilde{\text{\textbf{X}}}$ without the $k$-th column. We fit $p$ \fmten{elastic net} penalized regressions using the $k$-th component of the transformed vectors as dependent variable and the remaining components as predictors:
\begin{equation}\label{condreg}
  \hat{\text{\textbf{b}}}_k = \text{argmin}_{\text{\textbf{b}}_k}\Big{\{}\frac{1}{2n}||\widetilde{\text{\textbf{X}}}_k-\text{a}_k-\widetilde{\text{\textbf{X}}}_{-k}\text{\textbf{b}}_k||_2^2+\lambda[\alpha||\text{\textbf{b}}_k||_1+\frac{1}{2}(1-\alpha)||\text{\textbf{b}}_k||_2^2]\Big{\}}.
\end{equation}
with $k=1,...,p$. As in Meinshausen and B\"uhlmann \cite{Mein06}, we reconstruct the graph representing the connections (p$_{jk}\neq0$) among the components of $\mathbf{X}$ (and also of $\mathbf{Y}$). We include in the neighborhood of the node $k$ the node $j$ if the corresponding coefficient of the component $j$ in $\hat{\text{\textbf{b}}}_k$ is different from 0. Then, through the reconstructed neighborhoods of all nodes, we produce an estimate $\hat{\text{\textbf{E}}}$ of the edge set $\text{\textbf{E}}$. This procedure can lead to the situation where an edge ($j,k$) is included in $\hat{\text{\textbf{E}}}$ according to the neighborhood of $j$, ne($j$), but not accordingly to the neighborhood of $k$, ne($k$). To deal with such a situation, we can use two rules (see Meinshausen and B\"uhlmann \cite{Mein06}):
\begin{itemize}
    \item AND rule: edge $(j,k) \in \hat{\text{\textbf{E}}}$ if node $j \in \text{ne}(k)$ $\wedge$ node $k \in \text{ne}(j)$
    \item OR rule: edge $(j,k) \in \hat{\text{\textbf{E}}}$ if node $j \in \text{ne}(k)$ $\vee$ node $k \in \text{ne}(j)$
\end{itemize}
Once estimated $\hat{\text{\textbf{E}}}$, we use it to set the zero elements constraints in the current update $\hat{\mathbf{\Psi}}^{(t+1)}$. In particular, the update $\hat{\mathbf{\Psi}}^{(t+1)}$ is the maximizer of the following constrained optimization problem, with $\hat{\text{\textbf{S}}}^{(t+1)} = \frac{1}{n}\sum_{i=1}^n\hat{\tau}_i^{(t+1)}(\text{\textbf{x}}_i-\hat{\mathbf{\mu}}^{(t+1)})(\text{\textbf{x}}_i-\hat{\mathbf{\mu}}^{(t+1)})^\top$:
\begin{equation} \label{normformstep2}
\begin{split}
&\max_{\mathbf{\Psi}} \Big{\{} \text{log}(\text{det}(\mathbf{\Psi})) - \text{trace}(\hat{\text{\textbf{S}}}^{(t+1)}\mathbf{\Psi}) \Big{\}} \\
&\text{subject to} \\
&\psi_{jk} = \psi_{kj} = 0 \text{ if edge } (j,k) \notin\hat{\text{\textbf{E}}}.
\end{split}
\end{equation}
This problem can be rewritten in Lagrangian form as:
\begin{equation} \label{lagformstep2}
	\hat{\mathbf{\Psi}}^{(t+1)} = \argmax_{\mathbf{\Psi}} \Big{\{} \text{log}(\text{det}(\mathbf{\Psi})) - \text{trace}(\hat{\text{\textbf{S}}}^{(t+1)}\mathbf{\Psi}) - \sum_{(j,k)\notin\hat{\text{\textbf{E}}}}\gamma_{jk}\psi_{jk}\Big{\}}.
\end{equation}
where $\gamma_{ij}$ are Lagrange multipliers having nonzero values for all $\psi_{ij}$ constrained to 0. To solve this optimization problem (\ref{normformstep2}), we use the algorithm proposed by Hastie et al. \cite{bookESL2009} (Section 17.3.1, pp. 631-634) to maximize the constrained log-likelihood and produce an estimate $\hat{\mathbf{\Psi}}$, given the edge set $\hat{\text{\textbf{E}}}$ estimated in the previous step. As with \emph{2Sgelnet}, the existence of this estimator of $\mathbf{\Psi}$ is not guaranteed for all situations. When the number $n$ of observations is greater than the number $p$ of nodes, the estimator always exists (see Lauritzen \cite{GM1996}, Section 5.2.1). In the other situations, its existence depends on the structure of the estimated edge set $\hat{\text{\textbf{E}}}$ (see Lauritzen \cite{GM1996}, Section 5.3.2). In Appendix \ref{sec:app0} we reported a pseudo-code for the \emph{2Stelnet} algorithm.

\section{Simulations} \label{sec:sims}
\subsection{Simulation set-up}\label{sec:simsetup}
We rely on simulations to assess the performances of \emph{2Stelnet} (i.e. $\nu=3$ fixed a priori) and compare it with \emph{2Sgelnet} \cite{gelnet_ber} and with the well-known \emph{glasso} \cite{glasso08} and \emph{tlasso} \cite{tlassoFD11} algorithms. For sake of brevity, we report only the results obtained with AND rule as they are qualitatively similar to the ones obtained using OR rule. There are some specific situations where one rule is better than the other, but we do not find an overall winner. Results for OR rule are available upon request.

We consider the values $0.5$ and $1$ for $\alpha$, while for $\lambda$ we consider 100 exponentially spaced values between $e^{-6}$ and 2 (6.5 for \emph{tlasso}). We search for the optimal value of $\lambda$, given a value of $\alpha$ for \emph{2Sgelnet} and \emph{2Stelnet}, using BIC criterion (see \cite{ebic10}).

We randomly generate seven $50\times50$ precision matrices encoding different relationship structures among $50$ variables. We consider the following seven network topologies: scale-free, small-world, core-periphery, random, band, cluster and hub. Precision matrices embedding the randomly generated structures are reported in Figure \ref{fig:adjmats} of Appendix \ref{sec:app1}. For scale-free, random, band, cluster and hub we use the R package \emph{huge} (setting $v=0.3$, $u=0.1$) that allows to directly generate precision matrices with a given sparsity structure. Instead, for the small-world and core-periphery, we use the algorithms in \cite{smallworld98} and \cite{Torri18} respectively to generate the sparsity pattern, then we utilize the procedure suggested in \emph{huge} package to produce the precision matrices. Given a precision matrix $\mathbf{\Theta}$, we use it to generate $n$ vectors of observations from the following four multivariate distributions:
\begin{itemize}
    \item Multivariate normal: $\mathcal{N}_{50}(\mathbf{0},\mathbf{\Theta}^{-1})$.
    \item Multivariate t-Student: $t_{50}(\mathbf{0},\frac{\nu-2}{\nu}\mathbf{\Theta}^{-1},\nu)$, with $\nu=3$.
    \item Multivariate t-Student: $t_{50}(\mathbf{0},\frac{\nu-2}{\nu}\mathbf{\Theta}^{-1},\nu)$, with $\nu=20$.
    \item Contaminated normal: $\mathcal{N}_{50}(\mathbf{0},\mathbf{\Theta}^{-1})\cdot Ber + \mathcal{N}_{50}(\mathbf{0},\text{diag}(\mathbf{\Theta}^{-1}))\cdot (1-Ber)$, \\ with $Ber\sim \text{Bernoulli}(pd=0.85)$.
\end{itemize}
We consider three sample sizes, $n=100,250,500$. Thus, for each of the 14 couples network-distribution, we have $100\times50$, $250\times50$ and $500\times50$ datasets, respectively. We set up a Monte Carlo experiment with 100 runs for each sample size $n$. 

In order to compare the estimators in different settings, we test them on all couples network-distribution looking both at classification performance and numerical accuracy of the estimates for optimal $\lambda$. We use the $\text{F}_1\text{-score}=2\cdot\frac{\text{Precision}\cdot\text{Recall}}{\text{Precision}+\text{Recall}}$ as a measure of classification performance, where:
\begin{itemize}
    \item $\text{Precision}=\frac{\text{True positives}}{\text{True Positives}+\text{False Positives}}$.
    \item $\text{Recall}=\frac{\text{True positives}}{\text{True Positives}+\text{False Negatives}}$.
\end{itemize}
A true positive is a correctly identified edge, while a true negative is a missing edge which is correctly excluded from the edge set. A false positive is a missing edge which is erroneously included in the edge set, while a false negative is an edge which is wrongly excluded. The closer this measure is to 1, the better the classification. F$_1$-score is a better measure than accuracy when there is imbalance among classes (here existence or absence of an edge), as it happens in our simulation set-up.

In order to assess the numerical accuracy of the estimates, we compute the Frobenius distance between the theoretical and estimated partial correlation matrices, $\text{\textbf{P}}$ and $\hat{\text{\textbf{P}}}$ respectively:
\begin{equation*}
    ||\text{\textbf{P}}-\hat{\text{\textbf{P}}}||_\text{F} = \sqrt{\sum_{j=1}^p\sum_{k=1}^p |\text{p}_{jk}-\hat{\text{p}}_{jk}|^2}.
\end{equation*}
These matrices can be easily obtained through a proper scaling of the theoretical and estimated precision matrices, $\mathbf{\Theta}$ and $\hat{\mathbf{\Theta}}$, as in (\ref{formula:prec2pcorr}). We follow a common convention and set the diagonal elements $\text{p}_{jj}=\hat{\text{p}}_{jj}=0$, with $j=1,2,...,p$. When the Frobenius distance is 0, the estimate is exactly equal to the true values. Thus, the smaller the Frobenius distance, the better the estimate is from the numerical accuracy point of view.


\clearpage
\subsection{Simulation results} \label{sec:simresults}

In the following, we report the box plots for the performance measures and compare the estimators. A box plot represents the distribution of the performance measures of the optimal models obtained in 100 Monte Carlo runs for each combination of distribution and network. We test the performance of \emph{glasso}, \emph{tlasso}, \emph{2Sgelnet} and \emph{2Stelnet}. For the 2-stage estimators, we use the AND rule as a procedure to produce an estimate of the edge set. When $\alpha=1$, we have a pure LASSO penalty (see \ref{condreg}). For the sake of brevity, analyzing the simulation results, we refer to this situation using \emph{2Sglasso} and \emph{2Stlasso}, while \emph{2Sgelnet} and \emph{2Stelnet} refers to the case where $\alpha=0.5$.

\begin{figure}[h]
\begin{subfigure}{0.4\textwidth}
  \includegraphics[width=1\linewidth, angle = 0]{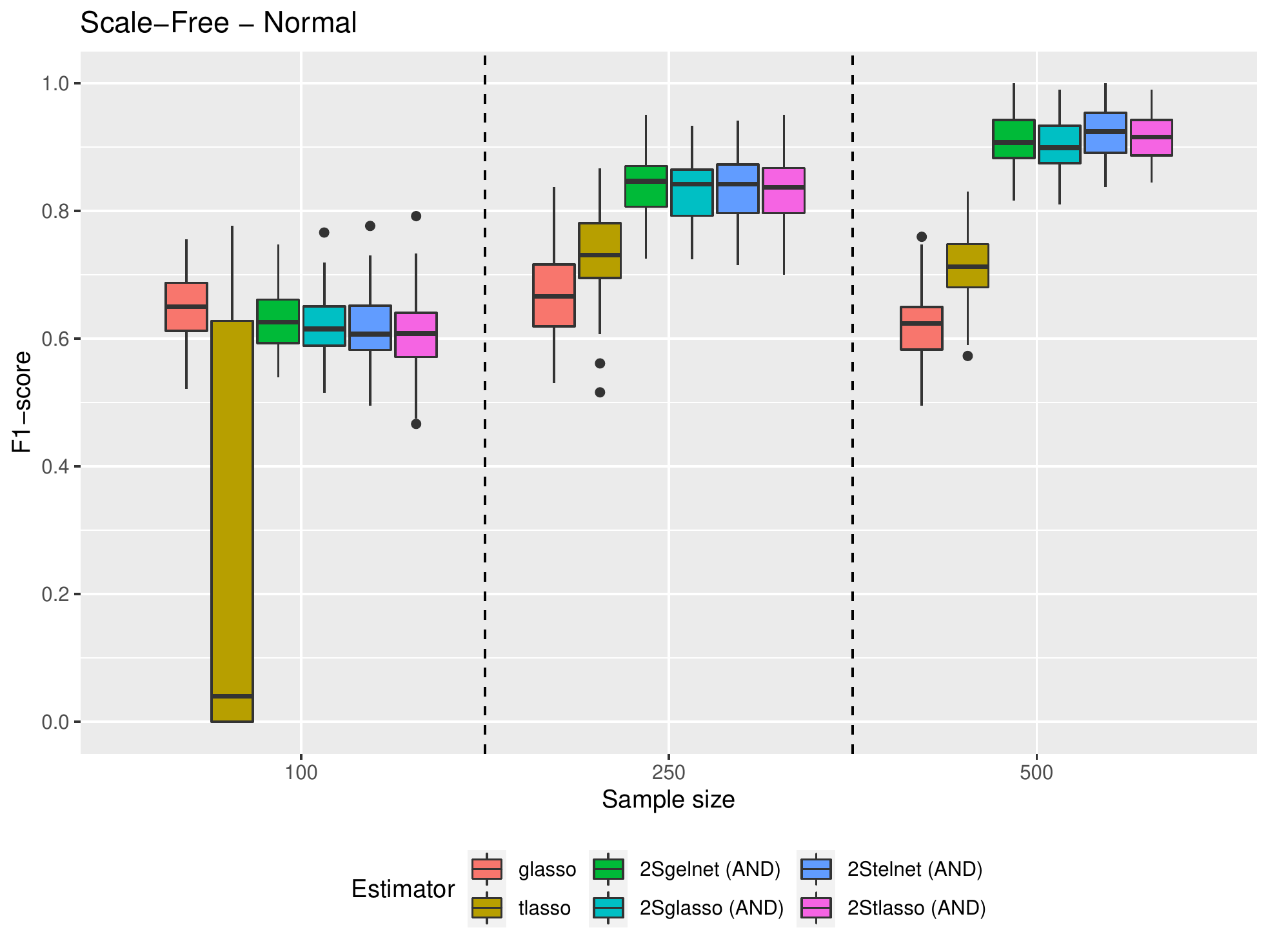}
\end{subfigure}
\begin{subfigure}{0.4\textwidth}
  \includegraphics[width=1\linewidth, angle = 0]{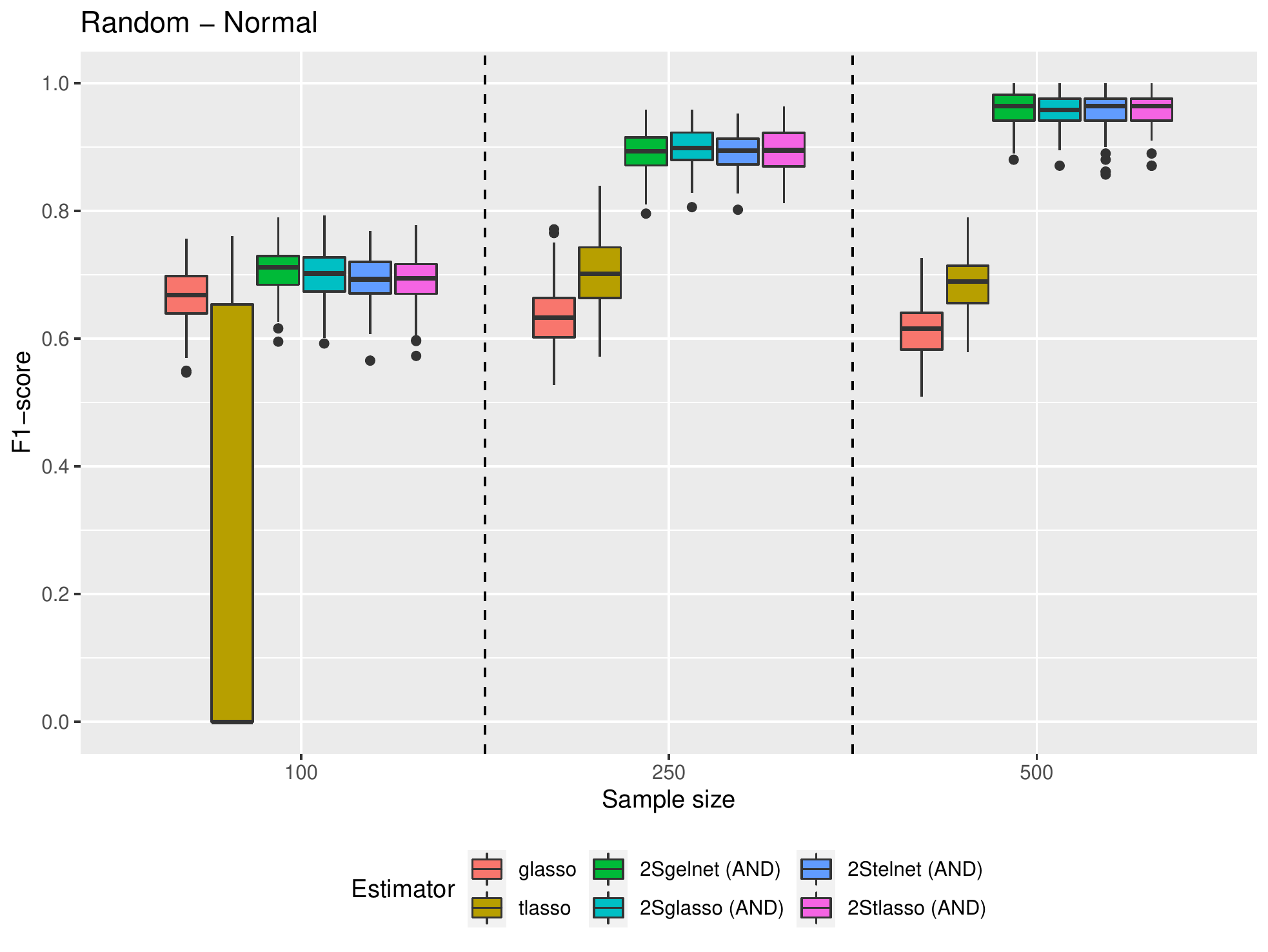}
\end{subfigure}
\newline
\begin{subfigure}{0.4\textwidth}
  \includegraphics[width=1\linewidth, angle = 0]{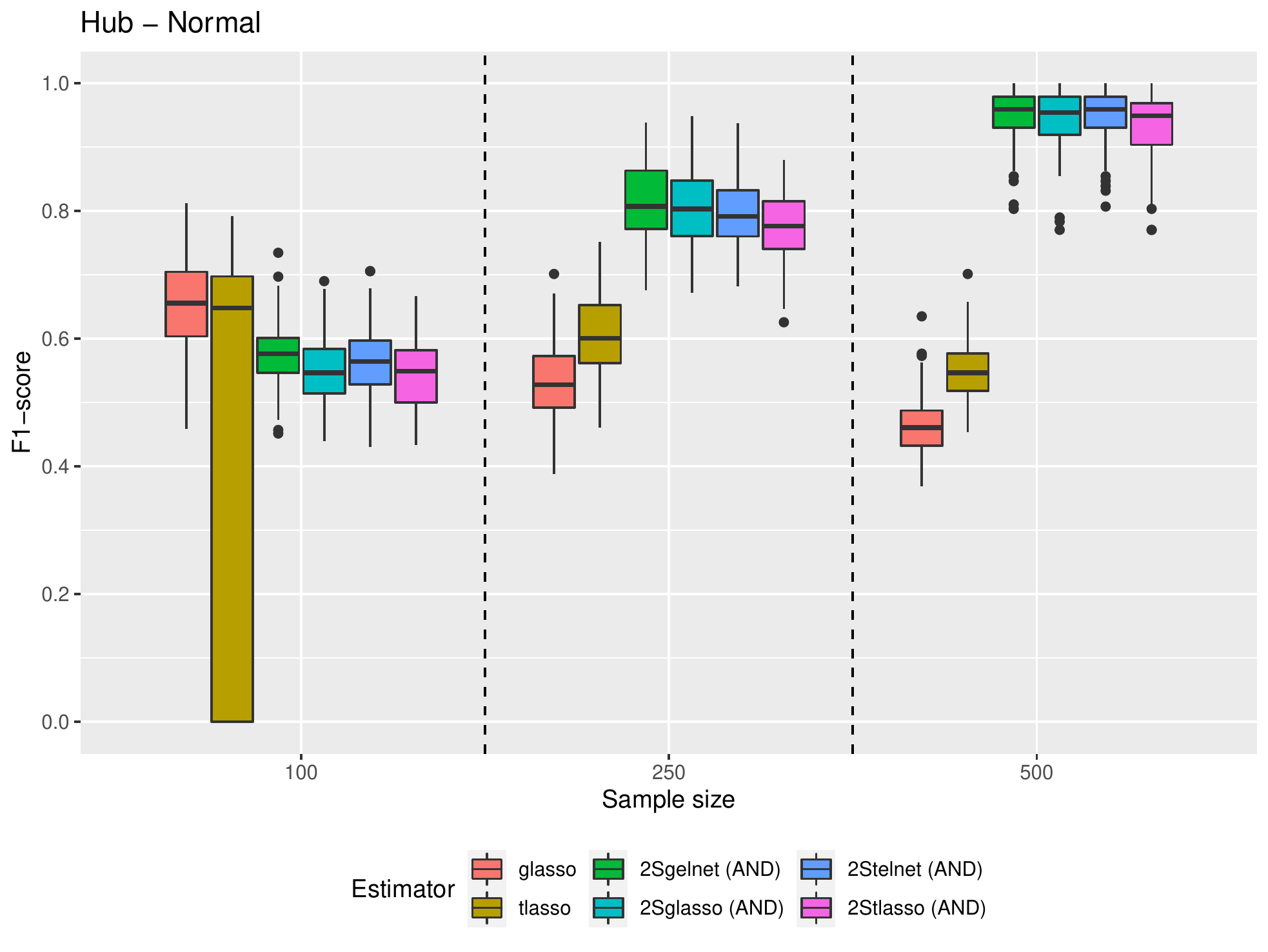}
\end{subfigure}
\begin{subfigure}{0.4\textwidth}
  \includegraphics[width=1\linewidth, angle = 0]{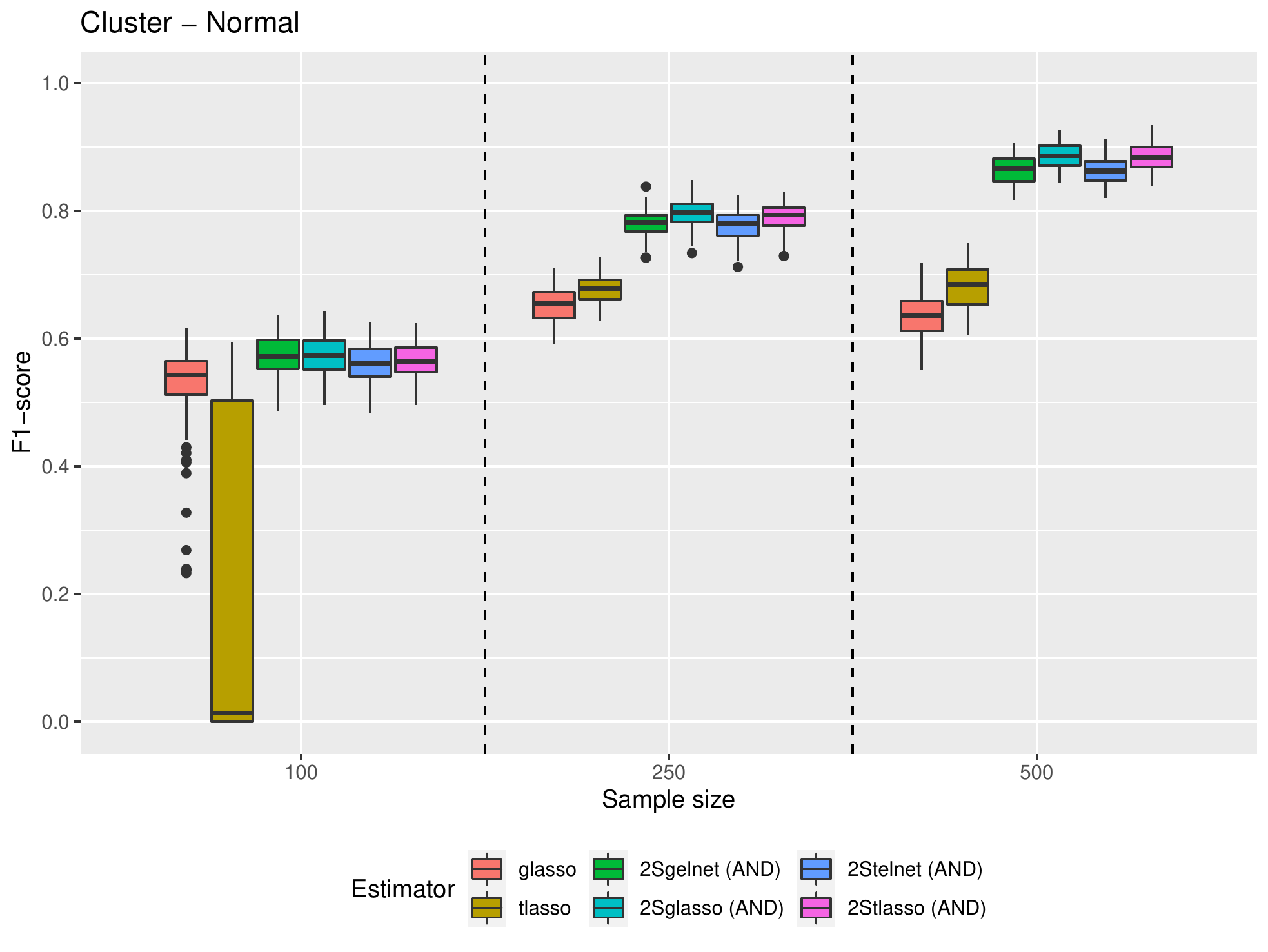}
\end{subfigure}
\newline
\begin{subfigure}{0.4\textwidth}
  \includegraphics[width=1\linewidth, angle = 0]{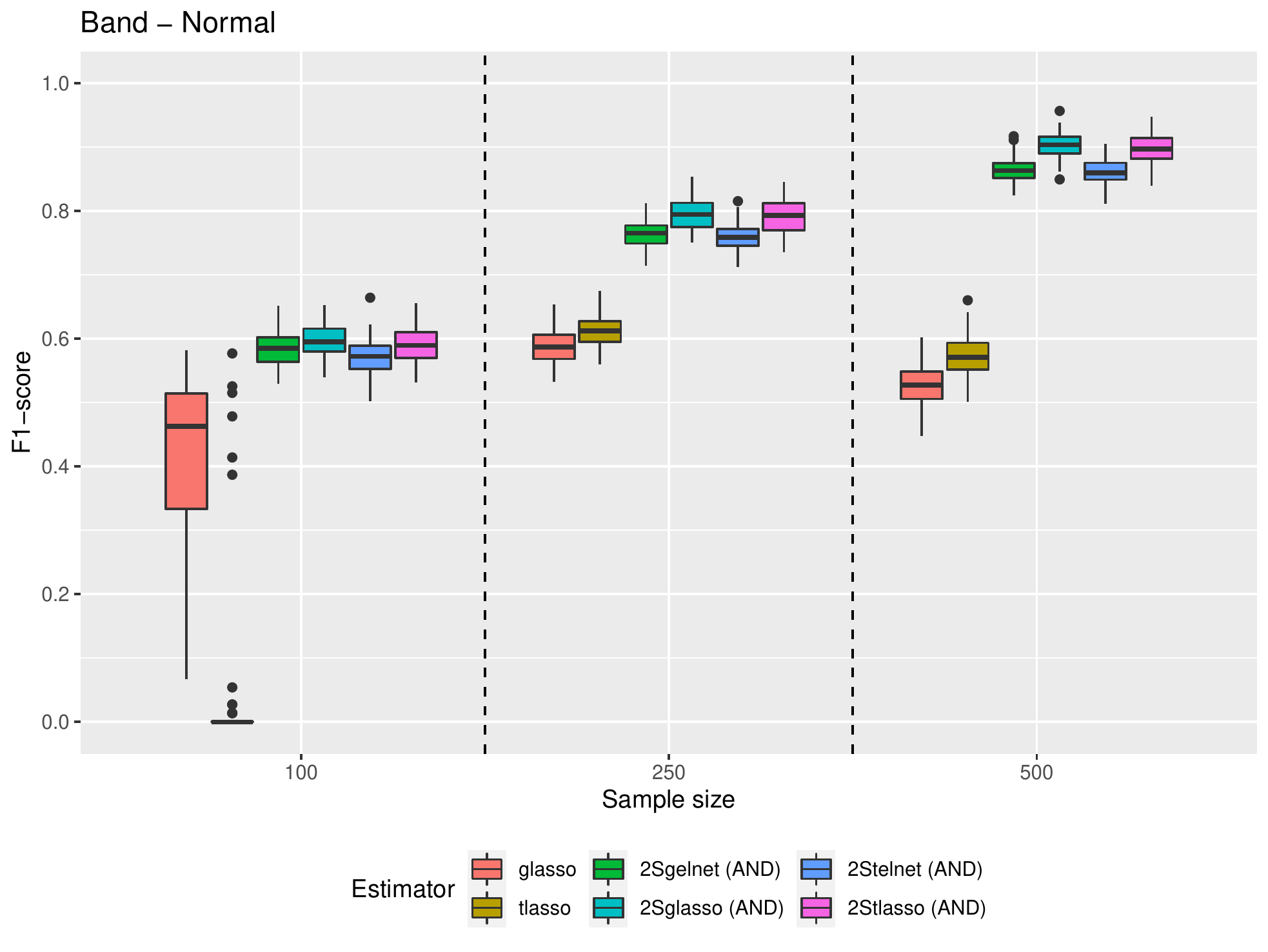}
\end{subfigure}
\begin{subfigure}{0.4\textwidth}
  \includegraphics[width=1\linewidth, angle = 0]{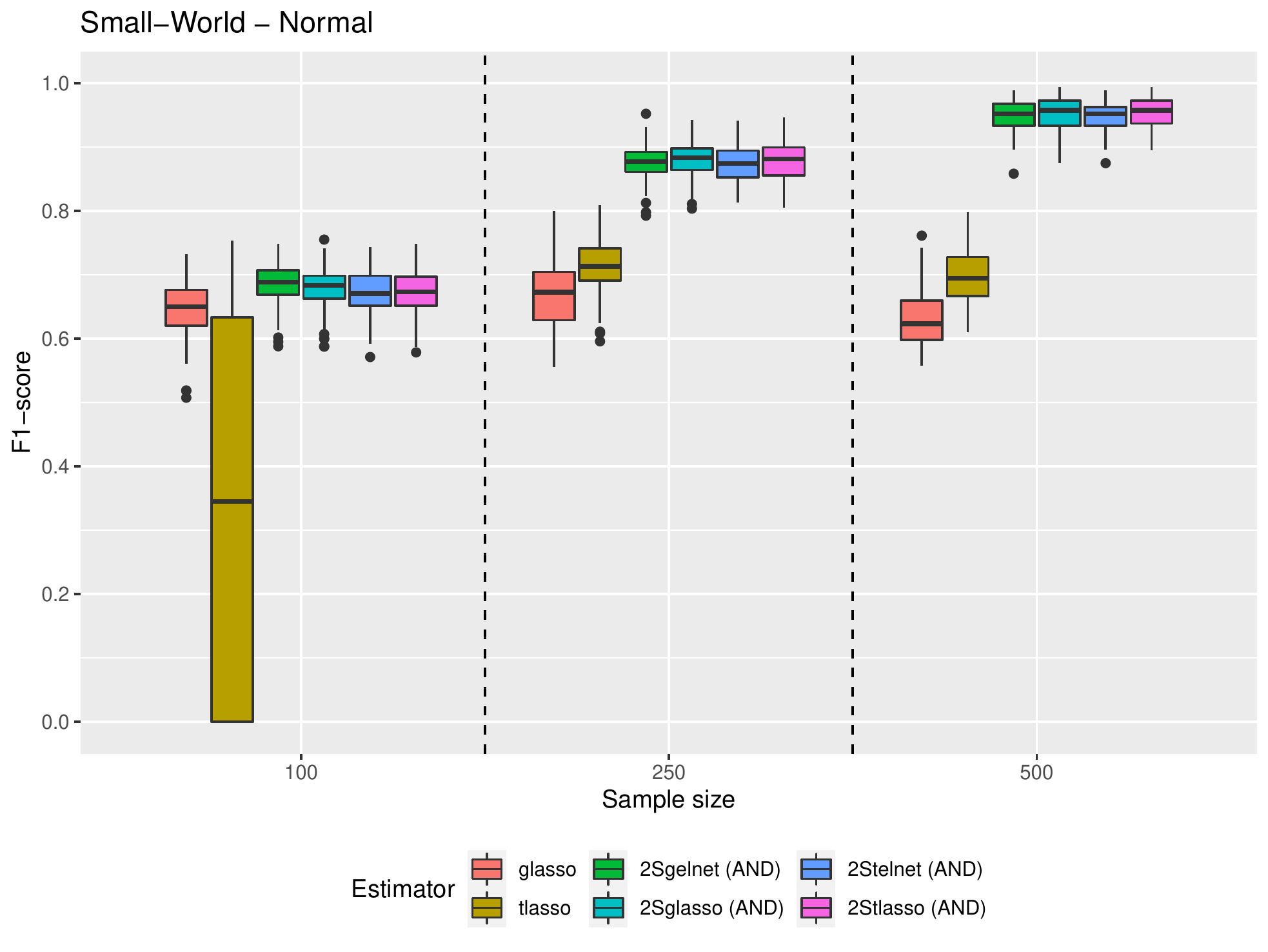}
\end{subfigure}
\newline
\centering
\begin{subfigure}{0.4\textwidth}
  \includegraphics[width=1\linewidth, angle = 0]{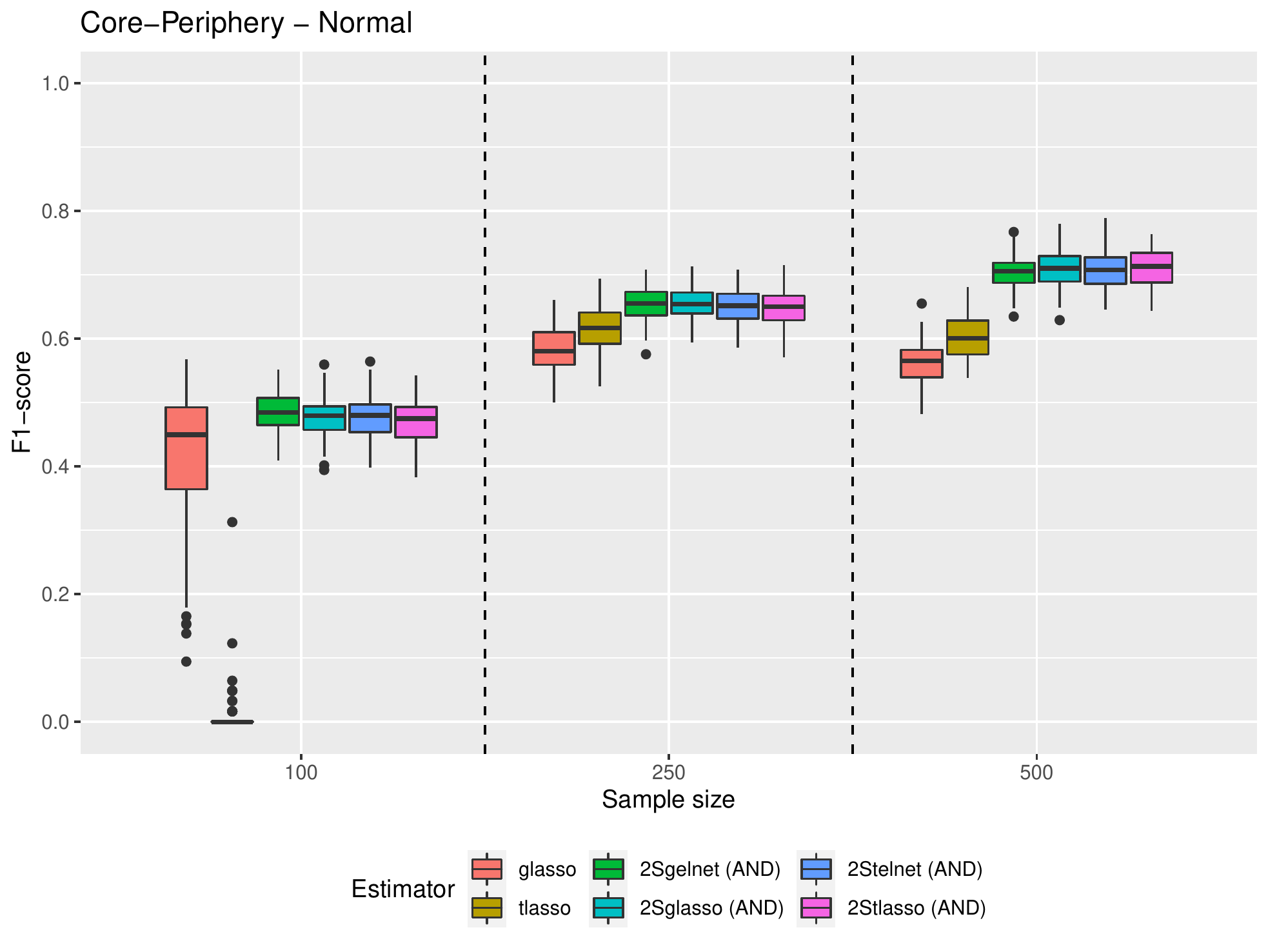}
\end{subfigure}
\caption{F$_1$-score - Normal distribution}
\label{fig:f1snorm}
\end{figure}

\begin{figure}[h]
\begin{subfigure}{0.4\textwidth}
  \includegraphics[width=1\linewidth, angle = 0]{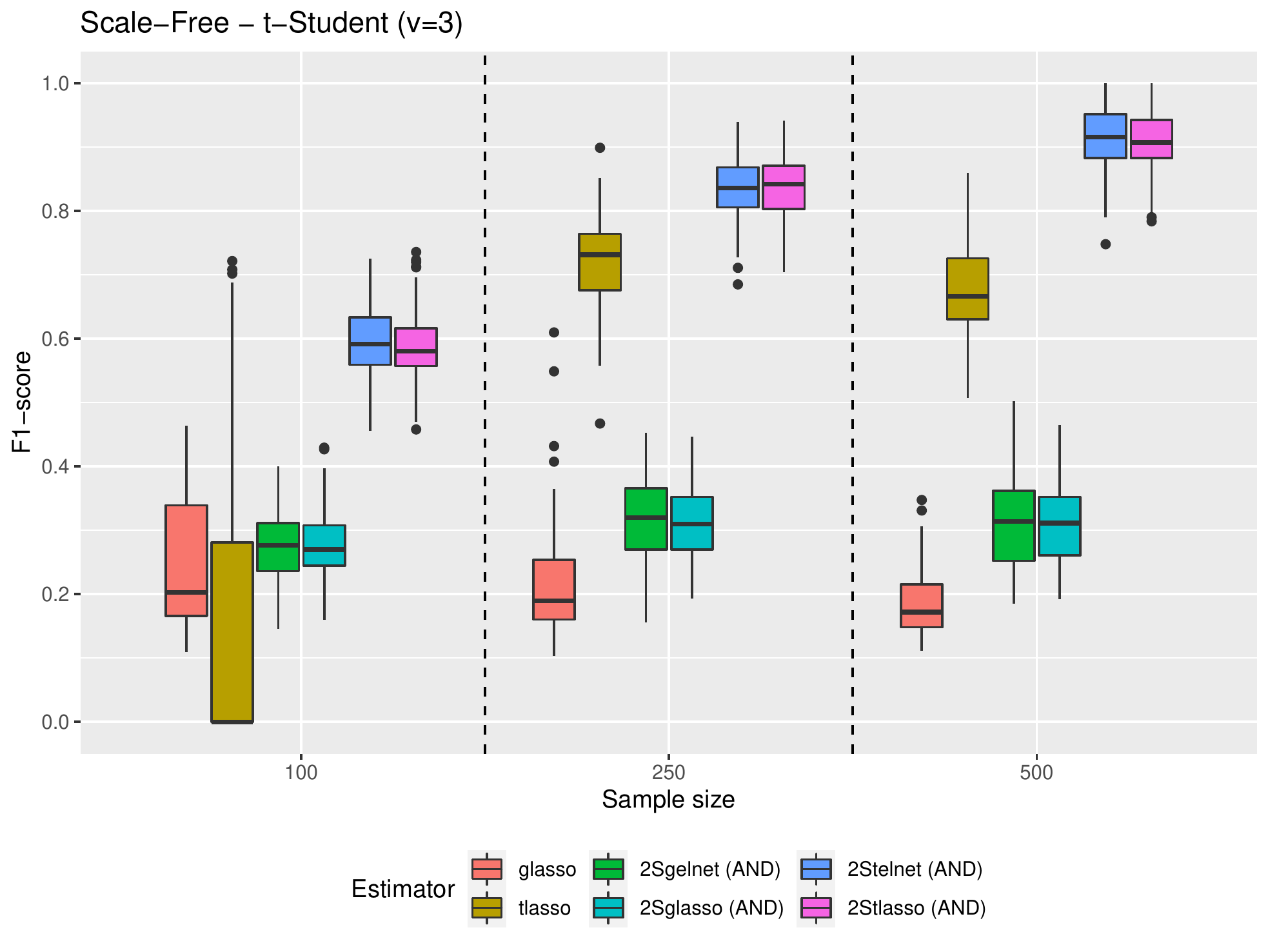}
\end{subfigure}
\begin{subfigure}{0.4\textwidth}
  \includegraphics[width=1\linewidth, angle = 0]{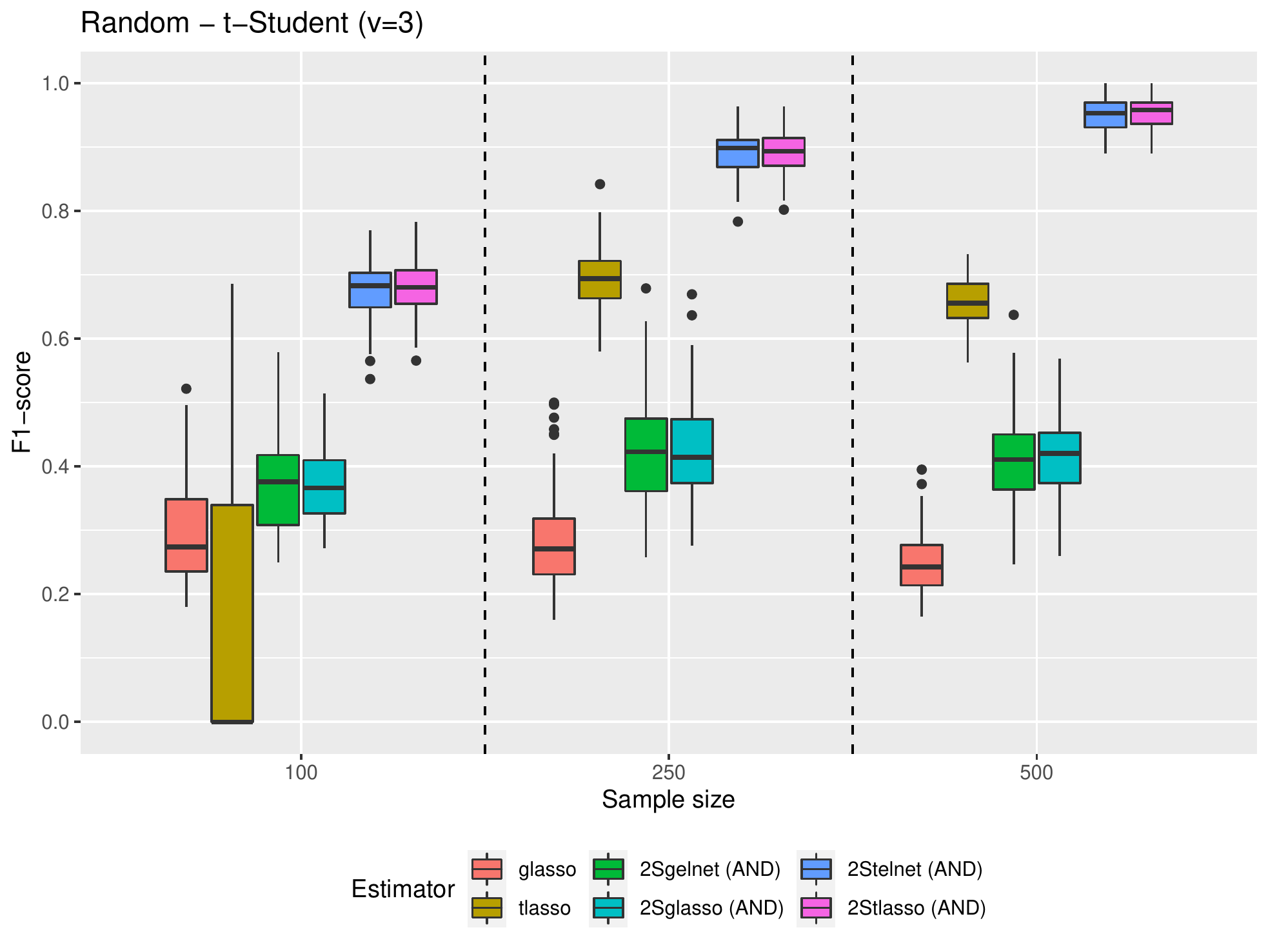}
\end{subfigure}
\newline
\begin{subfigure}{0.4\textwidth}
  \includegraphics[width=1\linewidth, angle = 0]{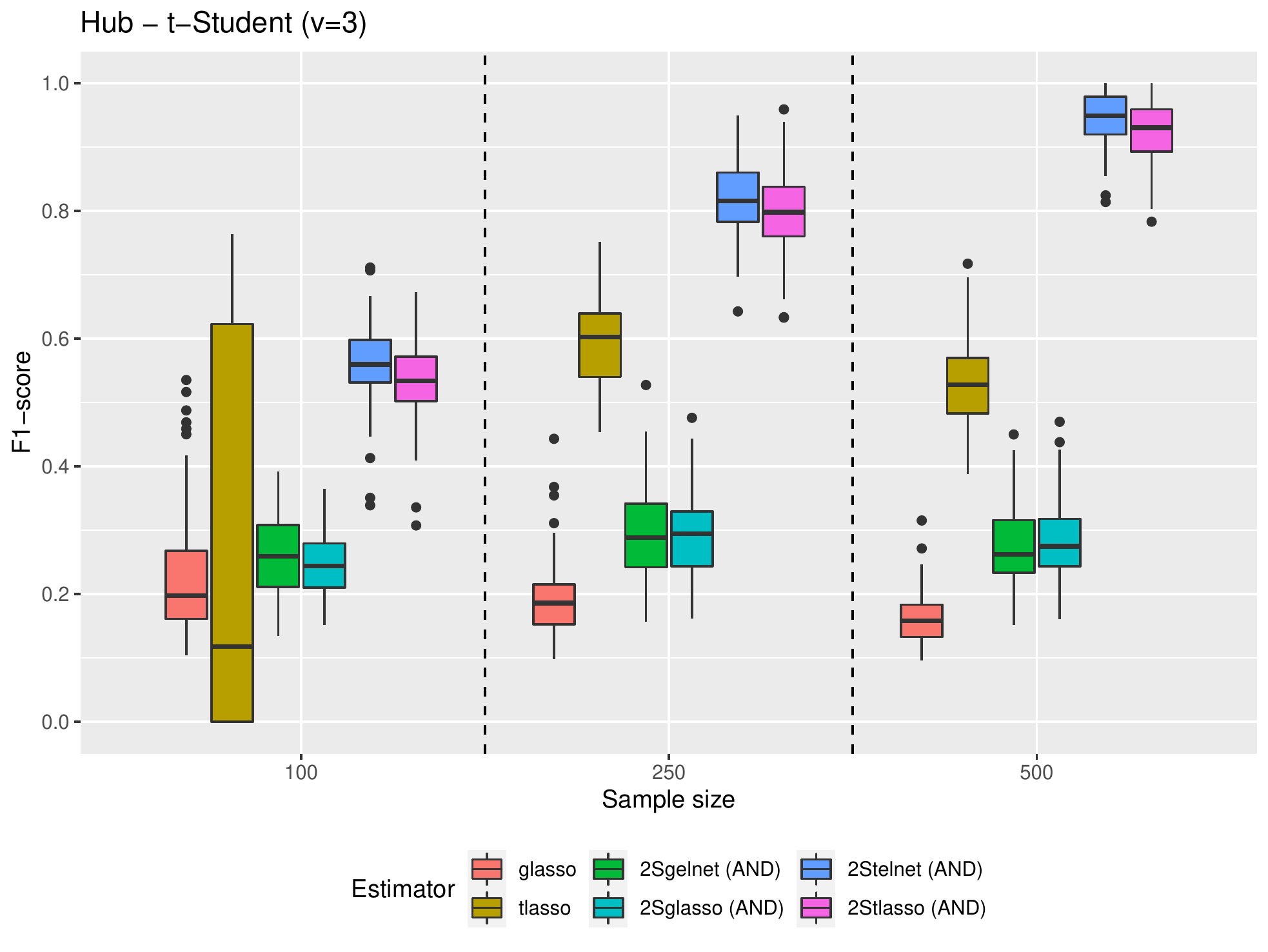}
\end{subfigure}
\begin{subfigure}{0.4\textwidth}
  \includegraphics[width=1\linewidth, angle = 0]{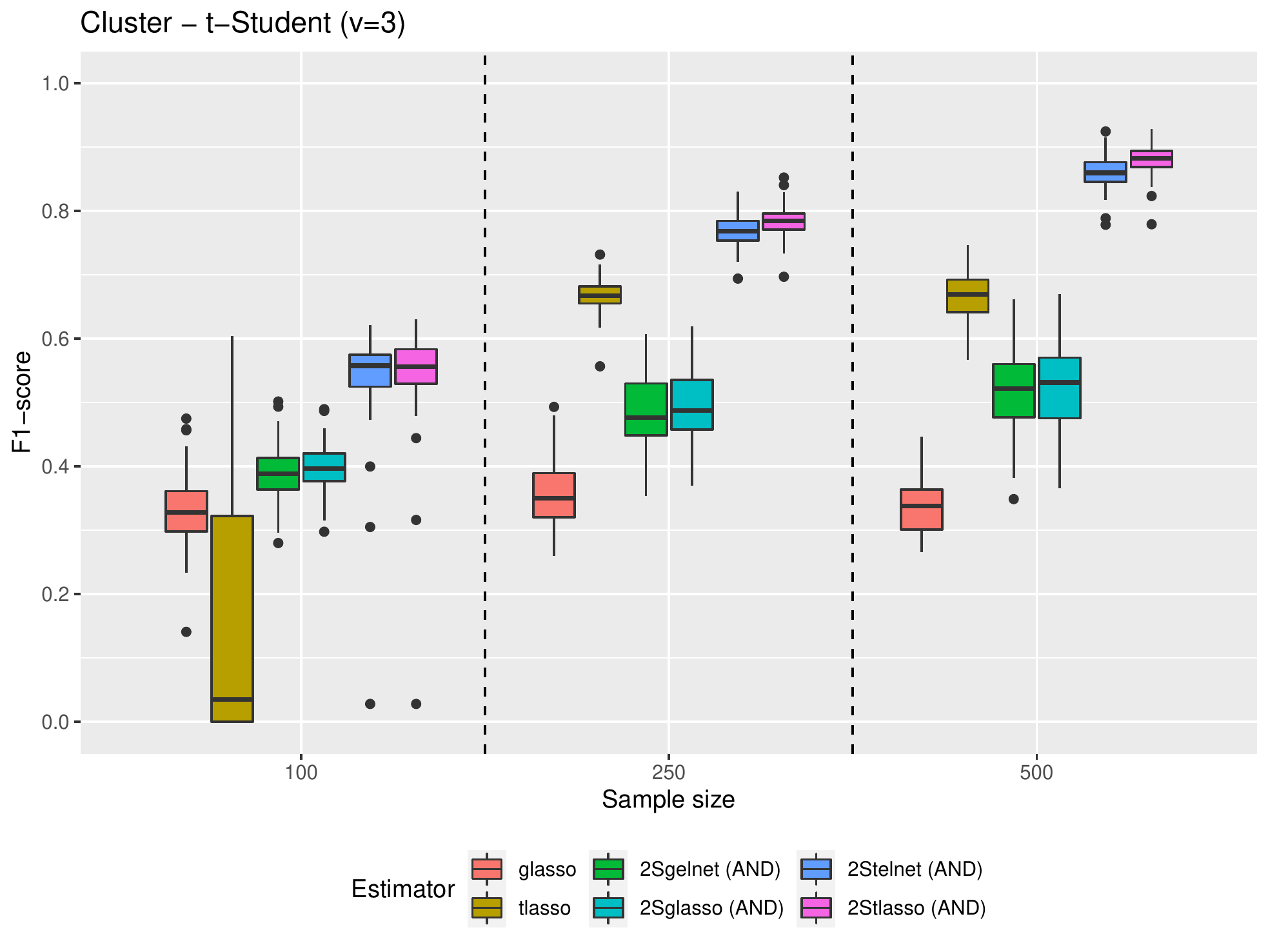}
\end{subfigure}
\newline
\begin{subfigure}{0.4\textwidth}
  \includegraphics[width=1\linewidth, angle = 0]{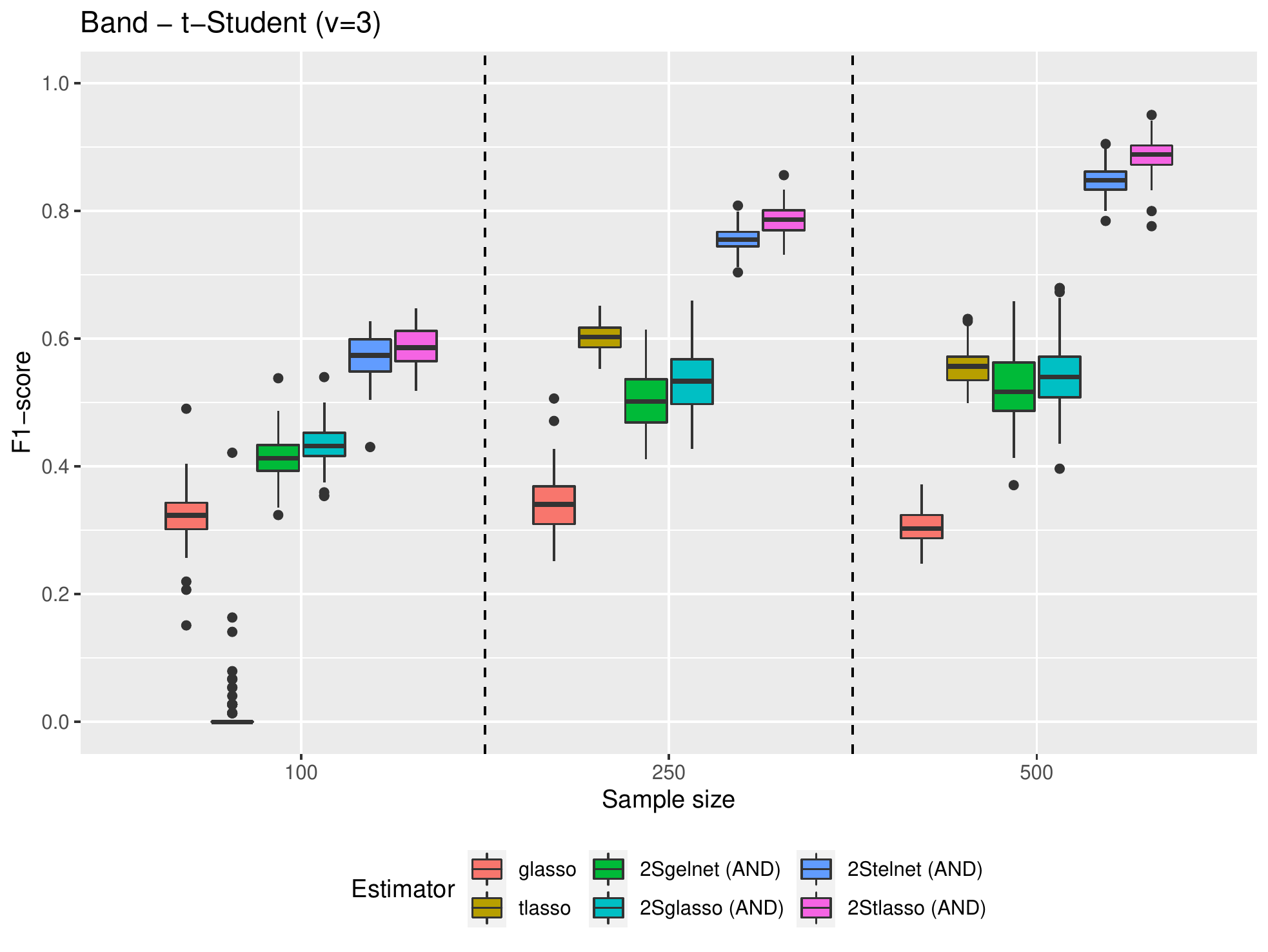}
\end{subfigure}
\begin{subfigure}{0.4\textwidth}
  \includegraphics[width=1\linewidth, angle = 0]{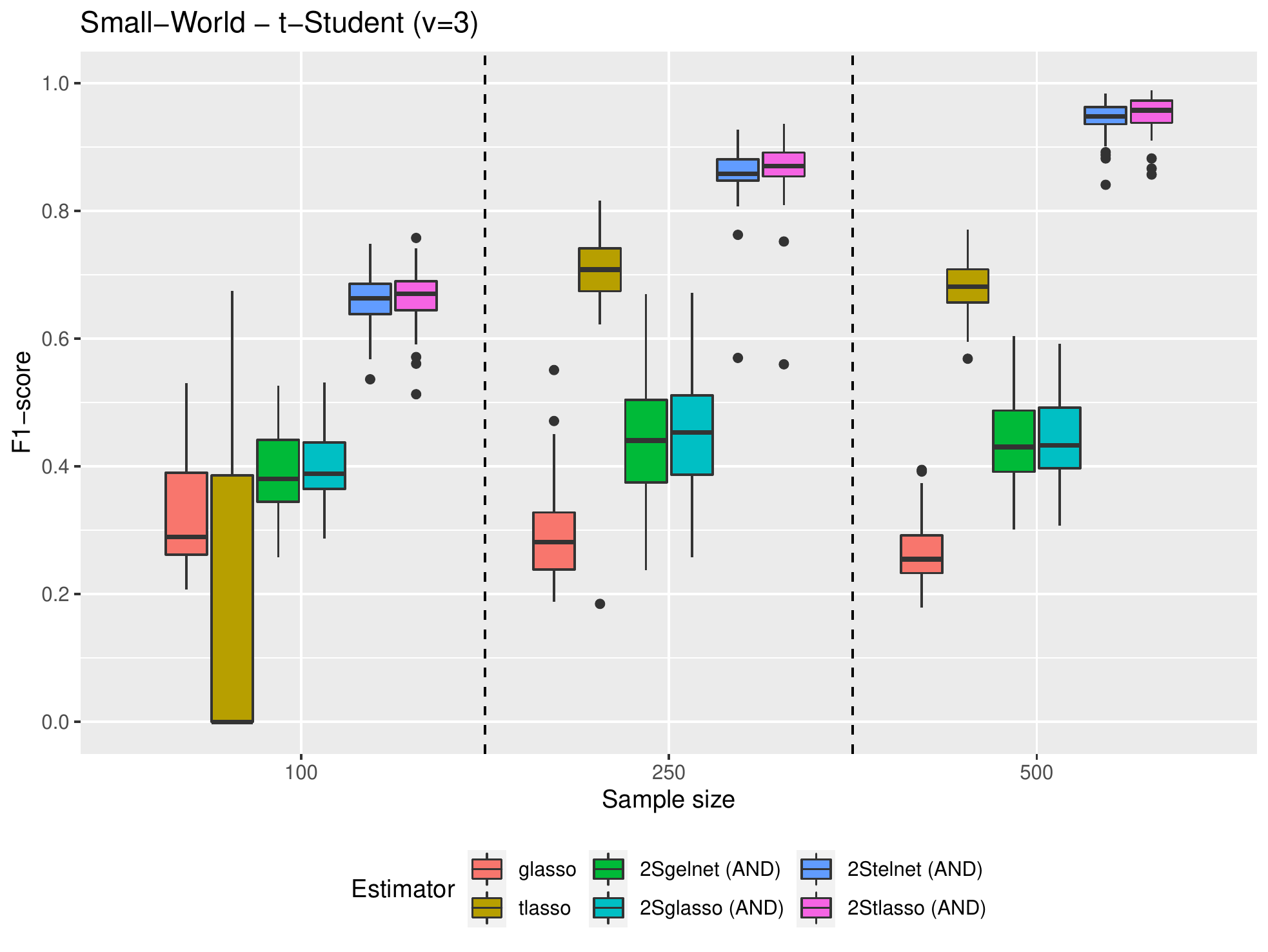}
\end{subfigure}
\newline
\centering
\begin{subfigure}{0.4\textwidth}
  \includegraphics[width=1\linewidth, angle = 0]{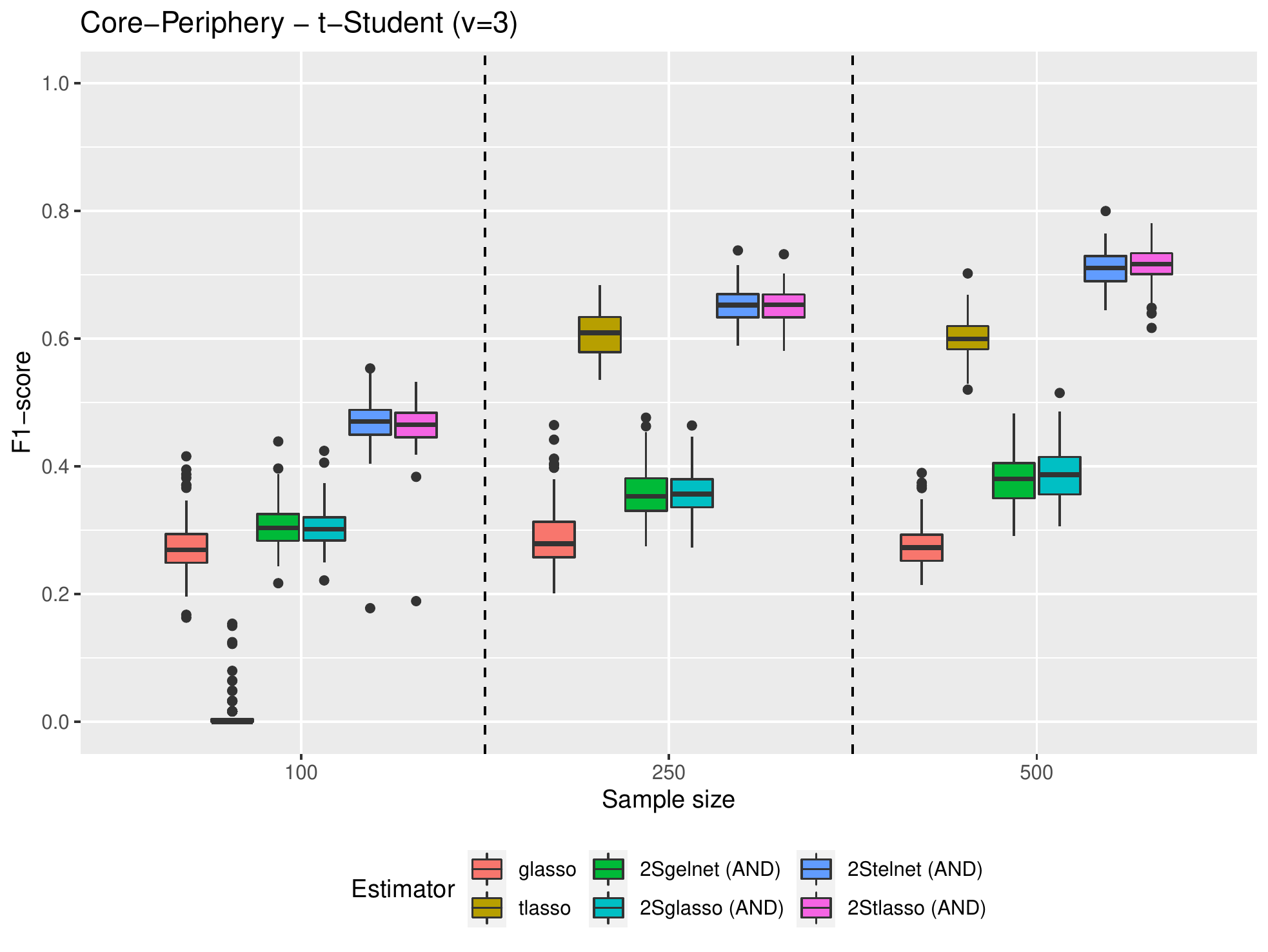}
\end{subfigure}
\caption{F$_1$-score - t-Student distribution with $\nu=3$}
\label{fig:f1ststd3}
\end{figure}

\begin{figure}[h]
\begin{subfigure}{0.4\textwidth}
  \includegraphics[width=1\linewidth, angle = 0]{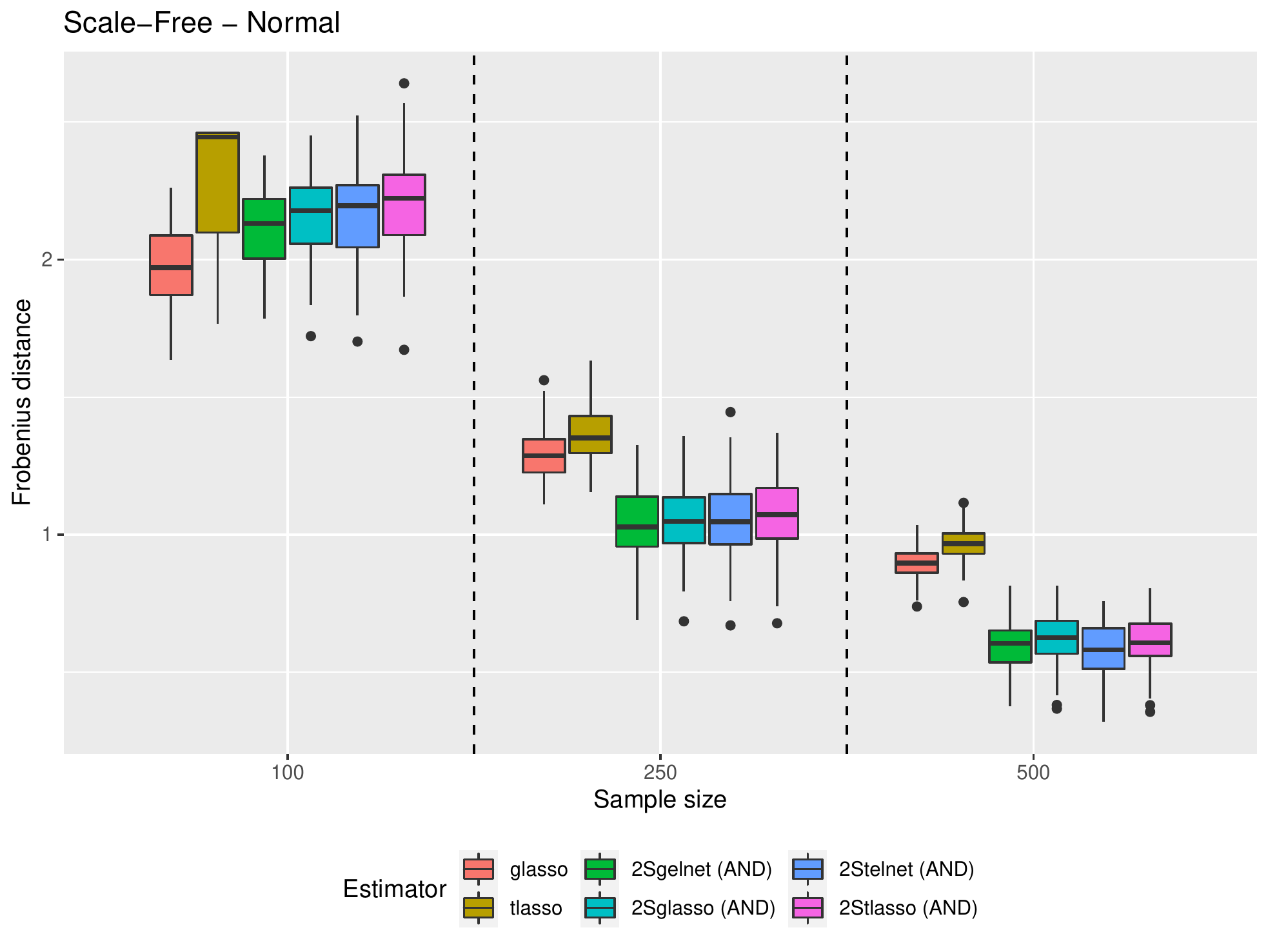}
\end{subfigure}
\begin{subfigure}{0.4\textwidth}
  \includegraphics[width=1\linewidth, angle = 0]{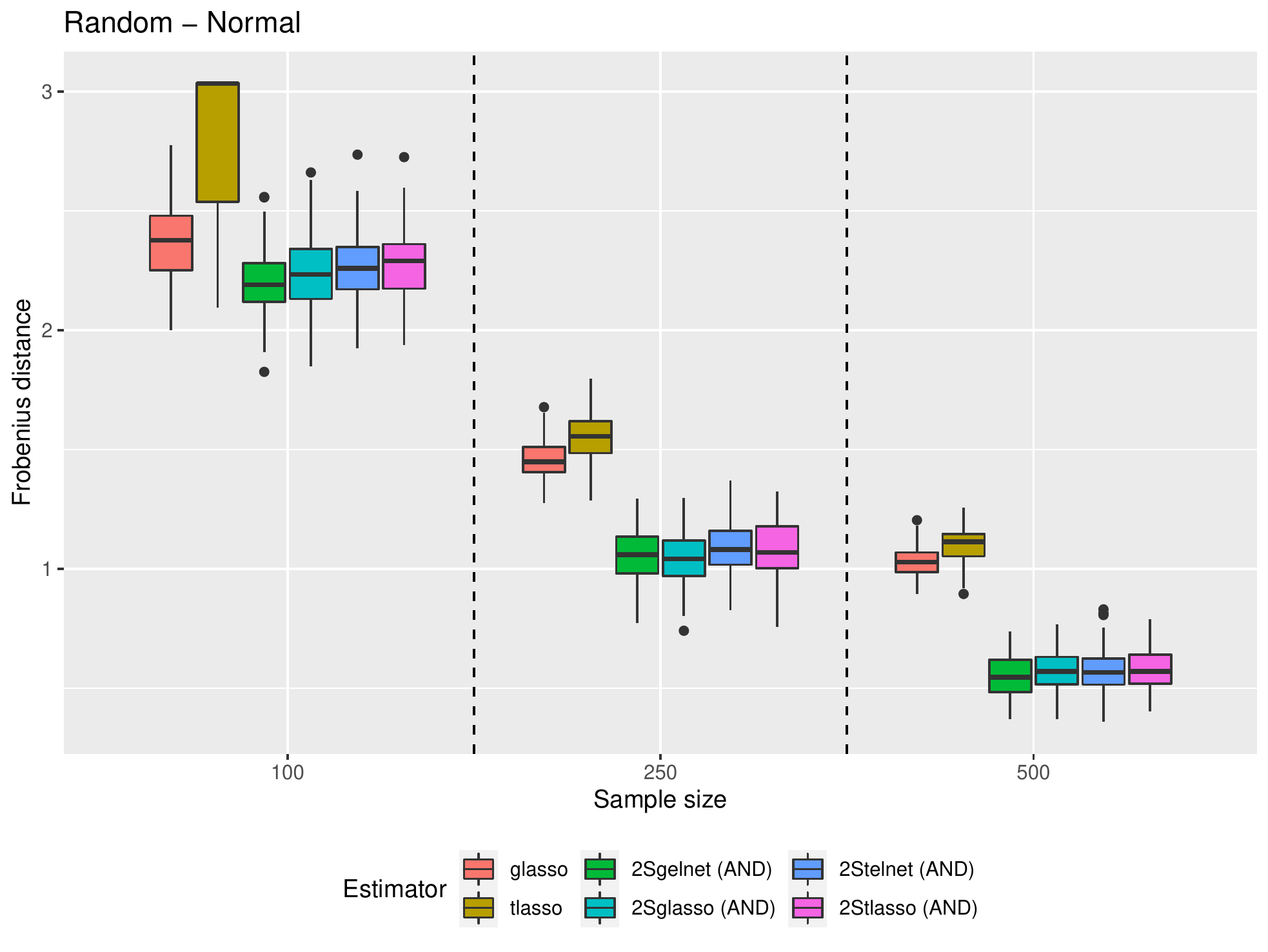}
\end{subfigure}
\newline
\begin{subfigure}{0.4\textwidth}
  \includegraphics[width=1\linewidth, angle = 0]{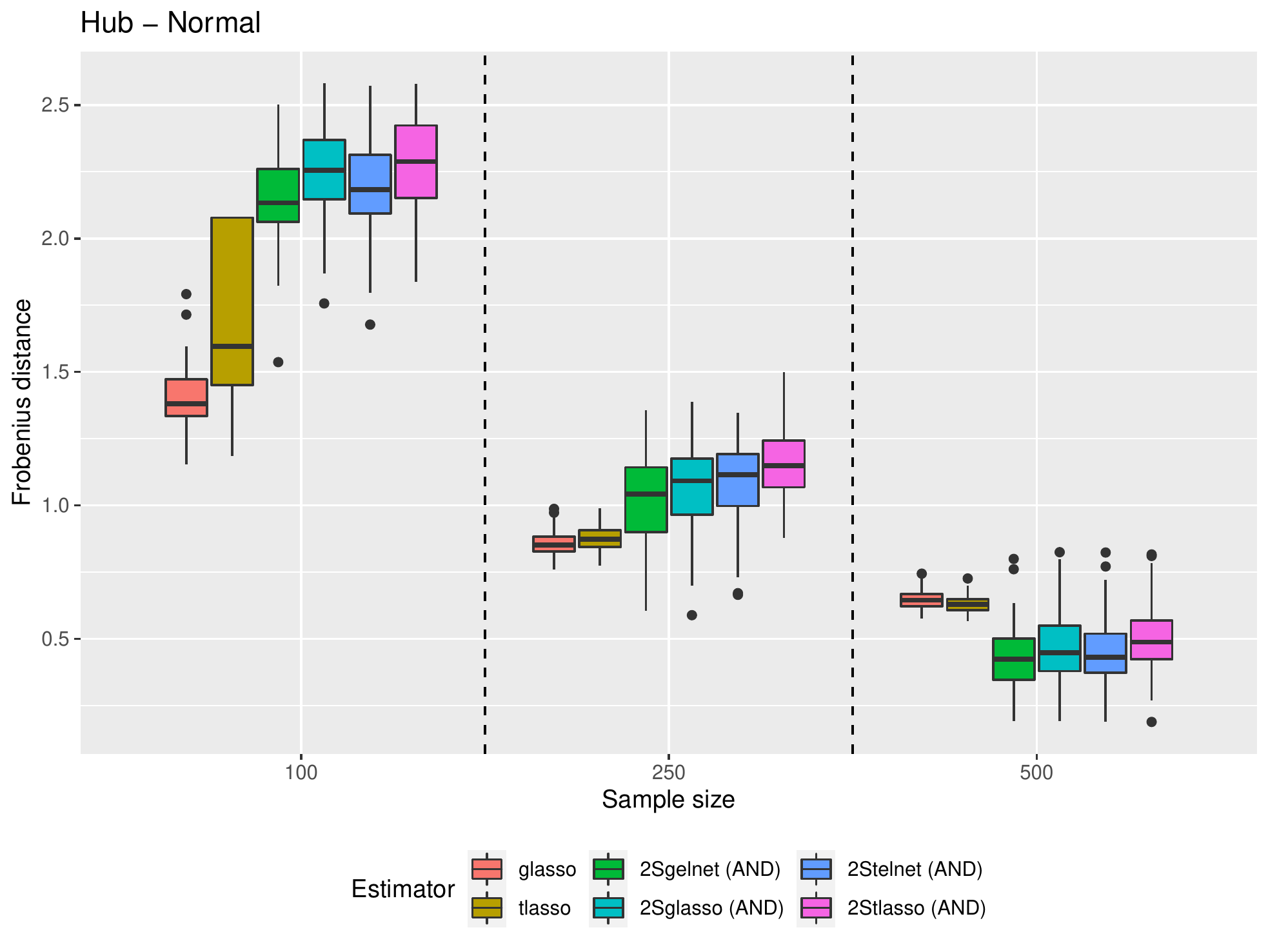}
\end{subfigure}
\begin{subfigure}{0.4\textwidth}
  \includegraphics[width=1\linewidth, angle = 0]{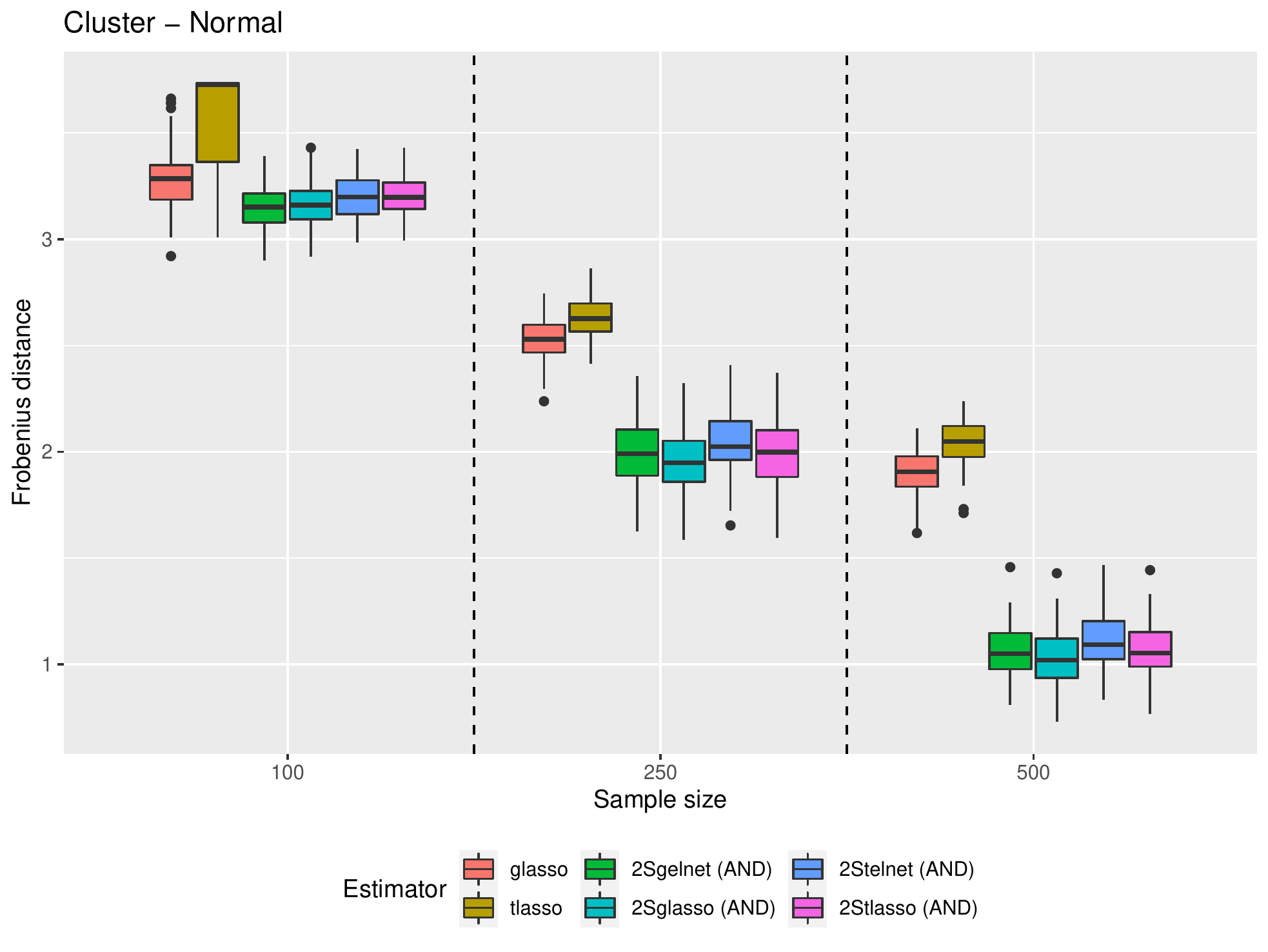}
\end{subfigure}
\newline
\begin{subfigure}{0.4\textwidth}
  \includegraphics[width=1\linewidth, angle = 0]{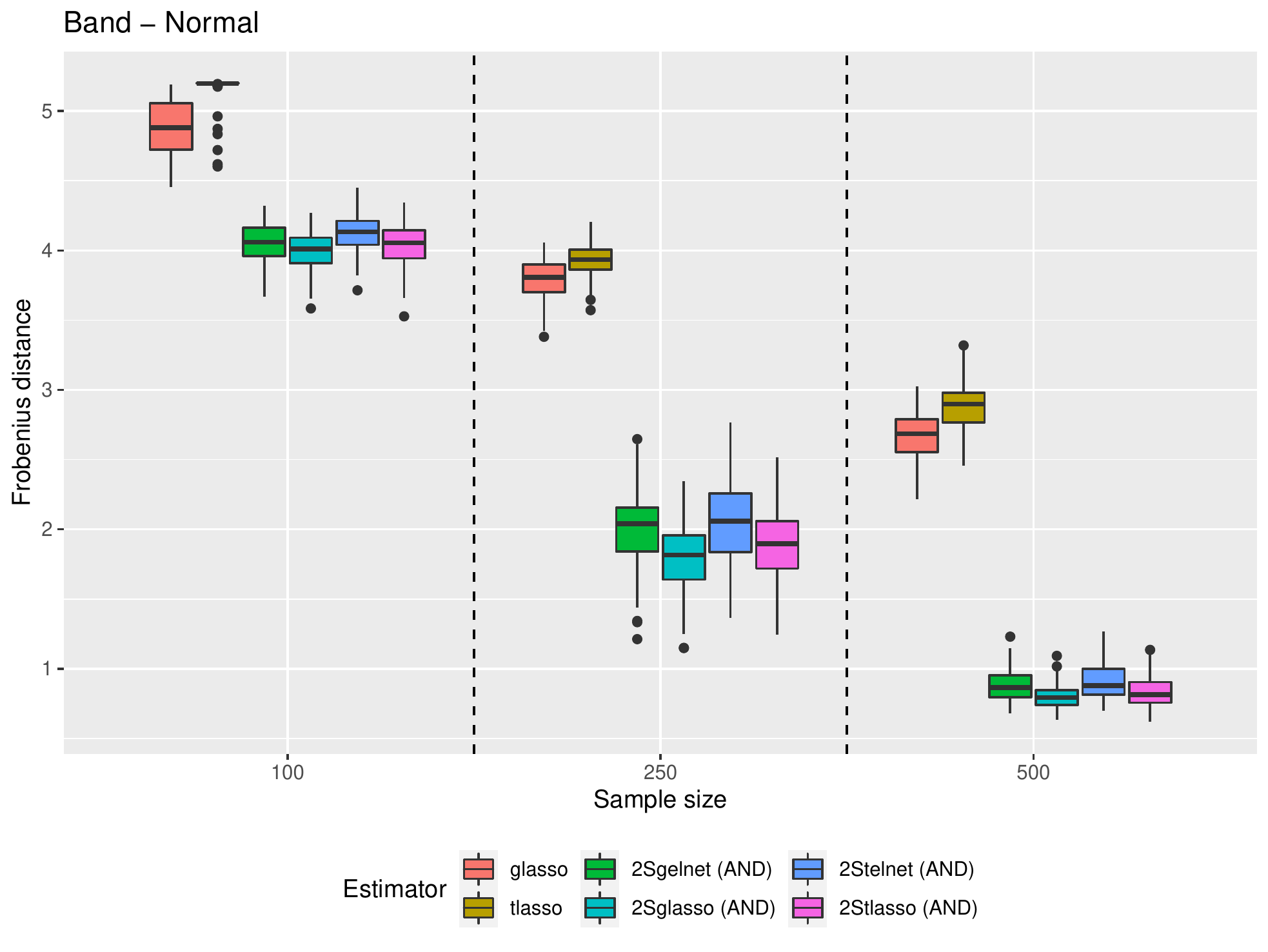}
\end{subfigure}
\begin{subfigure}{0.4\textwidth}
  \includegraphics[width=1\linewidth, angle = 0]{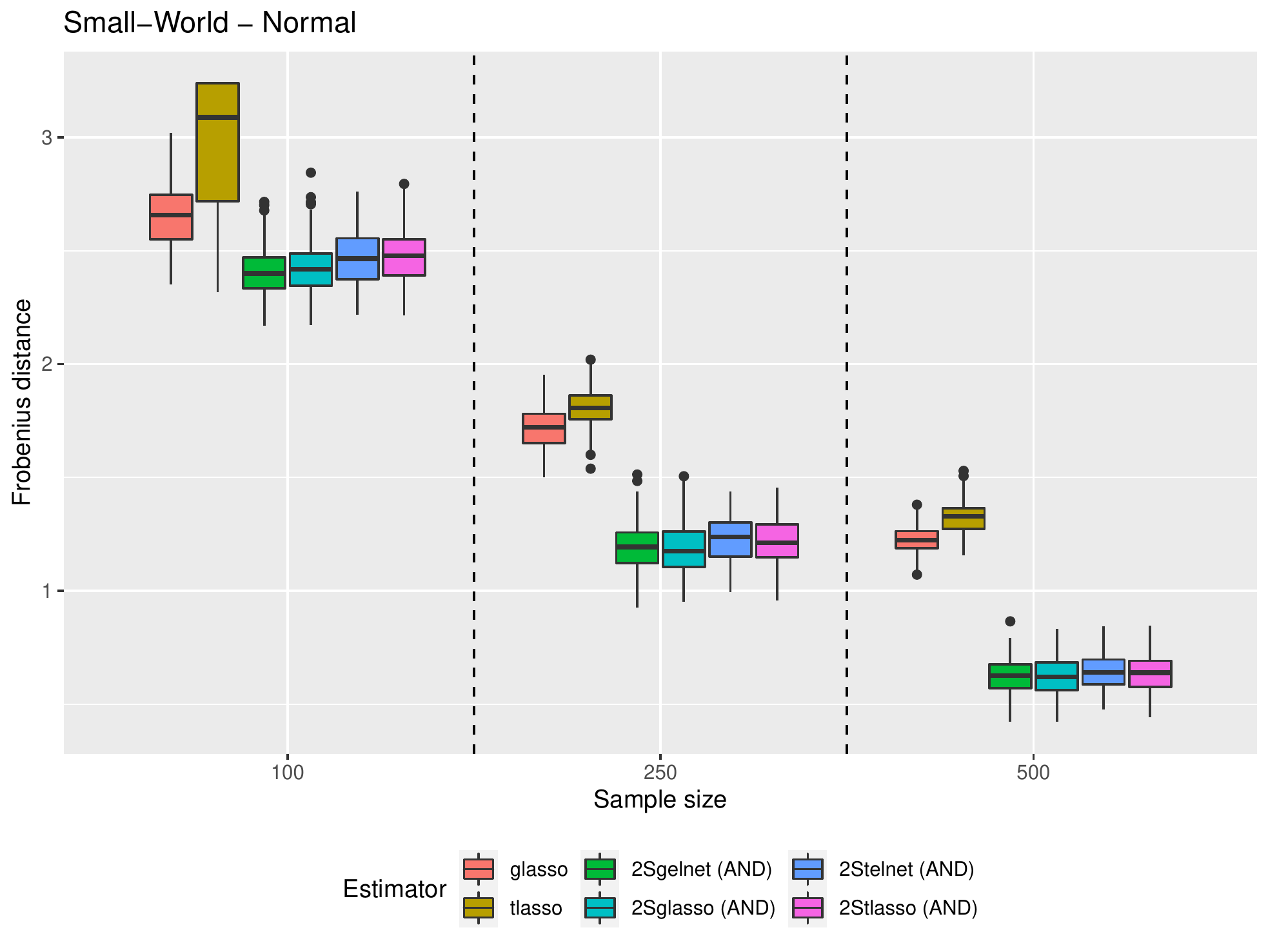}
\end{subfigure}
\newline
\centering
\begin{subfigure}{0.4\textwidth}
  \includegraphics[width=1\linewidth, angle = 0]{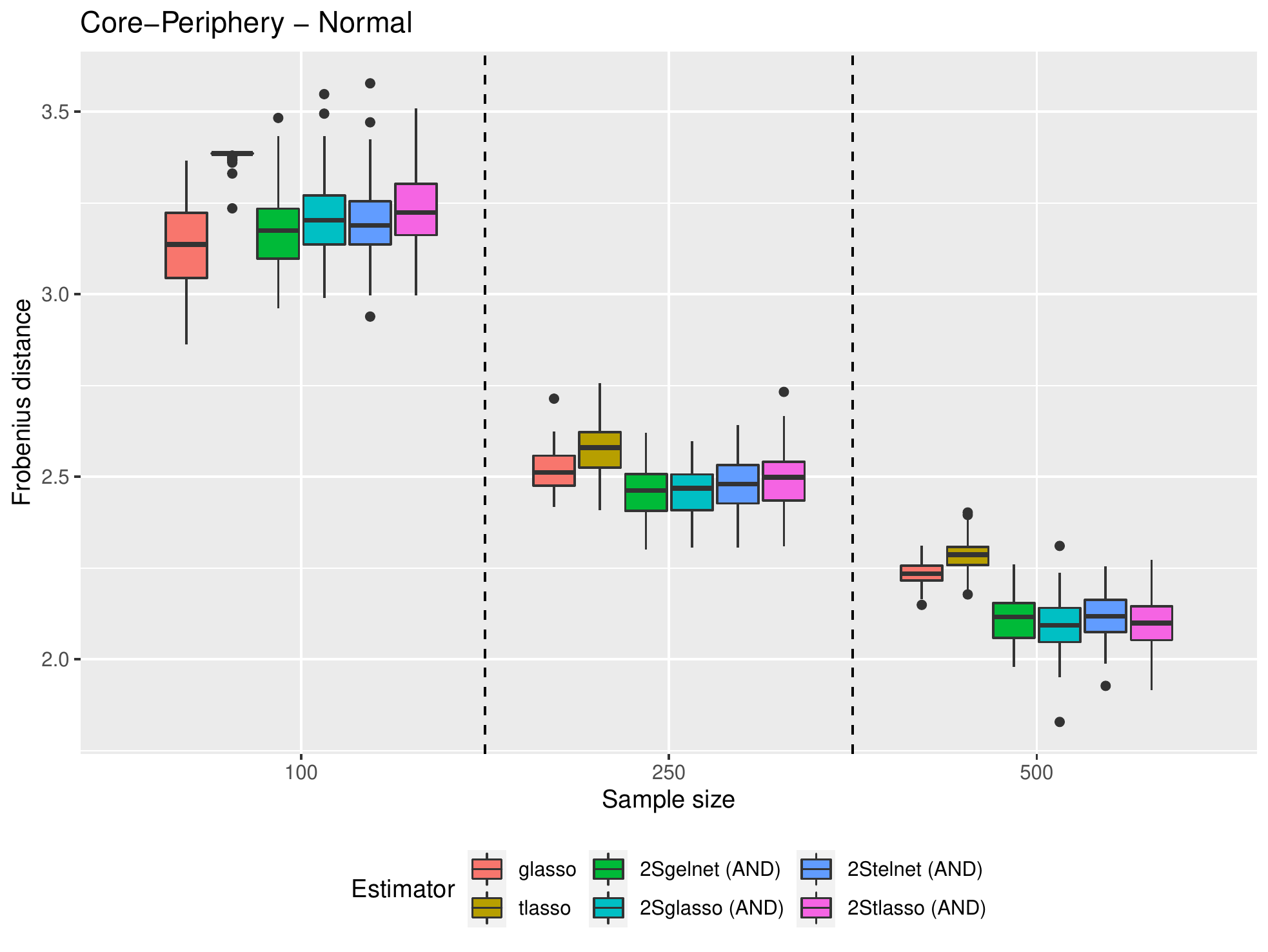}
\end{subfigure}
\caption{Frobenius distance - Normal distribution}
\label{fig:fdpcnorm}
\end{figure}

\begin{figure}[h]
\begin{subfigure}{0.4\textwidth}
  \includegraphics[width=1\linewidth, angle = 0]{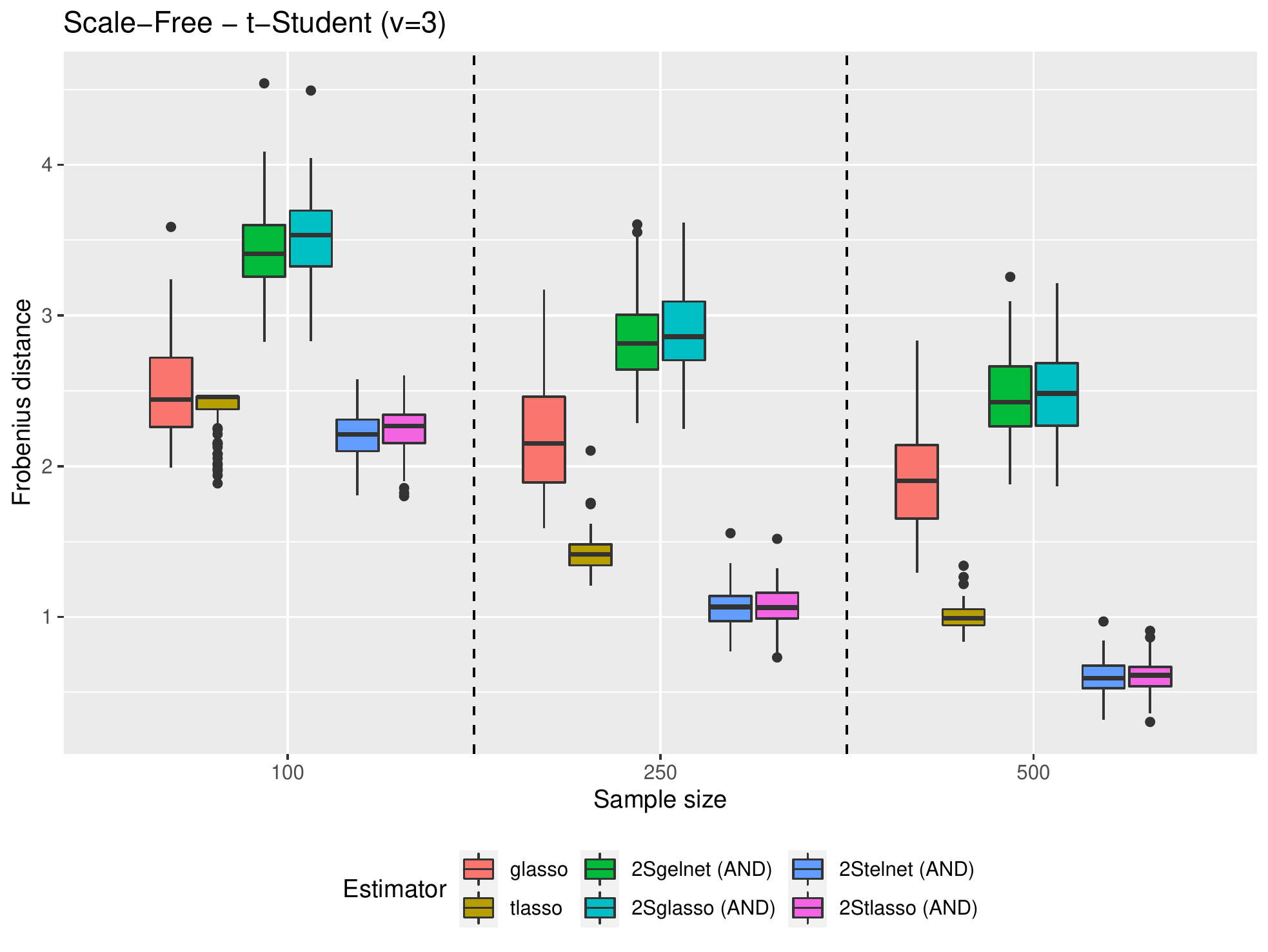}
\end{subfigure}
\begin{subfigure}{0.4\textwidth}
  \includegraphics[width=1\linewidth, angle = 0]{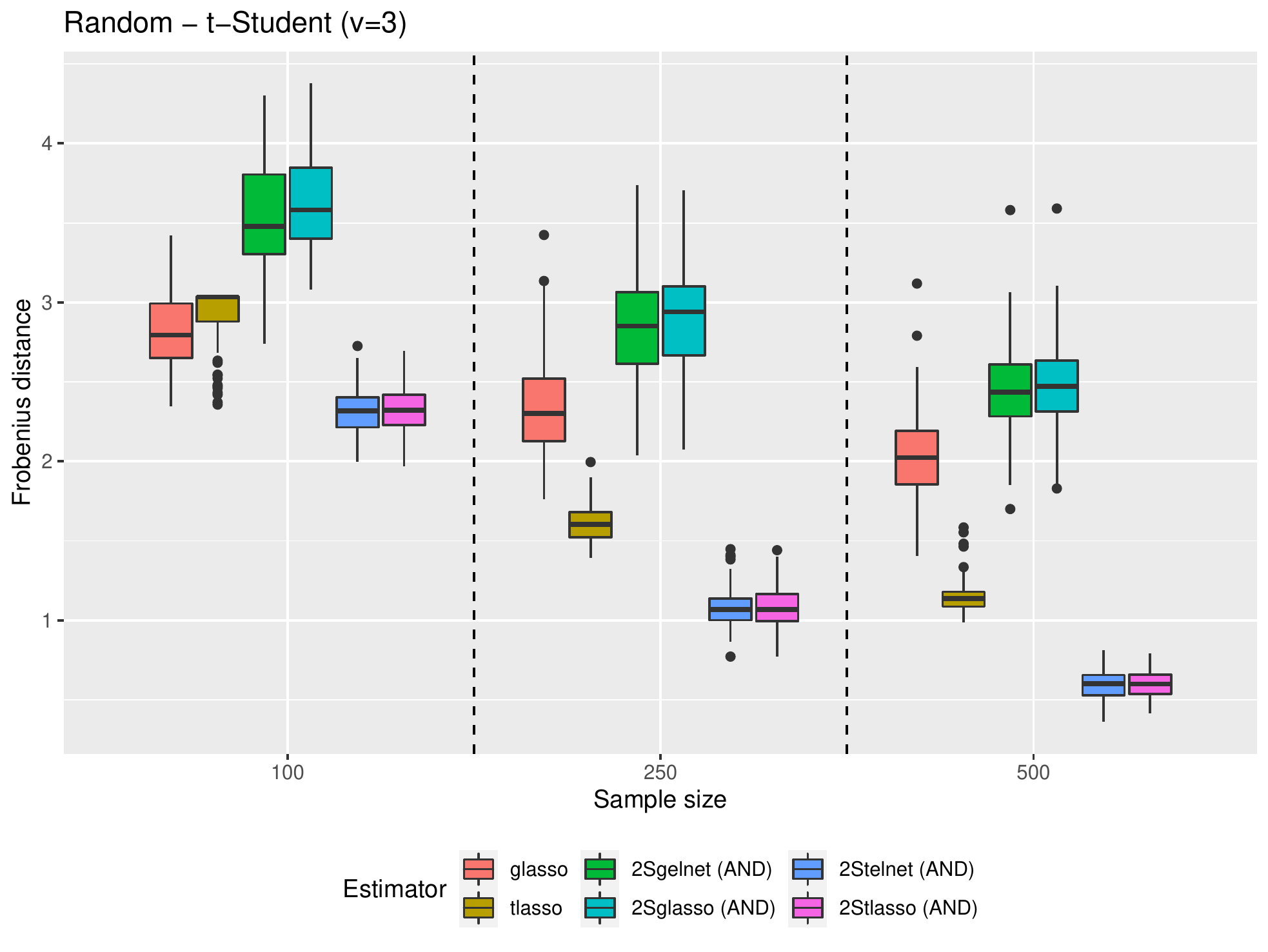}
\end{subfigure}
\newline
\begin{subfigure}{0.4\textwidth}
  \includegraphics[width=1\linewidth, angle = 0]{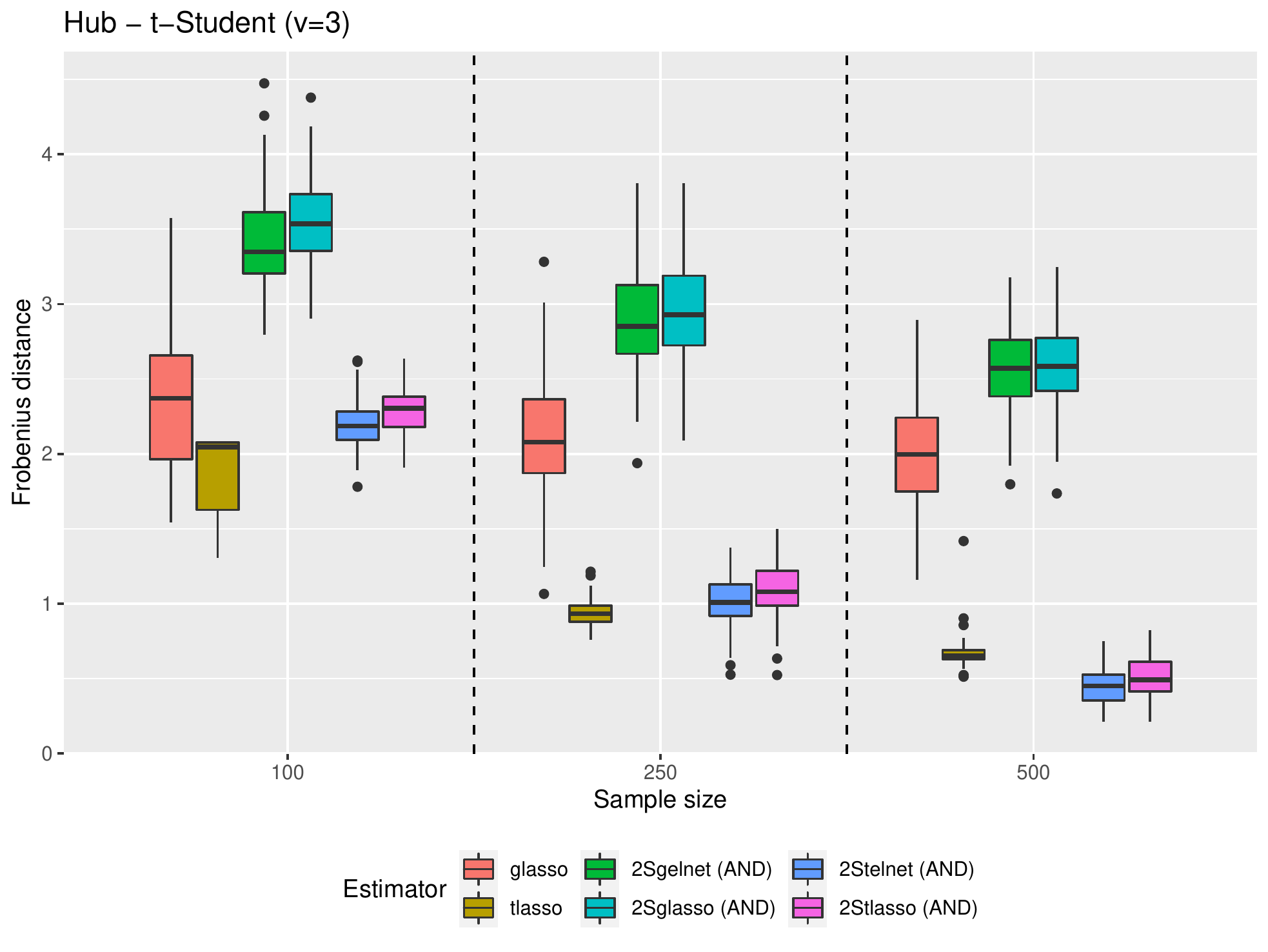}
\end{subfigure}
\begin{subfigure}{0.4\textwidth}
  \includegraphics[width=1\linewidth, angle = 0]{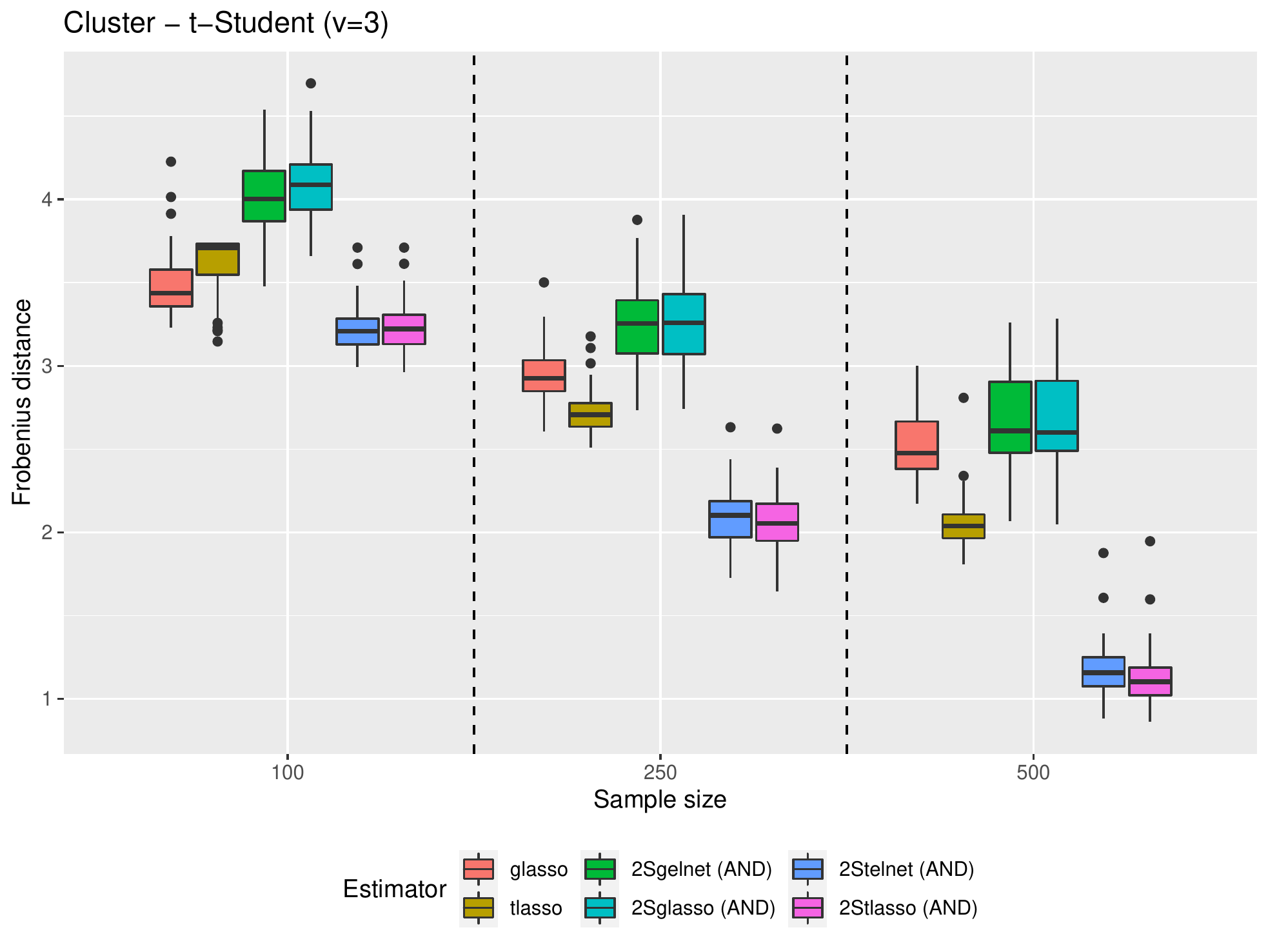}
\end{subfigure}
\newline
\begin{subfigure}{0.4\textwidth}
  \includegraphics[width=1\linewidth, angle = 0]{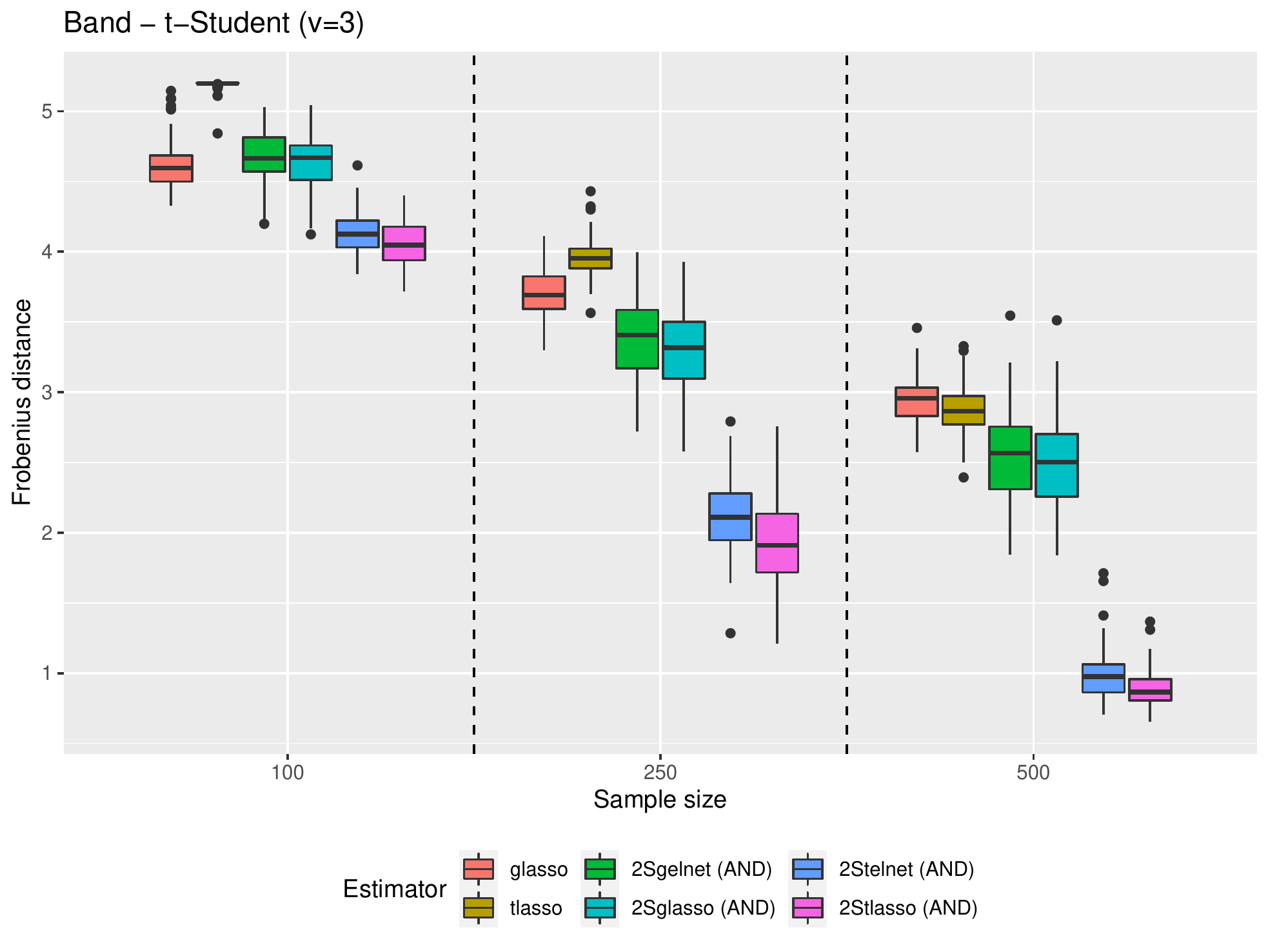}
\end{subfigure}
\begin{subfigure}{0.4\textwidth}
  \includegraphics[width=1\linewidth, angle = 0]{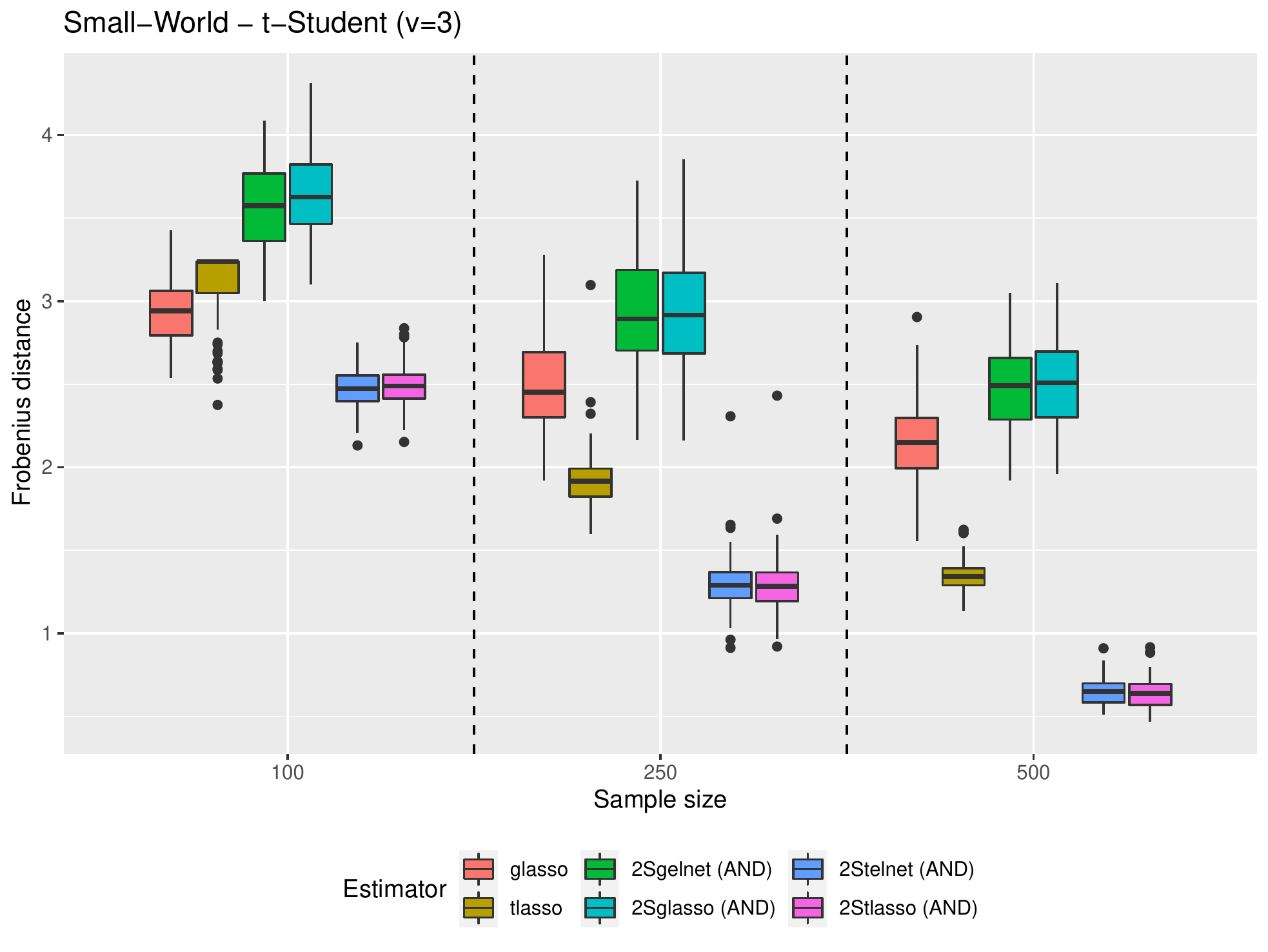}
\end{subfigure}
\newline
\centering
\begin{subfigure}{0.4\textwidth}
  \includegraphics[width=1\linewidth, angle = 0]{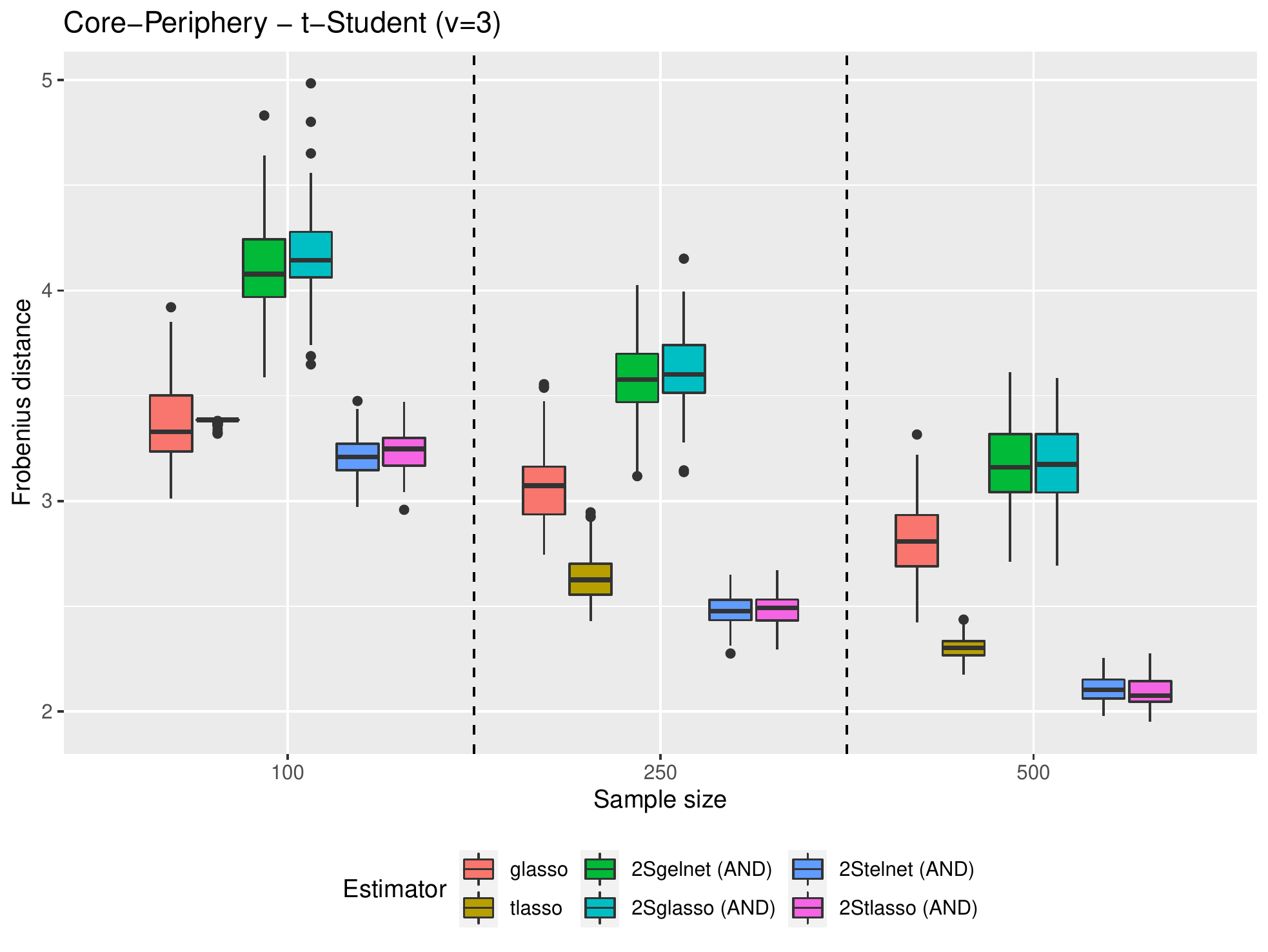}
\end{subfigure}
\caption{Frobenius distance - t-Student distribution with $\nu=3$}
\label{fig:fdpctstd3}
\end{figure}

The box plots of F$_1$-score are reported in Figures \ref{fig:f1snorm}, \ref{fig:f1ststd3}, \ref{fig:f1ststd20} and \ref{fig:f1scontnorm}. Figures \ref{fig:fdpcnorm}, \ref{fig:fdpctstd3}, \ref{fig:fdpctstd20} and \ref{fig:fdpccontnorm} show the box plots for the Frobenius distance between the theoretical and estimated partial correlation matrices. In Appendix \ref{sec:app2}, there are the average values of the optimal $\lambda$ selected. Notice that with \emph{2Sgelnet} and \emph{2Stelnet}, the average value of $\lambda$ with $\alpha=0.5$ is close to two times the mean value of with $\alpha=1$. This suggests that the overall strength of the penalty tends to be constant. It could also explain why the performance are quite similar regardless of the value of $\alpha$ chosen.

\clearpage
\clearpage

\clearpage

Figures \ref{fig:f1snorm}, \ref{fig:f1ststd3} and \ref{fig:f1ststd20}, \ref{fig:f1scontnorm} (in Appendix \ref{sec:app4}) show the box plots for F$_1$-score with normal, t-Student ($\nu=3$), t-Student ($\nu=20$) and contaminated normal distributions, respectively. The closer the value of the F$_1$-score is to 1, the better the classification performance of the estimator considered. Box plots are grouped by sample size as reported in the x-axis. Notice how the behavior of \emph{tlasso} when $n=100$ is quite different from other estimators and from the performance of \emph{tlasso} with $n=250,500$. This is due to the fact that the optimal model selected using BIC criterion with the smallest sample size is often the null model (or close to it), that is the models with zero edges or a diagonal precision matrix. The observed behavior suggests that the BIC criterion with \emph{tlasso} and a small sample is problematic. This is not to the case with the proposed estimator \emph{2Stelnet} (and \emph{2Stlasso}).

When the distribution of data is multivariate Gaussian (Figure \ref{fig:f1snorm}), we observe that the 2-stage procedures compete with \emph{glasso} when $n=100$. Depending on the underlying network structure, there are situations in which they perform better looking at median values. As $n$ gets larger, the 2-stage estimators tend to have a noticeable higher performance with respect to both \emph{glasso} and \emph{tlasso}. We also observe that the estimators based on t-Student distribution (with $\nu=3$) perform quite close to the ones based on multivariate Gaussian. Also \emph{tlasso} seems to be better in median terms than the \emph{glasso} while looking at classification performance. This is a notable fact because it suggests that they are quite robust to this distributional misspecification.

If datasets are randomly generated from multivariate t-Student with $\nu=3$ (Figure \ref{fig:f1ststd3}) the algorithms based on t-Student outperform the others based on Gaussian distribution, as expected. The only exception is the \emph{tlasso} when $n=100$. The \emph{2Stelnet} and \emph{2Stlasso} estimators perform the best, followed by \emph{tlasso} when $n=250,500$. We also observe that \emph{2Sgelnet} and \emph{2Sglasso} are better estimators than \emph{glasso} in this case. The relative increase in performance depends on the underlying structure, but we notice that these algorithms based on Gaussian distribution are not that robust when data are from an heavy-tailed distribution (i.e. t-Student distribution, $\nu=3$).

In general, when the degrees of freedom of the t-Student increase ($\nu=20$, see Figure \ref{fig:f1ststd20} in Appendix \ref{sec:app4}), the perfomances of \emph{glasso}, \emph{2Sgelnet} and \emph{2Sglasso} are, in terms of median, only slightly worse than the estimators based on t-Student. This is in line with the fact that the t-Student distribution tends to the Gaussian one as the degrees of freedom get larger. Indeed, with this setting, the behaviors of the considered estimators are similar to the ones in Figure \ref{fig:f1snorm}. However, note that \emph{tlasso} and \emph{2Stelnet}/\emph{2Stlasso} are still more robust even if they assume $\nu=3$.

The last test case is reported in Figure \ref{fig:f1scontnorm} (Appendix \ref{sec:app4}). Here the distribution is a mixture between two multivariate normal distributions with different variances and correlation structures. This is different from both the distributions assumed by all the algorithms considered. Simulation results suggest that \emph{tlasso}, \emph{2Stelnet} and \emph{2Stlasso} are much more robust to this misspecification than the counterparts based on Gaussian distribution. The best estimators in this situation are \emph{2Stelnet} and \emph{2Stlasso}, followed by \emph{tlasso} if we exclude the problematic case with $n=100$ when it performs the worst. Instead, there is not a clear winner between \emph{2Sgelnet/2Sglasso} and \emph{glasso}. In fact, it mostly depends on the underlying topology.

To sum up, we observe that \emph{2Stelnet} and \emph{2Stlasso} perform quite well in the situations analyzed through simulations, suggesting a good degree of robustness to distributional misspecifications. They also perform quite closely. This finding indicates that the value of $\alpha$, and thus the kind of penalty used, is not particularly relevant, at least in the simulation set-up considered. Depending on the network, one value of $\alpha$ can slightly outperform the other in median terms, but differences seem negligible and thus a particular winner does not emerge. Similar considerations hold true for the differences between \emph{2Sgelnet} and \emph{2Sglasso}. We notice also that some network structures are more difficult to extract. For example, when we look at F$_1$-score, the core-periphery topology is the most difficult to estimate in the majority of the situations. Band and cluster structures are also more difficult to retrieve if compared with the remaining ones.


\clearpage

\clearpage

In Figures \ref{fig:fdpcnorm}, \ref{fig:fdpctstd3} and \ref{fig:f1ststd20}, \ref{fig:f1scontnorm} (in Appendix \ref{sec:app4}) are reported the box plots of Frobenius distance between theoretical and estimated partial correlation matrices. The closer is to 0, the better the estimator's numerical accuracy. The observed behavior of \emph{tlasso} is the consequence of the fact discussed previously: using BIC criterion, the optimal model selected is often the empty, or almost empty, graph. Thus the distribution of this measure tends to be quite skewed for \emph{tlasso}.

Figure \ref{fig:fdpcnorm} shows the results obtained when the distribution of data is multivariate Gaussian. With sample size $n=100$, 2-stage estimators (i.e. \emph{2Sgelnet}/\emph{2Sglasso} and \emph{2Stelnet}/\emph{2Stlasso}) perform in general better or similarly to \emph{glasso} and \emph{tlasso}, with only one exception with the hub structure. The relative better performance of the 2-step procedures increases with $n=250$, even if with hub network they are still outperformed. Only when $n=500$, \emph{2Sgelnet}/\emph{2Sglasso} and \emph{2Stelnet}/\emph{2Stlasso} are always the best estimators. Notice that, also from the point of view of numerical accuracy, \emph{2Stelnet} and \emph{2Stlasso} (with $v=3$) are robust to multivariate normal data. In facts, they perform quite closely to \emph{2Sgelnet} and \emph{2Sglasso}.

In Figure \ref{fig:fdpctstd3}, there are the box plots of the Frobenius distance for datasets from t-Student distribution with $\nu=3$. In this case \emph{tlasso}, \emph{2Stelnet} and \emph{2Stlasso} assume correctly the underlying distribution. As expected, we observe that these procedures outperform the corresponding ones based on Gaussian distribution in general. This is not always the case with \emph{tlasso} when $n=100$ and also with $n=250$ and band topology. In almost all cases, \emph{2Stelnet}/\emph{2Stlasso} perform the best. There are only two exceptions with hub structure and $n=100,250$. Instead, \emph{2Sgelnet} and \emph{2Sglasso} perform the worst in the majority of the situations considered. Only with band network, they are competitive with \emph{glasso} and \emph{tlasso}. Thus, the 2-step procedures based on Gaussian distribution are not much robust to model misspecification.

Figure \ref{fig:fdpctstd20} (Appendix \ref{sec:app4}) reports the box plots for the t-Student distribution with $\nu=20$ degrees of freedom. Similarly to the situation of classification performance, as the number of degrees of freedom increases, the procedures based on Gaussian distribution achieve similar performances to the ones based on t-Student with $v=3$. We observe that \emph{2Stelnet} and \emph{2Stlasso} tend to slightly outperform, in median terms, \emph{2Sgelnet} and \emph{2Sglasso} respectively, while \emph{tlasso} is outperformed by \emph{glasso} in most of the cases. Only with hub structure and $n=250, 500$, it is slightly better if we look at median values.

Finally, in Figure \ref{fig:f1scontnorm} (Appendix \ref{sec:app4}) we compare the performance of the estimators when data are from a contaminated normal distribution. The procedures that assume multivariate t-Student distribution with $v=3$ are much more robust to the model misspecification than the counterparts based on multivariate Gaussian distribution. There is only one exception with $n=100$, where \emph{tlasso} performs similarly to \emph{glasso}. We observe that, for the largest sample size considered ($n=500$), \emph{2Stelnet}/\emph{2Stlasso} are always the most numerically accurate estimators. This is also true for $n=250$, with only one exception with the hub structure. For the smallest sample size ($n=100$), they are the best estimators in four out of seven networks (i.e. random, cluster, band, small-world). When comparing \emph{glasso} and \emph{2Sgelnet}/\emph{2Sglasso} there is not a clear winner in the cases analyzed. We only observe that the 2-stage procedures tend to improve their relative performances as $n$ grows. With $n=500$, they outperfom \emph{glasso} in the majority of cases. Still, they show worse performance than \emph{tlasso} in almost all the situations considered. Only with $n=100$ and band topology, they have comparable performance in terms of median.

In conclusion, similarly to classification performance, \emph{2Stlenet}/\emph{2Stlasso} show relatively good performances, especially with larger sample sizes. They are robust to different distributional misspecifications. Again, a winner between \fmten{elastic net} penalty and pure LASSO penalty in the 2-stage estimators does not emerge from our simulations, neither for the ones based on Gaussian distribution, nor for the ones based on t-Student. In general the performances observed are quite similar, only in few situations one penalty slightly outperforms the other looking at median values: which one is the best depends on sample size and on network.

\clearpage
\section{European banking network} \label{sec:appl}

Inspired by the work of Torri et al. \cite{Torri18}, we use the \emph{2Stelnet} estimator to reconstruct the European banking network in the period 2018-2020 by using daily stock prices of 36 large European banks. The period considered includes the Covid-19 pandemic, thus it is also possible to asses its impact on the evolution of this banking network and extend \cite{Torri18} to include the recent pandemic crisis. To deal with autocorrelation and heteroskedasticity, two common characteristics of financial time series, we fit an AR(1)-GARCH(1,1) model for each of the 36 time series of log-returns. Then, we use the residuals to estimate the partial correlation network both for the single years (i.e. 2018, 2019, 2020) and for the period 2018-2020, using a rolling window of one year with shifts of one month.  We set the value of $\alpha=0.5$ and consider a sequence of 100 exponentially spaced values between $e^{-6}$ and $1.5$ for $\lambda$. We select its best value using the BIC criterion. Figure \ref{fig:estimatedparcorr} in Appendix \ref{sec:app3} shows the estimated networks using Fruchterman-Reingold layout \cite{FRlayout91}. 

\begin{table}[ht]
\centering
\begin{tabular}{lcccccc}
  \hline
 & Degree & Eccentricity & Distance & Clustering & Strength & N°Edges \\ 
  \hline
2018 & 6.89 & 3.53 & 2.15 & 0.46 & 0.83 & 124 \\ 
  2019 & 7.00 & 3.69 & 2.24 & 0.47 & 0.87 & 126 \\ 
  2020 & 7.44 & 4.08 & 2.29 & 0.48 & 0.91 & 134 \\ 
   \hline
\end{tabular}
\caption{Network measures for the estimated networks in years 2018-2019-2020} \label{tab:netmeasures}
\end{table}

\noindent
Common networks measures\footnote{Only existing paths are considered for average distance. Eccentricity of individual nodes is set to 0.} (mean values) are reported in Table \ref{tab:netmeasures}, along with the total number of edges detected. We observe an increase in all the measure from 2018 to 2020. The mean values of eccentricity and distance give us an indication about the velocity of transmission of a shock in the network. The former is the minimum number of steps needed to reach the farthest node in terms of steps from a specific node, while the latter is length of the minimum path between two different nodes in the graph. Both values in all three years are quite small, suggesting that a shock could potentially diffuse pretty quickly in the system. The averages values of individual clustering coefficients (see \cite{smallworld98}) show a tendency of nodes to cluster together and are higher than the case of random graphs (0.19, 0.19 and 0.21 respectively in 2018, 2019 and 2020). This feature, together with the low values of distance, are two core characteristics of small-world graphs \cite{smallworld98}. The mean value of degree is the average number of banks connected to a bank\footnote{For weighted graphs it is common also to consider directly the weights and thus calculate a weighted degree. Here strength corresponds to this weighted degree measure.}. In our case we detect, on average, seven banks linked to each bank. Figure \ref{fig:degdists} in Appendix \ref{sec:app3} shows the degree distributions in all three years, which give us a more detailed pictures of individual degrees. In 2018 the distribution is more symmetric, while in 2019 and 2020 becomes multi-modal. In fact, there is a large number of banks with a number of connections below and above the average, but few banks with a number of connections around the mean value. Also, in the last column of Table \ref{tab:netmeasures} the total number of edges detected is reported. They are the 19.7\%, 20\% and 21.3\% of 630 possible edges, in 2018, 2019 and 2020, respectively. Furthermore, we also report the average strength of the nodes in each estimated network. This measure is an useful proxy of the intensity of the connections in the graph, that here are the relations among banks represented by partial correlations\footnote{Note that in a partial correlation network positive and negative values could average out. One can use absolute values instead, as suggested in \cite{anuf15}. Given the small percentage of negative partial correlation, 2\%, 10\% and 8\% in 2018, 2019 and 2020 respectively, we leave it for further research.}. Thus, a rise in this measure suggests that the intensity of connections in the inter-bank network increased in the time period considered.

\begin{figure}[ht]
\centering
 \centering
 \includegraphics[width=1\linewidth, angle = 0 ]{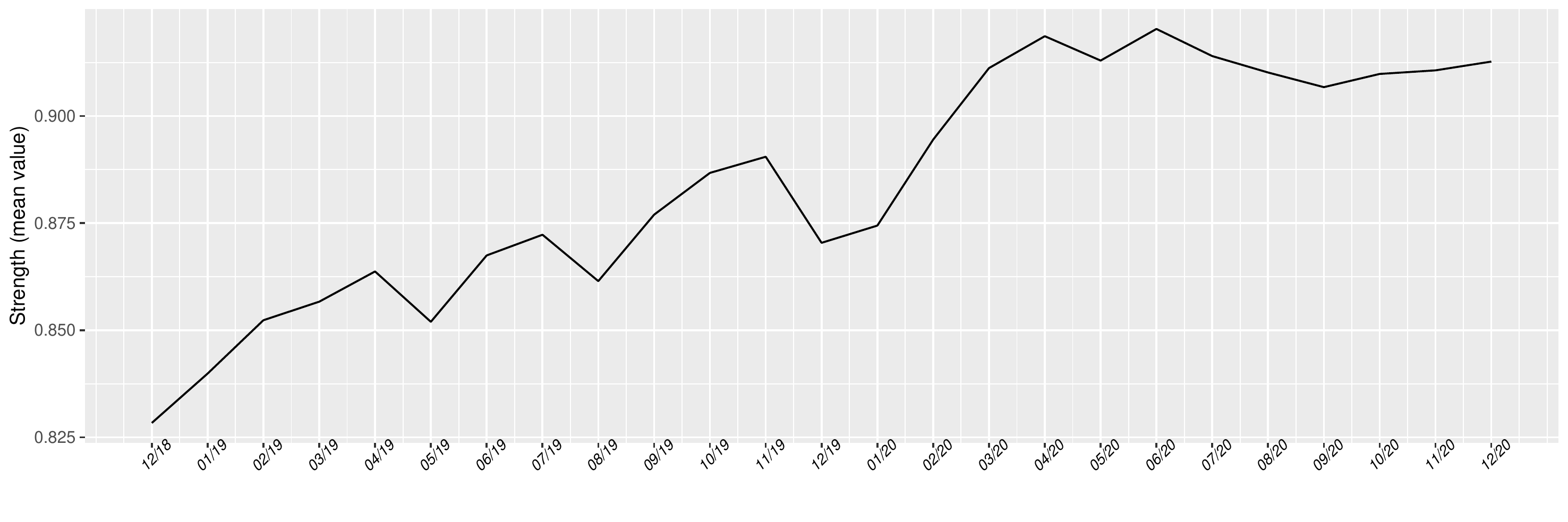}
\vspace{-1cm}
\caption{Strength evolution with rolling window of 1 year length}
\label{fig:str_dyn}
\end{figure}

This trend is also confirmed by the evolution of network strength evaluated using a rolling widow of one year and reported in Figure \ref{fig:str_dyn}. Notice the effect of the Covid-19 pandemic on the average network strength, with a sharp rise between February 2020 and April 2020 and the subsequent stabilization on an higher level for the remaining part of 2020. This trend could not only identify the presence, but also suggests the persistence of a potential crisis period.

\begin{figure}[h]
\centering
\begin{subfigure}{0.9\textwidth}
  \includegraphics[width=1\linewidth, angle = 0]{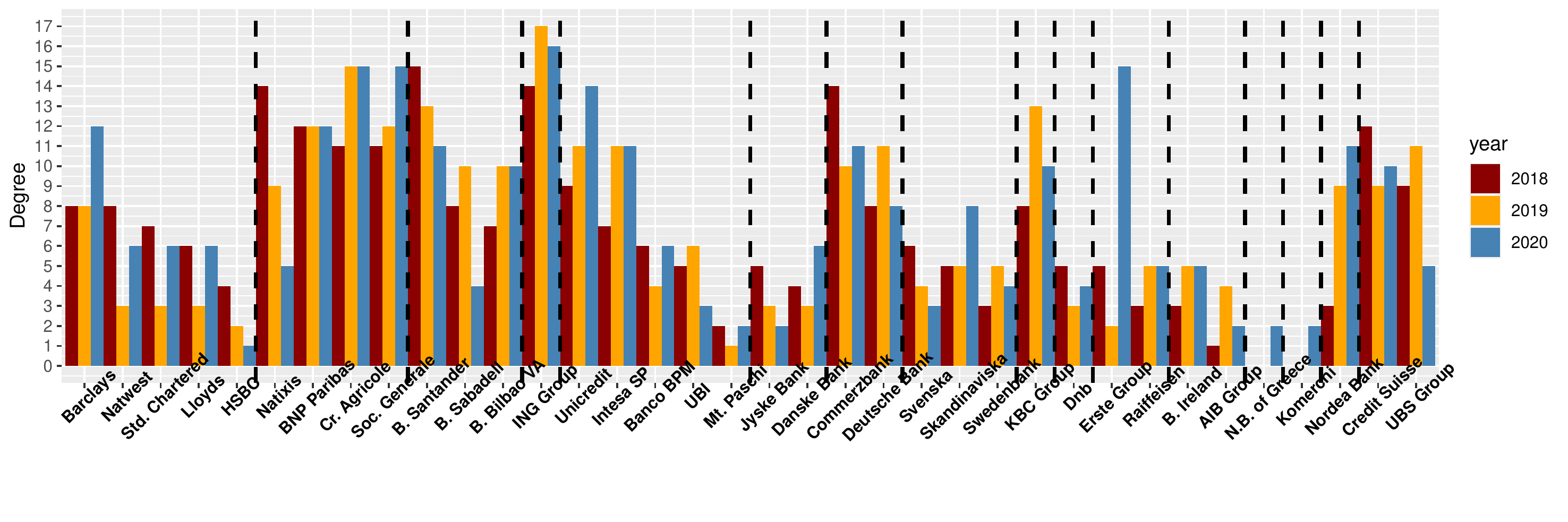}
\end{subfigure}
\vspace{-0.5cm}
\newline
\centering
\begin{subfigure}{0.9\textwidth}
  \includegraphics[width=1\linewidth, angle = 0]{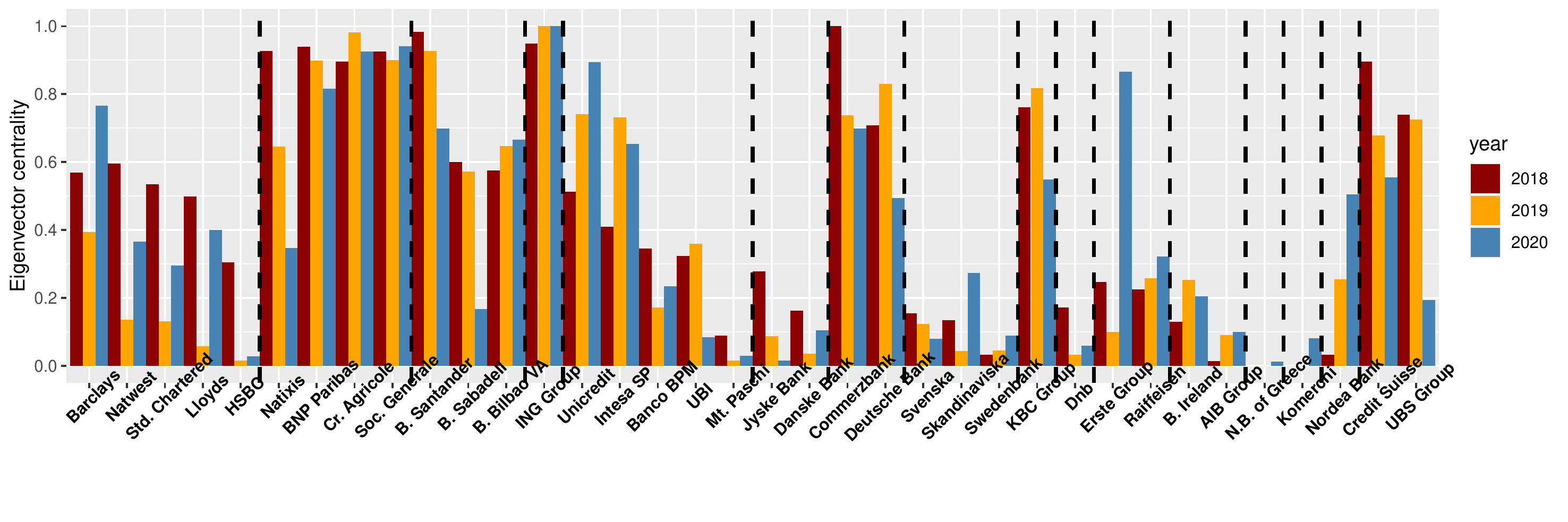}
\end{subfigure}
\vspace{-0.5cm}
\newline
\centering
\begin{subfigure}{0.9\textwidth}
  \includegraphics[width=1\linewidth, angle = 0]{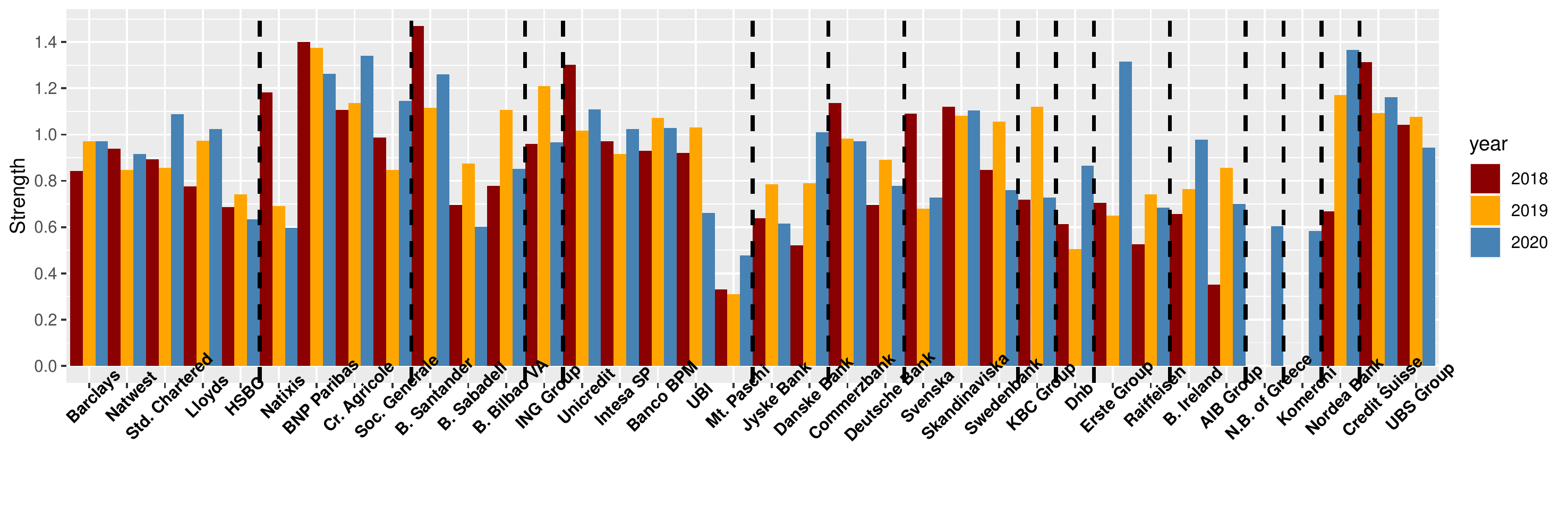}
\end{subfigure}
\vspace{-0.5cm}
\caption{Centrality measures for 2018-2020}
\label{fig:cent_measures}
\end{figure}

A network representation of the banking system also allows to detect the most important banks by looking at their position in the network instead of their market capitalization, for example. Thus, we compute three possible centrality measures and look at the first five banks according to each measure in order to detect the most important banks in the network. Values are reported in Figure \ref{fig:cent_measures} (banks ordered by country and degree centrality in 2018). Degree centrality identifies the most important banks according to the number of connections a specific bank has. Note that, for this measure, there are more than five banks in the first five positions. In fact, if we consider the first five larger values of degree, it could happen that there are multiple banks having the same number of connections. According to degree centrality, Credit Agricole, Banco Santander, Societé Generale, ING Group and Unicredit are among the five most important banks in all the three years considered, while BNP Paribas, Intesa San Paolo, Commerzbank and UBS Group appear twice within the first five. Strength centrality is based on the strength of each node in network. Thus, it considers not only the number of connections, but also their importance. According to strength, only BNP Paribas appears three times within the first five positions in 2018, 2019 and 2020. Credit Agricole, Banco Santander and Nordea Bank are instead included in two out of three years. The last measure is eigenvector centrality. It considers the importance, in terms of number of connections, of the neighbours of a node to quantify the importance of the node itself. Thus, from this point of view, the importance of a bank depends on the importance of other banks connected to that specific bank. Looking at eigenvector centrality, ING Group is the only bank that is among the first five banks in all the three years considered. Instead, BNP Paribas, Credit Agricole, Banco Santander and Societé Generale appear two times in the five most important banks. Summing up, we observe that BNP Paribas, Credit Agricole and Banco Santander are listed at least twice among the first five banks according to all the three centrality measures considered. This can suggest that they play an important role in the interbank network, at least in the time period considered.

\clearpage

\begin{figure}[h!]
\centering
 \centering
 \includegraphics[width=1\linewidth, angle = 0 ]{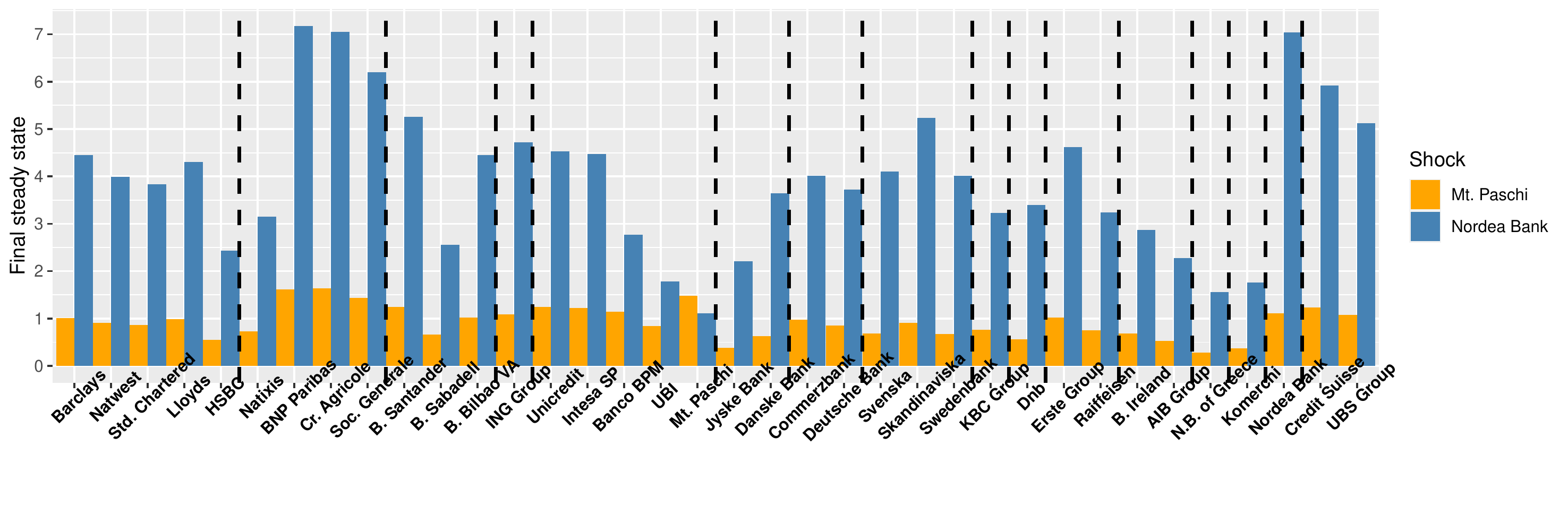}
\caption{Shock diffusion simulation in the 2020 estimated network. Final steady-states for all banks in the network after a unitary positive shock in Nordea Bank and Mt. Paschi.}\label{fig:netdiff}
\end{figure}

Finally, we also decide to investigate the effects of a shock in the banking network. We consider the network estimated in 2020, where all nodes are connected within a single huge component. The most and the least important banks according to strength centrality are Nordea Bank and Monte Paschi di Siena, respectively. We simulate the diffusion of a positive unitary shock in these two banks with the approach used by Anufriev and Panchenko \cite{anuf15} and observe how different are the effects on the whole banking network. Given the estimated partial correlation matrix in 2020, $\hat{\text{\textbf{P}}}_{2020}$, the direct, second and high-order effect of the shocks can be evaluated as follows:
\begin{equation} \label{diffformula}
    \text{\textbf{s}}_{i}^{\infty} = \text{\textbf{e}}_i + \hat{\text{\textbf{P}}}_{2020} \text{\textbf{e}}_i + \hat{\text{\textbf{P}}}_{2020}^2 \text{\textbf{e}}_i + ... = \sum_{t=0}^{\infty}\hat{\text{\textbf{P}}}_{2020}^t \text{\textbf{e}}_i = (\text{\textbf{I}}-\hat{\text{\textbf{P}}}_{2020})^{-1}\text{\textbf{e}}_i.
\end{equation}
where $\text{\textbf{e}}_i$ is the initial vector of shocks and \textbf{s}$_{i}^{\infty}$ is final steady-state induced by $\text{\textbf{e}}_i$. In our simulation, $\text{\textbf{e}}_i$ is a vector of $0$s with 1 in the $i$-th position corresponding to a positive unitary shock in the $i$-th bank (from Table \ref{tab:banksnames}, $\text{\textbf{e}}_{\text{Nordea}}=\text{\textbf{e}}_{31}$ and $\text{\textbf{e}}_{\text{Mt.Paschi}}=\text{\textbf{e}}_{24}$). The convergence of the sum in (\ref{diffformula}) is guaranteed when the spectral radius of $\hat{\text{\textbf{P}}}_{2020}$ is smaller than 1 (see Anufriev and Panchenko \cite{anuf15}). This is verified for our estimated partial correlation matrix $\hat{\text{\textbf{P}}}_{2020}$. In Figure \ref{fig:netdiff} we report the final steady-states of the two positive unitary shocks considered. It is clear how the whole system reacts differently to the two different vectors of initial shocks. An individual shock in a more central bank could have a much larger impact on the entire network. In our simulation, the final overall effects, measured as the sums of final steady-states of each bank, are 142.18 and 33.15 for initial shock in Nordea and Monte Paschi, respectively. That is, the overall effects of a shock are more than four times larger if it hits the most central bank.

\newpage
\section{Conclusions} \label{sec:conclusions}

We introduce a 2-stage estimator that brings together the \emph{tlasso} of Finegold and Drton \cite{tlassoFD11} and the \emph{2Sgelnet} of Bernardini et al. \cite{gelnet_ber}. This procedure is more flexible than \emph{2Sgelnet} and it is more suitable for situations when data exhibit heavy tails and model misspecification. The proposed estimator relies on the \fmten{elastic net} penalty and also allows to consider the LASSO penalty as a special case. Exploiting the scale-mixture representation of the multivariate t-Student distribution, we use the EM algorithm to estimate the precision matrix $\mathbf{\Theta}$. A sparse estimate is produced in the M-step through a 2-stage procedure using \fmten{elastic net} penalty. By running Monte-Carlo simulations, the proposed estimator is compared with \emph{glasso}, \emph{tlasso} and \emph{2Sgelnet} by looking both at classification performances and numerical accuracy of the optimal models selected using BIC criterion. Seven network topologies and four multivariate distributions have been considered. We observe that \emph{2Stelnet} performs quite well with respect to the other estimators considered, especially with the largest sample sizes. The results also suggest that \emph{2Stelnet} performs remarkably well even if there is a mismatch between the real distribution of data and the one assumed by the estimator (i.e. t-Student with $\nu=3$). 

Despite the good behavior of \emph{2Stelnet} in low-dimensional settings ($n>p$), severe limitations can arise in high-dimensional situations ($n<p$) where the existence and uniqueness of the estimator in the second stage (\ref{normformstep2}) is not guaranteed. 



Finally, an empirical application is proposed and \emph{2Stelnet} is used to estimate the European banking network from the share prices of a large set of European banks. We show the impact of the Covid-19 pandemic on network strength, which is as an indicator of potential crisis periods. Different centrality measures are also used to detect the most central banks in the network. To conclude our empirical analysis, we evaluate the effects of a shock in the most and least central banks, according to strength, by using the 2020 partial correlation network; not surprisingly, we found much larger effects if the shock hits the most central bank, suggesting that the degree of interconnectedness should play an important role in setting up adequate risk management and risk mitigation tools.

\bibliographystyle{plain}
\bibliography{biblio}

\begin{thebibliography}{10}

\bibitem{anuf15}
M.~Anufriev and V.~Panchenko.
\newblock Connecting the dots: Econometric methods for uncovering networks with
  an application to the australian financial institutions.
\newblock {\em European Journal of Banking and Finance}, 61:241--255, 2015.

\bibitem{Baba04}
K.~Baba, R.~Shibata, and M.~Sibuya.
\newblock Partial correlation and conditional correlation as measures of
  conditional independence.
\newblock {\em Australian \& New Zealand Journal of Statistics},
  46(4):657--664, 2004.

\bibitem{Bane08}
O.~Banerjee, L.~El Ghaoui, and A.~d'Aspremont.
\newblock Model selection through sparse maximum likelihood estimation for
  multivariate gaussian or binary data.
\newblock {\em Journal of Machine Learning Research}, 9:485--516, 2008.

\bibitem{FinNetSurvPhy}
M.~Bardoscia, P.~Barucca, S.~Battiston, F.~Caccioli, G.~Cimini,
  D.~Garlaschelli, F.~Saracco, T.~Squartini, and G.~Caldarelli.
\newblock The physics of financial networks.
\newblock {\em Nature Reviews Physics}, 3:490--507, 2021.

\bibitem{NETS19}
M.~Barigozzi and C.~Brownlees.
\newblock Nets: Network estimation for time series.
\newblock {\em Journal of Applied Econometrics}, 34:347--364, 2019.

\bibitem{gelnet_ber}
D.~Bernardini, S.~Paterlini, and E.~Taufer.
\newblock New estimation approaches for graphical models with elastic net
  penalty.
\newblock {\em arXiv:2102.01053}, 2021.

\bibitem{GrangerNet12}
M.~Billio, M.~Getmansky, A.~W. Lo, and L.~Pelizzon.
\newblock Econometric measures of connectedness and systemic risk in the
  finance and insurance sectors.
\newblock {\em Journal of Financial Economics}, 104:535--559, 2012.

\bibitem{Cai11}
T.~Cai, W.~Liu, and Xi~Luo.
\newblock A constrained l1 minimization approach to sparse precision matrix
  estimation.
\newblock {\em Journal of the American Statistical Association},
  106(494):594--607, 2011.

\bibitem{AREprod19}
V.~M. Carvalho and A.~Tahbaz-Salehi.
\newblock Production networks: A primer.
\newblock {\em Annual Review of Economics}, 11:635--663, 2019.

\bibitem{Cont01}
R.~Cont.
\newblock Empirical properties of asset returns: stylized facts and statistical
  issues.
\newblock {\em Quantitative Finance}, 1:223--236, 2001.

\bibitem{globalnet18}
M.~Demirer, F.~X. Diebold, L.~Liu, and K.~Yilmaz.
\newblock Estimating global bank network connectedness.
\newblock {\em Journal of Applied Econometrics}, 33:1--15, 2018.

\bibitem{Demp72}
A.~P. Dempster.
\newblock Covariance selection.
\newblock {\em Biometrics}, 28(1):157--175, 1972.

\bibitem{EMart77}
A.~P. Dempster, N.~M. Laird, and D.~B. Rubin.
\newblock Maximum likelihood from incomplete data via the em algorithm.
\newblock {\em Journal of the Royal Statistical Society - Series B
  (Methodological)}, 39:1--38, 1977.

\bibitem{finnet15}
F.~X. Diebold and K.~Yilmaz.
\newblock {\em Financial and Macroeconomic Connectedness: A Network Approach to
  Measurement and Monitoring}.
\newblock Oxford University Press, 2015.

\bibitem{tlassoFD11}
M.~Finegold and M.~Drton.
\newblock Robust graphical modeling of gene networks using classical and
  alternative t-distributions.
\newblock {\em The Annals of Applied Statistics}, 5:1057--1080, 2011.

\bibitem{ebic10}
R.~Foygel and M.~Drton.
\newblock Extended bayesian information criteria for gaussian graphical models.
\newblock {\em Advances in Neural Information Processing Systems (NIPS 2010)},
  23:604--612, 2010.

\bibitem{glasso08}
J.~Friedman, T.~Hastie, and R.~Tibshirani.
\newblock Sparse inverse covariance estimation with the graphical lasso.
\newblock {\em Biostatistics}, 9(3):432--441, 2008.

\bibitem{FRlayout91}
T.~M.~J. Fruchterman and E.~M. Reingold.
\newblock Graph drawing by force-directed placement.
\newblock {\em Software - Practice and Experience}, 11:1129--1164, 1991.

\bibitem{bookESL2009}
T.~Hastie, R.~Tibshirani, and J.~Friedman.
\newblock {\em The Elements of Statistical Learning: Data Mining, Inference,
  and Prediction}.
\newblock Springer, 2009.

\bibitem{AREnet09}
M.~O. Jackson.
\newblock Networks and economic behavior.
\newblock {\em Annual Review of Economics}, 1:489--511, 2009.

\bibitem{FinNetSurv20}
M.~O. Jackson and A.~Pernoud.
\newblock Systemic risk in financial networks: A survey.
\newblock {\em SSRN: http://dx.doi.org/10.2139/ssrn.3651864}, 2020.

\bibitem{GM2009}
D.~Koller and N.~Friedman.
\newblock {\em Probabilistic Graphical Models: Principles and Techniques}.
\newblock MIT Press, 2009.

\bibitem{gelnet_kov}
S.~Kovács, T.~Ruckstuhl, H.~Obrist, and Peter B{\"u}hlmann.
\newblock Graphical elastic net and target matrices: Fast algorithms and
  software for sparse precision matrix estimation.
\newblock {\em arXiv:2101.02148}, 2021.

\bibitem{GM1996}
S.~Lauritzen.
\newblock {\em Graphical Models}.
\newblock Oxford University Press, 1996.

\bibitem{gslope18}
S.~Lee, P.~Sobczyk, and M.~Bogdan.
\newblock Structure learning of {G}aussian markov random fields with false
  discovery rate control.
\newblock {\em Symmetry}, 11, 2019.

\bibitem{EMtdof}
C.~Liu and D.~B. Rubin.
\newblock Ml estimation of the t distribution using em and its extensions, ecm
  and ecme.
\newblock {\em Statistica Sinica}, 5:19--39, 1995.

\bibitem{glasso12}
R.~Mazumder and T.~Hastie.
\newblock The graphical lasso: New insights and alternatives.
\newblock {\em Electronic Journal of Statistics}, 6:2125--2149, 2012.

\bibitem{Mein06}
N.~Meinshausen and P.~B{\"u}hlmann.
\newblock High-dimensional graphs and variable selection with the lasso.
\newblock {\em The Annals of Statistics}, 34(3):1436--1462, 2006.

\bibitem{Tib96}
R.~Tibshirani.
\newblock Regression shrinkage and selection via the lasso.
\newblock {\em Journal of the Royal Statistical Society. Series B
  (Methodological)}, 58(1):267--288, 1996.

\bibitem{Torri18}
G.~Torri, R.~Giacometti, and S.~Paterlini.
\newblock Robust and sparse banking network estimation.
\newblock {\em European Journal of Operational Research}, 270(1):51--65, 2018.

\bibitem{smallworld98}
D.~J. Watts and S.~H. Strogatz.
\newblock Collective dynamics of 'small-world' networks.
\newblock {\em Nature}, 393:440--442, 1998.

\bibitem{Yuan10}
M.~Yuan.
\newblock High dimensional inverse covariance matrix estimation via linear
  programming.
\newblock {\em Journal of Machine Learning Research}, 11:2261--2286, 2010.

\bibitem{gelato}
S.~Zhou, P.~R{\"u}timann, M.~Xu, and P.~B{\"u}hlmann.
\newblock High-dimensional covariance estimation based on gaussian graphical
  models.
\newblock {\em Journal of Machine Learning Research}, 12:2975--3026, 2011.

\bibitem{ElNet05}
H.~Zou and T.~Hastie.
\newblock Regularization and variable selection via the elastic-net.
\newblock {\em Journal of the Royal Statistical Society (Series B)},
  67:301--320, 2005.

\end{thebibliography}
\newpage

\appendix
\section{Pseudo-code \emph{2Stelnet}} \label{sec:app0}

\begin{algorithm} \label{pseudocode2Stelnet}
\caption{\emph{2Stelnet} (Section \ref{sec:2Stelnet})} \label{2Stelnetalg}
\begin{algorithmic}
\STATE Set $\hat{\mathbf{\mu}}^{(0)}$ = $\frac{1}{n}\sum_{i=1}^n\text{\textbf{x}}_i$, $\hat{\text{\textbf{S}}}^{(0)}$ = $\frac{1}{n}\sum_{i=1}^n(\text{\textbf{x}}_i-\hat{\mathbf{\mu}}^{(0)})(\text{\textbf{x}}_i-\hat{\mathbf{\mu}}^{(0)})^{\top}$ and $\hat{\mathbf{\Psi}}^{(0)}$ = $\frac{\nu}{\nu-2}\hat{\text{\textbf{S}}}^{(0)}$
\STATE Set a threshold value $\delta$
\STATE Set Convergence = FALSE
\STATE Set $t = 0$
\WHILE{Convergence==FALSE}
\STATE \begin{itemize}
    \itemsep-0.5em
    \item \textbf{E-Step:}
       \begin{itemize}
       \vspace{-0.4cm}
           \item Compute the updated estimate $\hat{\tau}_i^{(t+1)}$ as in (\ref{updatetau})
       \end{itemize}
    \vspace{-0.3cm}
    \item \textbf{M-Step:}
        \begin{itemize}
        \vspace{-0.4cm}
        \itemsep-0.5em
            \item Compute the updated estimate $\hat{\mathbf{\mu}}^{(t+1)}$ as in (\ref{updatemu})
            \item Transform each observation $\text{\textbf{x}}_i$, for $i=1,...,p$ as in (\ref{datatransformed})
            \item Estimate $\hat{\text{\textbf{b}}}_k$ using $p$ elastic net regressions as in (\ref{condreg})
            \item Using each $\hat{\text{\textbf{b}}}_k$, reconstruct an estimated edge set edge set $\hat{\text{\textbf{E}}}$
            \item Given $\hat{\text{\textbf{E}}}$ compute the updated estimate $\hat{\mathbf{\Psi}}^{(t+1)}$ as the maximizer of (\ref{normformstep2})
        \end{itemize}
    \vspace{-0.3cm}
    \STATE Set $t = t + 1$
\end{itemize}
\vspace{0.3cm}
\IF{max$_{j,k}(|\hat{\psi}^{(t+1)}_{j,k} - \hat{\psi}^{(t)}_{j,k}|)$ $<$ $\delta$}
\STATE Convergence=TRUE
\ENDIF
\ENDWHILE
\end{algorithmic}
\end{algorithm}
\clearpage
\section{Synthetic precision matrices} \label{sec:app1}

\begin{figure}[ht]
\vspace{-0.5cm}
\begin{subfigure}{.33\textwidth}
  \centering
  \includegraphics[width=1\linewidth]{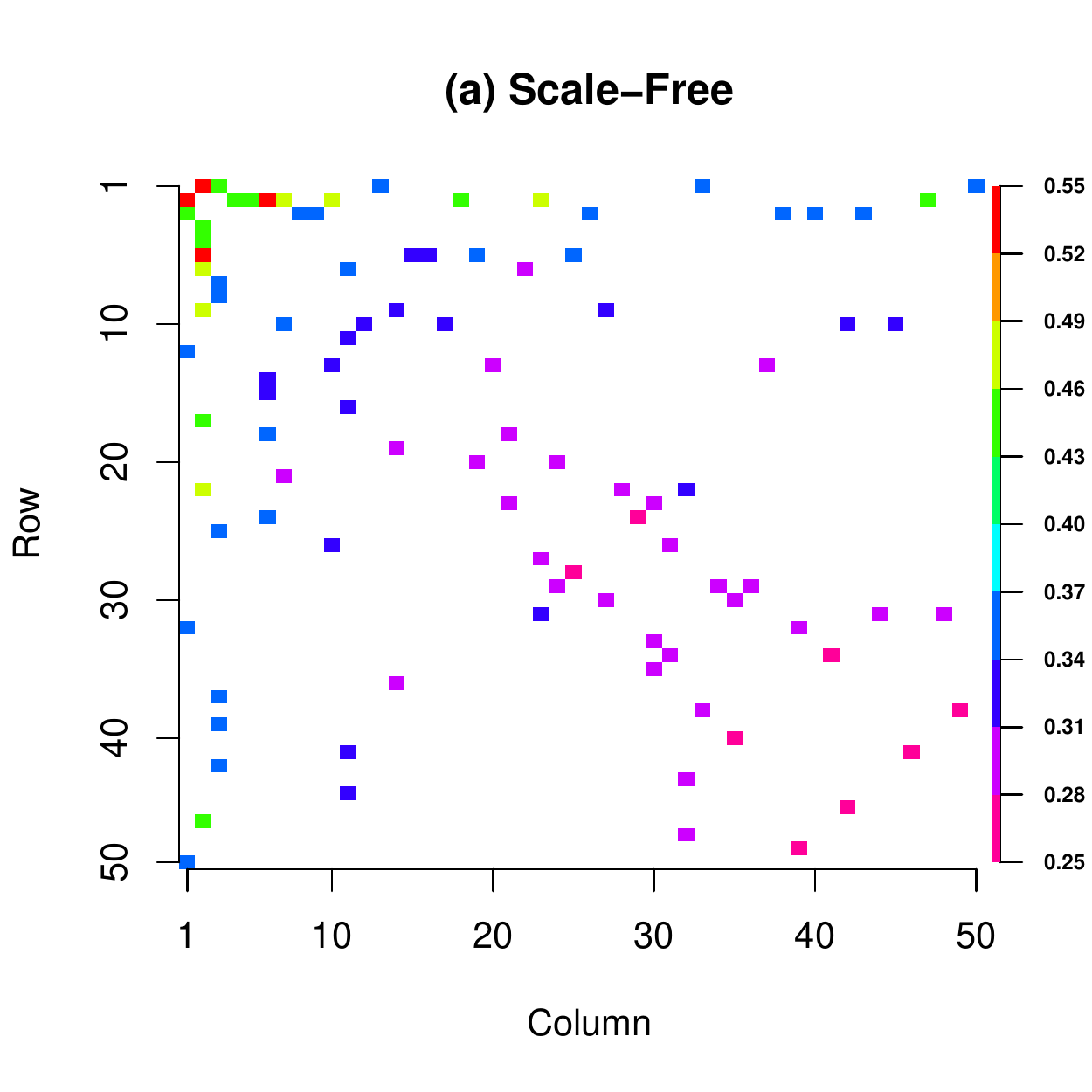}
\end{subfigure}%
\begin{subfigure}{.33\textwidth}
  \centering
  \includegraphics[width=1\linewidth]{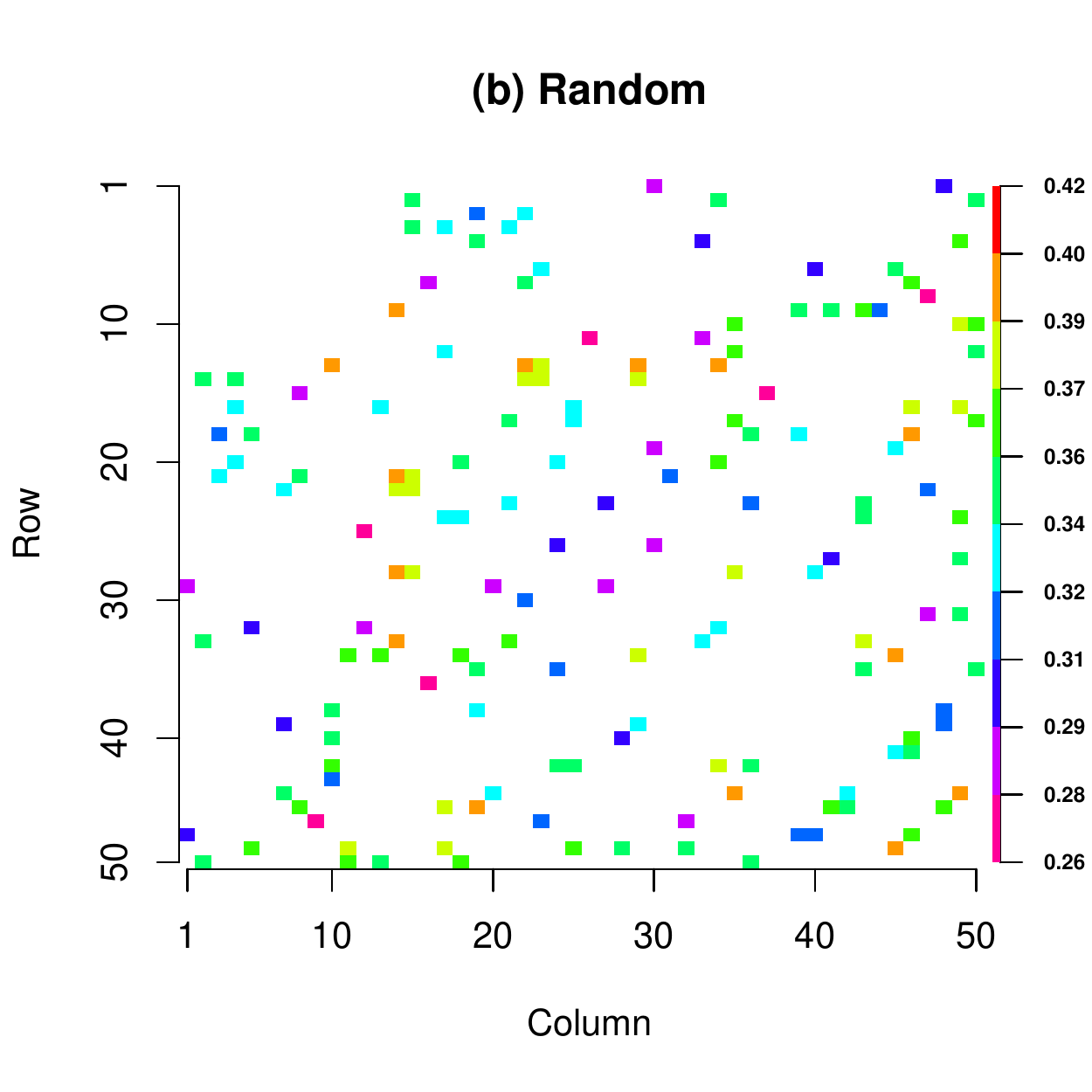}
\end{subfigure}
\begin{subfigure}{.33\textwidth}
  \centering
  \includegraphics[width=1\linewidth]{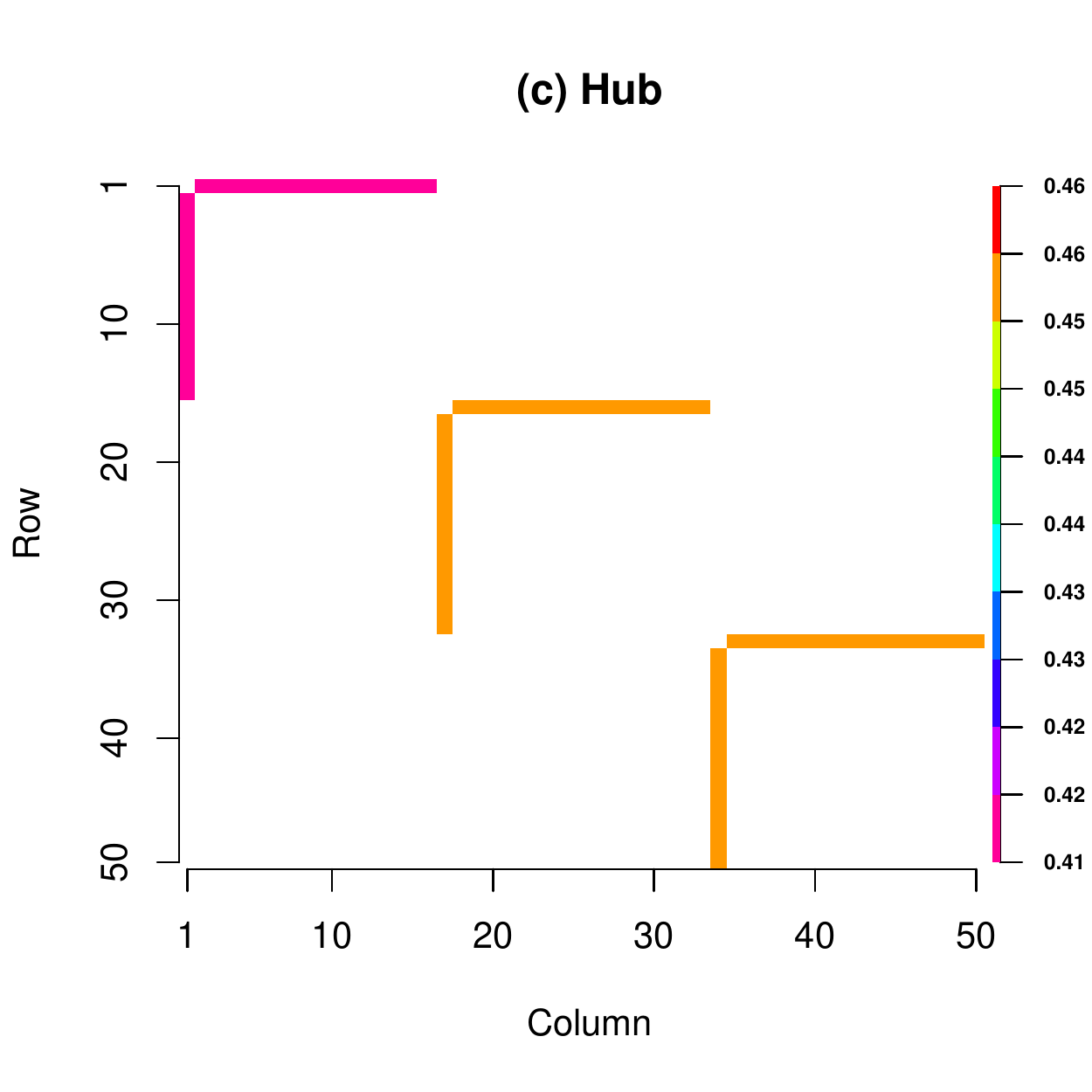}
\end{subfigure}
\newline
\begin{subfigure}{.33\textwidth}
  \centering
  \includegraphics[width=1\linewidth]{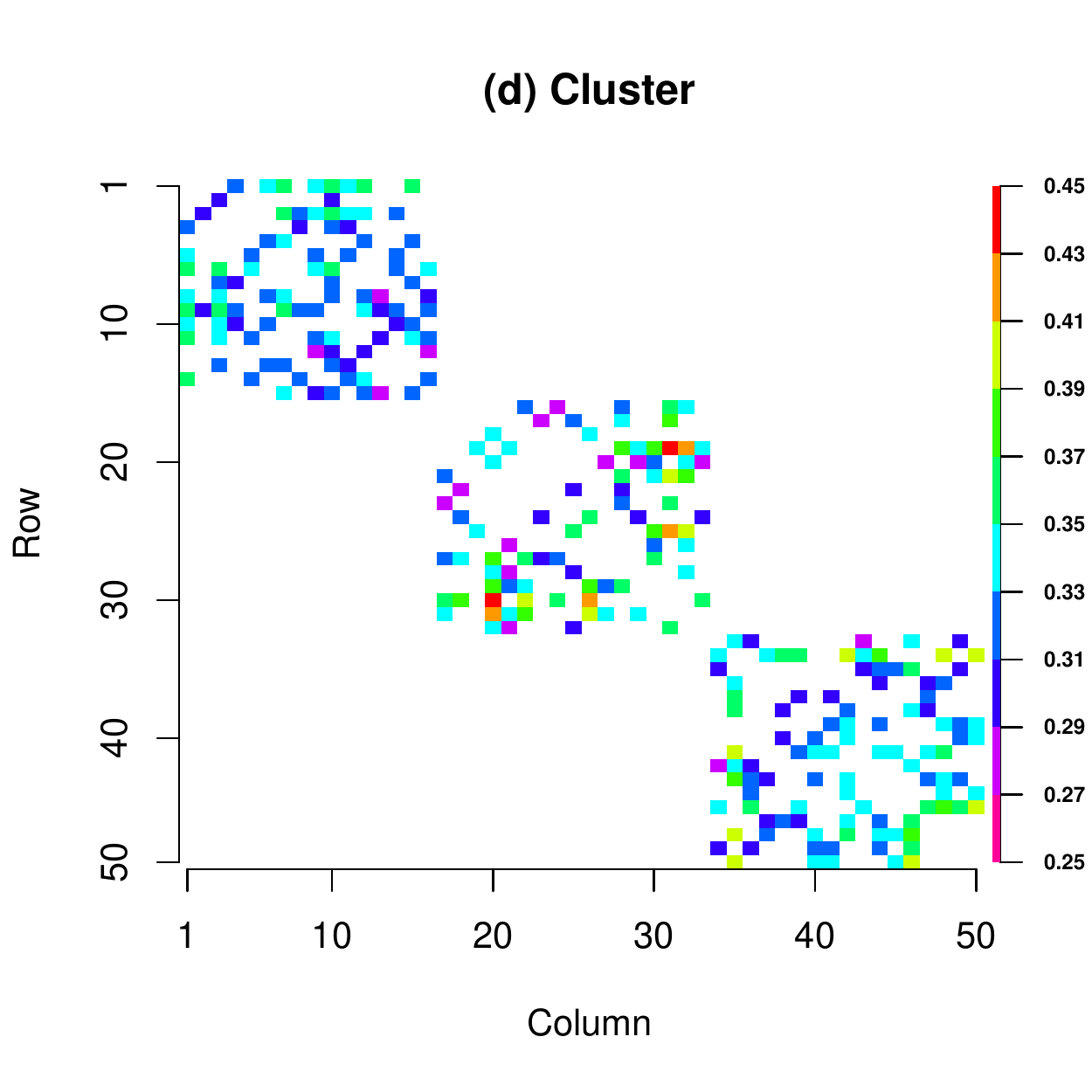}
\end{subfigure}%
\begin{subfigure}{.33\textwidth}
  \centering
  \includegraphics[width=1\linewidth]{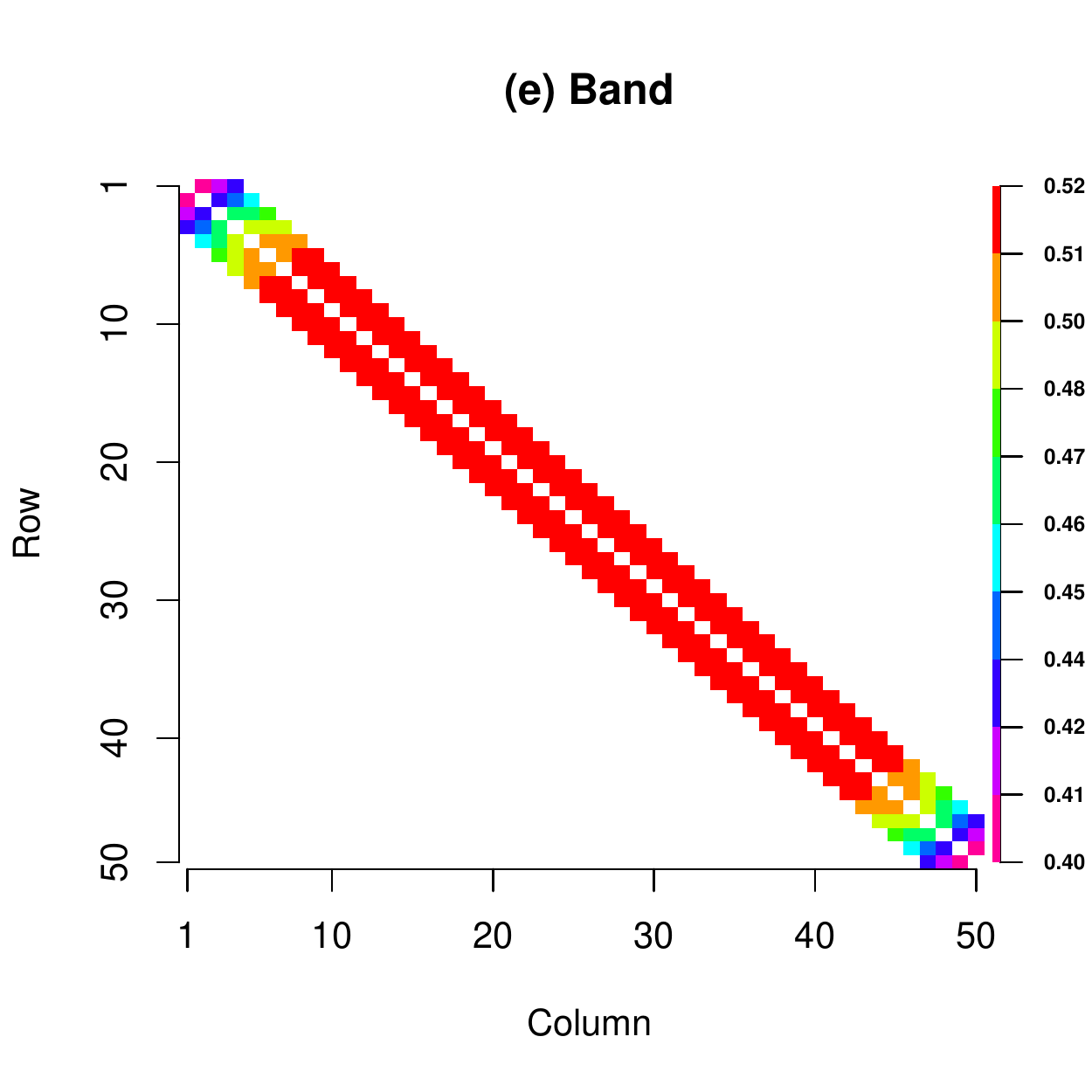}
\end{subfigure}
\begin{subfigure}{.33\textwidth}
  \centering
  \includegraphics[width=1\linewidth]{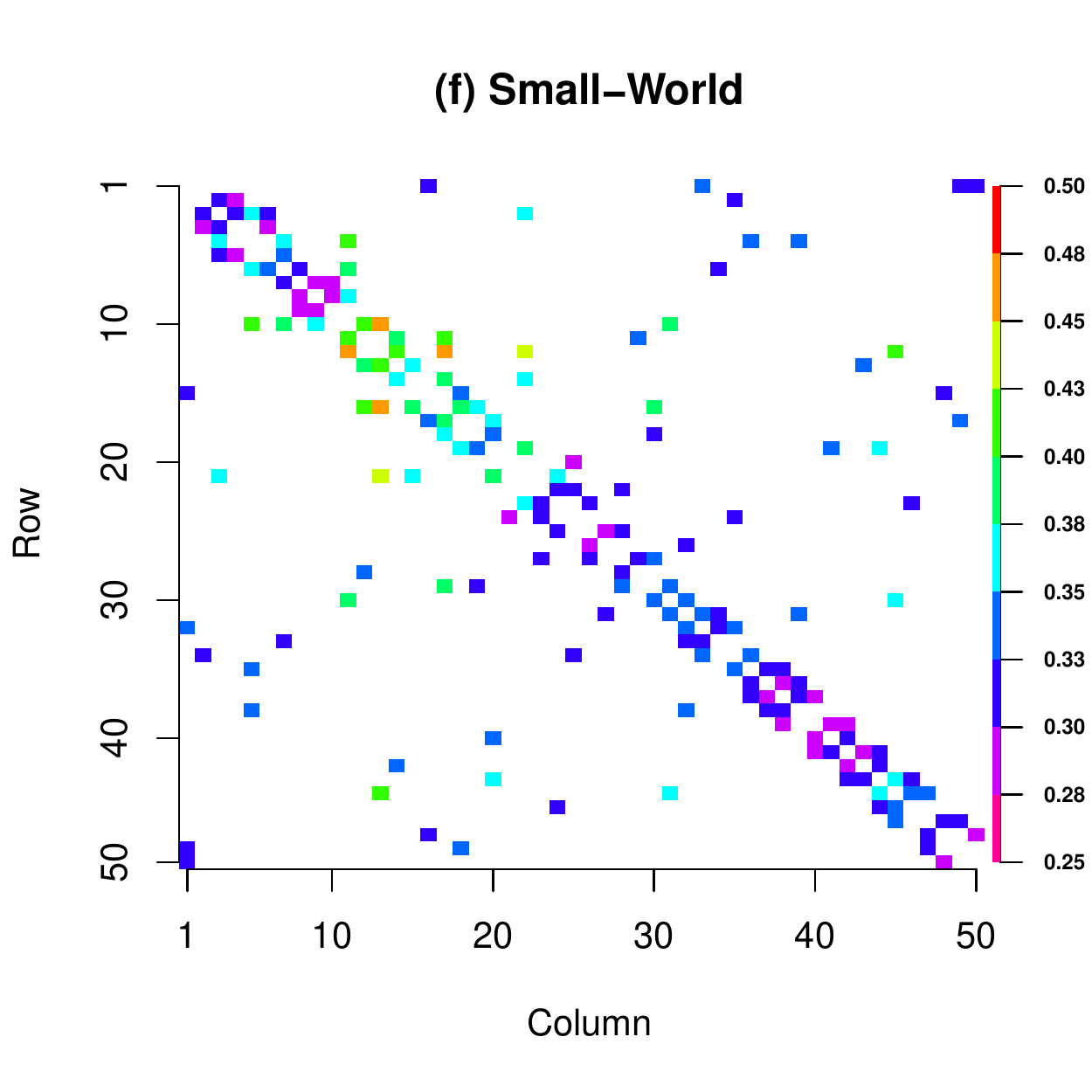}
\end{subfigure}
\newline
\centering
\begin{subfigure}{.33\textwidth}
  \centering
  \includegraphics[width=1\linewidth]{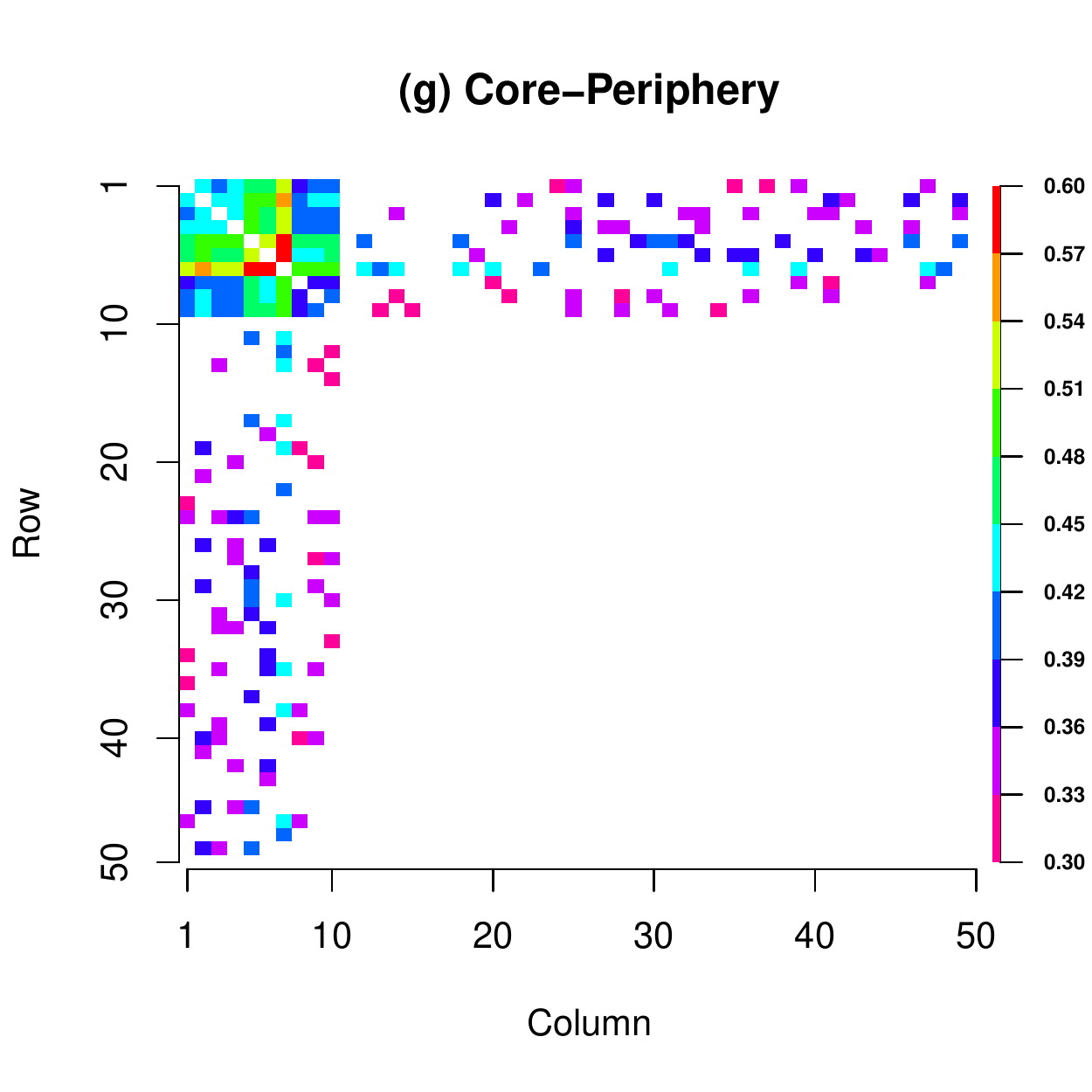}
\end{subfigure}
\caption{Precision matrices of simulated data (diagonal elements ignored)}
\label{fig:adjmats}
\end{figure}

\clearpage

\section{Optimal values of $\lambda$} \label{sec:app2}

\begin{table}[ht]
\centering

\begin{subfigure}[b]{0.45\textwidth}
\resizebox{1\textwidth}{!}{
\begin{tabular}{rcccccc}
\hline
& \emph{glasso} & \emph{tlasso} & \emph{2Sgelnet} & \emph{2Sglasso} & \emph{2Stelnet} & \emph{2Stlasso} \\ 
\hline
n=100 & 0.253 & 2.319 & 0.403 & 0.194 & 0.384 & 0.186 \\ 
 & (0.024) & (0.928) & (0.019) & (0.009) & (0.020) & (0.009) \\ 
n=250 & 0.139 & 0.548 & 0.289 & 0.139 & 0.278 & 0.134 \\ 
 & (0.010) & (0.049) & (0.018) & (0.008) & (0.017) & (0.008) \\ 
n=500 & 0.094 & 0.327 & 0.225 & 0.106 & 0.225 & 0.104 \\ 
 & (0.006) & (0.029) & (0.025) & (0.009) & (0.030) & (0.010) \\ 
\hline
\end{tabular}
}
\caption{Scale-free}
\end{subfigure}
\hspace{0.5cm}
\begin{subfigure}[b]{0.45\textwidth}
\resizebox{1\textwidth}{!}{
\begin{tabular}{rcccccc}
\hline
& \emph{glasso} & \emph{tlasso} & \emph{2Sgelnet} & \emph{2Sglasso} & \emph{2Stelnet} & \emph{2Stlasso} \\ 
\hline
n=100 & 0.238 & 2.509 & 0.403 & 0.189 & 0.385 & 0.183 \\ 
 & (0.027) & (1.043) & (0.026) & (0.011) & (0.025) & (0.011) \\ 
n=250 & 0.120 & 0.521 & 0.284 & 0.136 & 0.274 & 0.130 \\ 
 & (0.010) & (0.053) & (0.024) & (0.009) & (0.022) & (0.009) \\ 
n=500 & 0.083 & 0.314 & 0.244 & 0.110 & 0.231 & 0.108 \\ 
 & (0.006) & (0.029) & (0.035) & (0.013) & (0.034) & (0.015) \\ 
\hline
\end{tabular}
}
\caption{Random}
\end{subfigure}

\vspace{1cm}

\begin{subfigure}[b]{0.45\textwidth}
\resizebox{1\textwidth}{!}{
\begin{tabular}{rcccccc}
\hline
& \emph{glasso} & \emph{tlasso} & \emph{2Sgelnet} & \emph{2Sglasso} & \emph{2Stelnet} & \emph{2Stlasso} \\ 
\hline
n=100 & 0.247 & 1.960 & 0.409 & 0.190 & 0.391 & 0.182 \\ 
 & (0.024) & (0.948) & (0.026) & (0.011) & (0.025) & (0.011) \\ 
n=250 & 0.141 & 0.620 & 0.321 & 0.146 & 0.302 & 0.135 \\ 
 & (0.011) & (0.059) & (0.031) & (0.012) & (0.028) & (0.011) \\ 
n=500 & 0.095 & 0.379 & 0.309 & 0.134 & 0.302 & 0.127 \\ 
 & (0.008) & (0.035) & (0.032) & (0.014) & (0.039) & (0.016) \\ 
\hline
\end{tabular}
}
\caption{Hub}
\end{subfigure}
\hspace{0.5cm}
\begin{subfigure}[b]{0.45\textwidth}
\resizebox{1\textwidth}{!}{
\begin{tabular}{rcccccc}
\hline
& \emph{glasso} & \emph{tlasso} & \emph{2Sgelnet} & \emph{2Sglasso} & \emph{2Stelnet} & \emph{2Stlasso} \\ 
\hline
n=100 & 0.247 & 2.513 & 0.386 & 0.183 & 0.373 & 0.175 \\ 
 & (0.035) & (1.074) & (0.033) & (0.017) & (0.029) & (0.015) \\ 
n=250 & 0.115 & 0.512 & 0.229 & 0.111 & 0.222 & 0.107 \\ 
 & (0.011) & (0.054) & (0.019) & (0.008) & (0.021) & (0.008) \\ 
n=500 & 0.068 & 0.259 & 0.164 & 0.082 & 0.157 & 0.078 \\ 
 & (0.006) & (0.030) & (0.010) & (0.005) & (0.011) & (0.005) \\ 
\hline
\end{tabular}
}
\caption{Cluster}
\end{subfigure}

\vspace{1cm}

\begin{subfigure}[b]{0.45\textwidth}
\resizebox{1\textwidth}{!}{
\begin{tabular}{rcccccc}
\hline
& \emph{glasso} & \emph{tlasso} & \emph{2Sgelnet} & \emph{2Sglasso} & \emph{2Stelnet} & \emph{2Stlasso} \\ 
\hline
n=100 & 0.284 & 3.024 & 0.354 & 0.165 & 0.349 & 0.157 \\ 
 & (0.048) & (0.550) & (0.033) & (0.015) & (0.031) & (0.014) \\ 
n=250 & 0.107 & 0.494 & 0.189 & 0.092 & 0.179 & 0.088 \\ 
 & (0.013) & (0.060) & (0.013) & (0.006) & (0.014 )& (0.006) \\ 
n=500 & 0.053 & 0.197 & 0.166 & 0.086 & 0.156 & 0.080 \\ 
 & (0.006) & (0.026) & (0.012) & (0.006) & (0.010) & (0.006) \\ 
\hline
\end{tabular}
}
\caption{Band}
\end{subfigure}
\hspace{0.5cm}
\begin{subfigure}[b]{0.45\textwidth}
\resizebox{1\textwidth}{!}{
\begin{tabular}{rcccccc}
\hline
& \emph{glasso} & \emph{tlasso} & \emph{2Sgelnet} & \emph{2Sglasso} & \emph{2Stelnet} & \emph{2Stlasso} \\ 
\hline
n=100 & 0.242 & 2.228 & 0.391 & 0.185 & 0.375 & 0.178 \\ 
 & (0.024) & (1.017) & (0.023) & (0.013) & (0.025) & (0.011) \\ 
n=250 & 0.119 & 0.496 & 0.262 & 0.126 & 0.253 & 0.123 \\ 
 & (0.009) & (0.047) & (0.017) & (0.008) & (0.017) & (0.008) \\ 
n=500 & 0.079 & 0.295 & 0.220 & 0.104 & 0.209 & 0.101 \\ 
 & (0.006) & (0.029) & (0.023) & (0.010) & (0.018) & (0.010) \\ 
\hline
\end{tabular}
}
\caption{Small-world}
\end{subfigure}

\vspace{1cm}

\begin{subfigure}[b]{0.45\textwidth}
\resizebox{1\textwidth}{!}{
\begin{tabular}{rcccccc}
\hline
& \emph{glasso} & \emph{tlasso} & \emph{2Sgelnet} & \emph{2Sglasso} & \emph{2Stelnet} & \emph{2Stlasso} \\ 
\hline
n=100 & 0.287 & 3.036 & 0.403 & 0.190 & 0.384 & 0.181 \\ 
 & (0.040) & (0.463) & (0.024) & (0.012) & (0.022) & (0.010) \\ 
n=250 & 0.133 & 0.567 & 0.266 & 0.124 & 0.254 & 0.120 \\ 
 & (0.011) & (0.061) & (0.021) & (0.010) & (0.020) & (0.010) \\ 
n=500 & 0.088 & 0.328 & 0.176 & 0.084 & 0.171 & 0.080 \\ 
 & (0.007) & (0.034) & (0.017) & (0.008) & (0.017) & (0.007) \\ 
\hline
\end{tabular}
}
\caption{Core-periphery}
\end{subfigure}
\vspace{0.5cm}
\caption{Mean values of optimal $\lambda$ (std. dev. in brackets) - Normal} 
\label{lambdas_norm}
\end{table}

\clearpage

\begin{table}[ht]
\centering

\begin{subfigure}[b]{0.45\textwidth}
\resizebox{1\textwidth}{!}{
\begin{tabular}{rcccccc}
\hline
& \emph{glasso} & \emph{tlasso} & \emph{2Sgelnet} & \emph{2Sglasso} & \emph{2Stelnet} & \emph{2Stlasso} \\ 
\hline
n=100 & 0.198 & 1.504 & 0.313 & 0.139 & 0.240 & 0.115 \\ 
 & (0.098) & (0.793) & (0.082) & (0.033) & (0.027) & (0.014) \\ 
n=250 & 0.096 & 0.251 & 0.226 & 0.102 & 0.170 & 0.082 \\ 
 & (0.037) & (0.042) & (0.045) & (0.019) & (0.022) & (0.009) \\ 
n=500 & 0.064 & 0.135 & 0.170 & 0.081 & 0.135 & 0.064 \\ 
 & (0.020) & (0.024) & (0.030) & (0.013) & (0.019) & (0.007) \\ 
\hline
\end{tabular}
}
\caption{Scale-free}
\end{subfigure}
\hspace{0.5cm}
\begin{subfigure}[b]{0.45\textwidth}
\resizebox{1\textwidth}{!}{
\begin{tabular}{rcccccc}
\hline
& \emph{glasso} & \emph{tlasso} & \emph{2Sgelnet} & \emph{2Sglasso} & \emph{2Stelnet} & \emph{2Stlasso} \\ 
\hline
n=100 & 0.152 & 1.599 & 0.276 & 0.117 & 0.239 & 0.114 \\ 
 & (0.080) & (0.718) & (0.083) & (0.030) & (0.032) & (0.015) \\ 
n=250 & 0.079 & 0.230 & 0.193 & 0.090 & 0.164 & 0.079 \\ 
 & (0.029) & (0.042) & (0.041) & (0.019) & (0.021) & (0.010) \\ 
n=500 & 0.050 & 0.129 & 0.144 & 0.069 & 0.134 & 0.064 \\ 
 & (0.015) & (0.022) & (0.027) & (0.012) & (0.020) & (0.010) \\ 
\hline
\end{tabular}
}
\caption{Random}
\end{subfigure}

\vspace{1cm}

\begin{subfigure}[b]{0.45\textwidth}
\resizebox{1\textwidth}{!}{
\begin{tabular}{rcccccc}
\hline
& \emph{glasso} & \emph{tlasso} & \emph{2Sgelnet} & \emph{2Sglasso} & \emph{2Stelnet} & \emph{2Stlasso} \\ 
\hline
n=100 & 0.164 & 1.443 & 0.312 & 0.134 & 0.247 & 0.114 \\ 
 & (0.078) & (0.769) & (0.086) & (0.030) & (0.034) & (0.016) \\ 
n=250 & 0.087 & 0.276 & 0.216 & 0.101 & 0.188 & 0.084 \\ 
 & (0.029) & (0.051) & (0.050) & (0.021) & (0.026) & (0.011) \\ 
n=500 & 0.057 & 0.151 & 0.162 & 0.078 & 0.175 & 0.074 \\ 
 & (0.018) & (0.030) & (0.031) & (0.014) & (0.028) & (0.012) \\ 
\hline
\end{tabular}
}
\caption{Hub}
\end{subfigure}
\hspace{0.5cm}
\begin{subfigure}[b]{0.45\textwidth}
\resizebox{1\textwidth}{!}{
\begin{tabular}{rcccccc}
\hline
& \emph{glasso} & \emph{tlasso} & \emph{2Sgelnet} & \emph{2Sglasso} & \emph{2Stelnet} & \emph{2Stlasso} \\ 
\hline
n=100 & 0.132 & 1.501 & 0.239 & 0.105 & 0.227 & 0.108 \\ 
 & (0.081) & (0.842) & (0.088) & (0.033) & (0.035) & (0.017) \\ 
n=250 & 0.067 & 0.241 & 0.172 & 0.081 & 0.137 & 0.066 \\ 
 & (0.025) & (0.042) & (0.037) & (0.018) & (0.018) & (0.009) \\ 
n=500 & 0.039 & 0.104 & 0.129 & 0.062 & 0.098 & 0.048 \\ 
 & (0.014) & (0.020) & (0.026) & (0.012) & (0.012) & (0.005) \\ 
\hline
\end{tabular}
}
\caption{Cluster}
\end{subfigure}

\vspace{1cm}

\begin{subfigure}[b]{0.45\textwidth}
\resizebox{1\textwidth}{!}{
\begin{tabular}{rcccccc}
\hline
& \emph{glasso} & \emph{tlasso} & \emph{2Sgelnet} & \emph{2Sglasso} & \emph{2Stelnet} & \emph{2Stlasso} \\ 
\hline
n=100 & 0.138 & 1.847 & 0.212 & 0.096 & 0.216 & 0.099 \\ 
 & (0.089) & (0.541) & (0.068) & (0.029) & (0.030) & (0.014) \\ 
n=250 & 0.051 & 0.225 & 0.141 & 0.068 & 0.110 & 0.054 \\ 
 & (0.022) & (0.040) & (0.032) & (0.014) & (0.013) & (0.006) \\ 
n=500 & 0.026 & 0.077 & 0.107 & 0.053 & 0.091 & 0.046 \\ 
 & (0.009) & (0.013) & (0.023) & (0.011) & (0.013) & (0.007) \\ 
\hline
\end{tabular}
}
\caption{Band}
\end{subfigure}
\hspace{0.5cm}
\begin{subfigure}[b]{0.45\textwidth}
\resizebox{1\textwidth}{!}{
\begin{tabular}{rcccccc}
\hline
& \emph{glasso} & \emph{tlasso} & \emph{2Sgelnet} & \emph{2Sglasso} & \emph{2Stelnet} & \emph{2Stlasso} \\ 
\hline
n=100 & 0.166 & 1.631 & 0.273 & 0.121 & 0.236 & 0.113 \\ 
 & (0.087) & (0.867) & (0.083) & (0.031) & (0.033) & (0.015) \\ 
n=250 & 0.072 & 0.232 & 0.187 & 0.088 & 0.154 & 0.075 \\ 
 & (0.029) & (0.046) & (0.051) & (0.021) & (0.023) & (0.011) \\ 
n=500 & 0.048 & 0.119 & 0.143 & 0.067 & 0.128 & 0.062 \\ 
 & (0.015) & (0.019) & (0.023) & (0.010) & (0.016) & (0.008) \\ 
\hline
\end{tabular}
}
\caption{Small-world}
\end{subfigure}

\vspace{1cm}

\begin{subfigure}[b]{0.45\textwidth}
\resizebox{1\textwidth}{!}{
\begin{tabular}{rcccccc}
\hline
& \emph{glasso} & \emph{tlasso} & \emph{2Sgelnet} & \emph{2Sglasso} & \emph{2Stelnet} & \emph{2Stlasso} \\ 
\hline
n=100 & 0.168 & 1.936 & 0.290 & 0.126 & 0.238 & 0.113 \\ 
 & (0.097) & (0.715) & (0.079) & (0.030) & (0.029) & (0.014) \\ 
n=250 & 0.079 & 0.255 & 0.191 & 0.089 & 0.156 & 0.074 \\ 
 & (0.029) & (0.041) & (0.042) & (0.019) & (0.021) & (0.010) \\ 
n=500 & 0.047 & 0.137 & 0.141 & 0.068 & 0.104 & 0.049 \\ 
 & (0.016) & (0.024) & (0.025) & (0.012) & (0.012) & (0.006) \\ 
\hline
\end{tabular}
}
\caption{Core-periphery}
\end{subfigure}
\vspace{0.5cm}
\caption{Mean values of optimal $\lambda$ (std. dev. in brackets) - t-Student ($v=3$)} 
\label{lambdas_tstudv3}
\end{table}

\clearpage

\begin{table}[ht]
\centering

\begin{subfigure}[b]{0.45\textwidth}
\resizebox{1\textwidth}{!}{
\begin{tabular}{rcccccc}
\hline
& \emph{glasso} & \emph{tlasso} & \emph{2Sgelnet} & \emph{2Sglasso} & \emph{2Stelnet} & \emph{2Stlasso} \\ 
\hline
n=100 & 0.265 & 2.172 & 0.403 & 0.192 & 0.365 & 0.176 \\ 
 & (0.026) & (0.871) & (0.019) & (0.010) & (0.017) & (0.009) \\ 
n=250 & 0.142 & 0.501 & 0.286 & 0.139 & 0.266 & 0.129 \\ 
 & (0.010) & (0.046) & (0.015) & (0.009) & (0.017) & (0.008) \\ 
n=500 & 0.098 & 0.301 & 0.224 & 0.105 & 0.216 & 0.099 \\ 
 & (0.007) & (0.026) & (0.021) & (0.007) & (0.029) & (0.008) \\ 
\hline
\end{tabular}
}
\caption{Scale-free}
\end{subfigure}
\hspace{0.5cm}
\begin{subfigure}[b]{0.45\textwidth}
\resizebox{1\textwidth}{!}{
\begin{tabular}{rcccccc}
\hline
& \emph{glasso} & \emph{tlasso} & \emph{2Sgelnet} & \emph{2Sglasso} & \emph{2Stelnet} & \emph{2Stlasso} \\ 
\hline
n=100 & 0.246 & 2.440 & 0.397 & 0.187 & 0.364 & 0.173 \\ 
 & (0.033) & (1.022) & (0.026) & (0.012) & (0.023) & (0.011) \\ 
n=250 & 0.123 & 0.480 & 0.280 & 0.135 & 0.258 & 0.123 \\ 
 & (0.011) & (0.043) & (0.021) & (0.010) & (0.019) & (0.010) \\ 
n=500 & 0.083 & 0.287 & 0.230 & 0.107 & 0.224 & 0.102 \\ 
 & (0.006) & (0.032) & (0.025) & (0.011) & (0.028) & (0.012) \\ 
\hline
\end{tabular}
}
\caption{Random}
\end{subfigure}

\vspace{1cm}

\begin{subfigure}[b]{0.45\textwidth}
\resizebox{1\textwidth}{!}{
\begin{tabular}{rcccccc}
\hline
& \emph{glasso} & \emph{tlasso} & \emph{2Sgelnet} & \emph{2Sglasso} & \emph{2Stelnet} & \emph{2Stlasso} \\ 
\hline
n=100 & 0.254 & 1.923 & 0.412 & 0.190 & 0.374 & 0.173 \\ 
 & (0.025) & (1.012) & (0.030) & (0.014) & (0.022) & (0.011) \\ 
n=250 & 0.141 & 0.577 & 0.317 & 0.143 & 0.293 & 0.131 \\ 
 & (0.011) & (0.062) & (0.029) & (0.011) & (0.027) & (0.010) \\ 
n=500 & 0.097 & 0.345 & 0.303 & 0.126 & 0.288 & 0.122 \\ 
 & (0.008) & (0.039) & (0.033) & (0.013) & (0.038) & (0.018) \\ 
\hline
\end{tabular}
}
\caption{Hub}
\end{subfigure}
\hspace{0.5cm}
\begin{subfigure}[b]{0.45\textwidth}
\resizebox{1\textwidth}{!}{
\begin{tabular}{rcccccc}
\hline
& \emph{glasso} & \emph{tlasso} & \emph{2Sgelnet} & \emph{2Sglasso} & \emph{2Stelnet} & \emph{2Stlasso} \\ 
\hline
n=100 & 0.250 & 2.285 & 0.390 & 0.182 & 0.362 & 0.169 \\ 
 & (0.033) & (1.096) & (0.033) & (0.015) & (0.034) & (0.015) \\ 
n=250 & 0.118 & 0.482 & 0.229 & 0.111 & 0.208 & 0.101 \\ 
 & (0.011) & (0.058) & (0.021) & (0.009) & (0.015) & (0.007) \\ 
n=500 & 0.070 & 0.234 & 0.168 & 0.082 & 0.152 & 0.076 \\ 
 & (0.007) & (0.026) & (0.014) & (0.006) & (0.012) & (0.005) \\ 
\hline
\end{tabular}
}
\caption{Cluster}
\end{subfigure}

\vspace{1cm}

\begin{subfigure}[b]{0.45\textwidth}
\resizebox{1\textwidth}{!}{
\begin{tabular}{rcccccc}
\hline
& \emph{glasso} & \emph{tlasso} & \emph{2Sgelnet} & \emph{2Sglasso} & \emph{2Stelnet} & \emph{2Stlasso} \\ 
\hline
n=100 & 0.284 & 2.954 & 0.354 & 0.158 & 0.324 & 0.148 \\ 
 & (0.049) & (0.562) & (0.036) & (0.017) & (0.028) & (0.014) \\ 
n=250 & 0.108 & 0.443 & 0.187 & 0.091 & 0.170 & 0.083 \\ 
 & (0.014) & (0.056) & (0.016) & (0.007) & (0.012) & (0.006) \\ 
n=500 & 0.054 & 0.176 & 0.163 & 0.082 & 0.147 & 0.076 \\ 
 & (0.005) & (0.021) & (0.011) & (0.006) & (0.010) & (0.005) \\ 
\hline
\end{tabular}
}
\caption{Band}
\end{subfigure}
\hspace{0.5cm}
\begin{subfigure}[b]{0.45\textwidth}
\resizebox{1\textwidth}{!}{
\begin{tabular}{rcccccc}
\hline
& \emph{glasso} & \emph{tlasso} & \emph{2Sgelnet} & \emph{2Sglasso} & \emph{2Stelnet} & \emph{2Stlasso} \\ 
\hline
n=100 & 0.255 & 2.317 & 0.388 & 0.185 & 0.352 & 0.167 \\ 
 & (0.028) & (1.008) & (0.027) & (0.014) & (0.024) & (0.011) \\ 
n=250 & 0.122 & 0.471 & 0.265 & 0.128 & 0.242 & 0.116 \\ 
 & (0.011) & (0.045) & (0.018) & (0.008) & (0.016) & (0.007) \\ 
n=500 & 0.081 & 0.271 & 0.215 & 0.102 & 0.200 & 0.096 \\ 
 & (0.006) & (0.029) & (0.022) & (0.009) & (0.021) & (0.012) \\ 
\hline
\end{tabular}
}
\caption{Small-world}
\end{subfigure}

\vspace{1cm}

\begin{subfigure}[b]{0.45\textwidth}
\resizebox{1\textwidth}{!}{
\begin{tabular}{rcccccc}
\hline
& \emph{glasso} & \emph{tlasso} & \emph{2Sgelnet} & \emph{2Sglasso} & \emph{2Stelnet} & \emph{2Stlasso} \\ 
\hline
n=100 & 0.290 & 2.875 & 0.397 & 0.186 & 0.369 & 0.172 \\ 
 & (0.035) & (0.391) & (0.026) & (0.012) & (0.023) & (0.010) \\ 
n=250 & 0.135 & 0.524 & 0.266 & 0.126 & 0.243 & 0.114 \\ 
 & (0.013) & (0.058) & (0.020) & (0.009) & (0.020) & (0.009) \\ 
n=500 & 0.089 & 0.304 & 0.178 & 0.085 & 0.160 & 0.076 \\ 
 & (0.008) & (0.037) & (0.018) & (0.007) & (0.017) & (0.008) \\ 
\hline
\end{tabular}
}
\caption{Core-periphery}
\end{subfigure}
\vspace{0.5cm}
\caption{Mean values of optimal $\lambda$ (std. dev. in brackets) - t-Student ($v=20$)} 
\label{lambdas_tstudv20}
\end{table}


\begin{table}[ht]
\centering

\begin{subfigure}[b]{0.45\textwidth}
\resizebox{1\textwidth}{!}{
\begin{tabular}{rcccccc}
\hline
& \emph{glasso} & \emph{tlasso} & \emph{2Sgelnet} & \emph{2Sglasso} & \emph{2Stelnet} & \emph{2Stlasso} \\ 
\hline
n=100 & 0.319 & 2.221 & 0.397 & 0.192 & 0.344 & 0.167 \\ 
 & (0.022) & (0.589) & (0.021) & (0.010) & (0.018) & (0.007) \\ 
n=250 & 0.191 & 0.452 & 0.285 & 0.140 & 0.243 & 0.119 \\ 
 & (0.018) & (0.050) & (0.009) & (0.005) & (0.012) & (0.006) \\ 
n=500 & 0.125 & 0.251 & 0.216 & 0.108 & 0.194 & 0.093 \\ 
 & (0.009) & (0.025) & (0.008) & (0.004) & (0.017) & (0.006) \\ 
\hline
\end{tabular}
}
\caption{Scale-free}
\end{subfigure}
\hspace{0.5cm}
\begin{subfigure}[b]{0.45\textwidth}
\resizebox{1\textwidth}{!}{
\begin{tabular}{rcccccc}
\hline
& \emph{glasso} & \emph{tlasso} & \emph{2Sgelnet} & \emph{2Sglasso} & \emph{2Stelnet} & \emph{2Stlasso} \\ 
\hline
n=100 & 0.326 & 2.383 & 0.391 & 0.185 & 0.344 & 0.166 \\ 
 & (0.027) & (0.438) & (0.023) & (0.012) & (0.020) & (0.010) \\ 
n=250 & 0.190 & 0.425 & 0.284 & 0.139 & 0.240 & 0.115 \\ 
 & (0.022) & (0.043) & (0.014) & (0.007) & (0.015) & (0.007) \\ 
n=500 & 0.113 & 0.235 & 0.217 & 0.107 & 0.193 & 0.091 \\ 
 & (0.010) & (0.023) & (0.009) & (0.005) & (0.018) & (0.007) \\ 
\hline
\end{tabular}
}
\caption{Random}
\end{subfigure}

\vspace{1cm}

\begin{subfigure}[b]{0.45\textwidth}
\resizebox{1\textwidth}{!}{
\begin{tabular}{rcccccc}
\hline
& \emph{glasso} & \emph{tlasso} & \emph{2Sgelnet} & \emph{2Sglasso} & \emph{2Stelnet} & \emph{2Stlasso} \\ 
\hline
n=100 & 0.327 & 2.326 & 0.398 & 0.189 & 0.350 & 0.165 \\ 
 & (0.025) & (0.650) & (0.025) & (0.012) & (0.017) & (0.010) \\ 
n=250 & 0.190 & 0.481 & 0.291 & 0.141 & 0.258 & 0.121 \\ 
 & (0.020) & (0.055) & (0.012) & (0.006) & (0.016) & (0.008) \\ 
n=500 & 0.120 & 0.290 & 0.225 & 0.110 & 0.233 & 0.104 \\ 
 & (0.011) & (0.034) & (0.010) & (0.006) & (0.022) & (0.009) \\ 
\hline
\end{tabular}
}
\caption{Hub}
\end{subfigure}
\hspace{0.5cm}
\begin{subfigure}[b]{0.45\textwidth}
\resizebox{1\textwidth}{!}{
\begin{tabular}{rcccccc}
\hline
& \emph{glasso} & \emph{tlasso} & \emph{2Sgelnet} & \emph{2Sglasso} & \emph{2Stelnet} & \emph{2Stlasso} \\ 
\hline
n=100 & 0.326 & 2.524 & 0.388 & 0.185 & 0.349 & 0.164 \\ 
 & (0.027) & (0.521) & (0.028) & (0.012) & (0.024) & (0.010) \\ 
n=250 & 0.181 & 0.437 & 0.278 & 0.135 & 0.223 & 0.108 \\ 
 & (0.021) & (0.048) & (0.015) & (0.007) & (0.019) & (0.008) \\ 
n=500 & 0.112 & 0.218 & 0.212 & 0.105 & 0.150 & 0.074 \\ 
 & (0.012) & (0.027) & (0.012) & (0.006) & (0.010) & (0.005) \\ 
\hline
\end{tabular}
}
\caption{Cluster}
\end{subfigure}

\vspace{1cm}

\begin{subfigure}[b]{0.45\textwidth}
\resizebox{1\textwidth}{!}{
\begin{tabular}{rcccccc}
\hline
& \emph{glasso} & \emph{tlasso} & \emph{2Sgelnet} & \emph{2Sglasso} & \emph{2Stelnet} & \emph{2Stlasso} \\ 
\hline
n=100 & 0.329 & 2.491 & 0.376 & 0.179 & 0.330 & 0.158 \\ 
 & (0.025) & (0.597) & (0.026) & (0.015) & (0.022) & (0.010) \\ 
n=250 & 0.194 & 0.407 & 0.275 & 0.134 & 0.194 & 0.091 \\ 
 & (0.025) & (0.047) & (0.014) & (0.007) & (0.019) & (0.007) \\ 
n=500 & 0.110 & 0.199 & 0.204 & 0.101 & 0.133 & 0.067 \\ 
 & (0.012) & (0.023) & (0.011) & (0.005) & (0.009) & (0.004) \\ 
\hline
\end{tabular}
}
\caption{Band}
\end{subfigure}
\hspace{0.5cm}
\begin{subfigure}[b]{0.45\textwidth}
\resizebox{1\textwidth}{!}{
\begin{tabular}{rcccccc}
\hline
& \emph{glasso} & \emph{tlasso} & \emph{2Sgelnet} & \emph{2Sglasso} & \emph{2Stelnet} & \emph{2Stlasso} \\ 
\hline
n=100 & 0.319 & 2.345 & 0.397 & 0.186 & 0.339 & 0.164 \\ 
 & (0.027) & (0.580) & (0.025) & (0.012) & (0.019) & (0.011) \\ 
n=250 & 0.184 & 0.420 & 0.281 & 0.137 & 0.231 & 0.112 \\ 
 & (0.018) & (0.052) & (0.012) & (0.006) & (0.015) & (0.007) \\ 
n=500 & 0.114 & 0.225 & 0.214 & 0.105 & 0.182 & 0.088 \\ 
 & (0.010) & (0.023) & (0.010) & (0.004) & (0.015) & (0.007) \\ 
\hline
\end{tabular}
}
\caption{Small-world}
\end{subfigure}

\vspace{1cm}

\begin{subfigure}[b]{0.45\textwidth}
\resizebox{1\textwidth}{!}{
\begin{tabular}{rcccccc}
\hline
& \emph{glasso} & \emph{tlasso} & \emph{2Sgelnet} & \emph{2Sglasso} & \emph{2Stelnet} & \emph{2Stlasso} \\ 
\hline
n=100 & 0.327 & 2.341 & 0.396 & 0.187 & 0.342 & 0.166 \\ 
 & (0.023) & (0.535) & (0.024) & (0.012) & (0.017) & (0.010) \\ 
n=250 & 0.197 & 0.565 & 0.283 & 0.137 & 0.238 & 0.114 \\ 
 & (0.017) & (0.260) & (0.013) & (0.006) & (0.016) & (0.007) \\ 
n=500 & 0.122 & 0.256 & 0.215 & 0.105 & 0.172 & 0.082 \\ 
 & (0.014) & (0.027) & (0.010) & (0.005) & (0.013) & (0.005) \\ 
\hline
\end{tabular}
}
\caption{Core-periphery}
\end{subfigure}
\vspace{0.5cm}
\caption{Mean values of optimal $\lambda$ (std. dev. in brackets) - Contaminated normal} 
\label{lambdas_contnorm}
\end{table}
\clearpage
\section{Additional simulation results} \label{sec:app4}
\vspace{-0.5cm}
\begin{figure}[H]
\begin{subfigure}{0.4\textwidth}
  \includegraphics[width=1\linewidth, angle = 0]{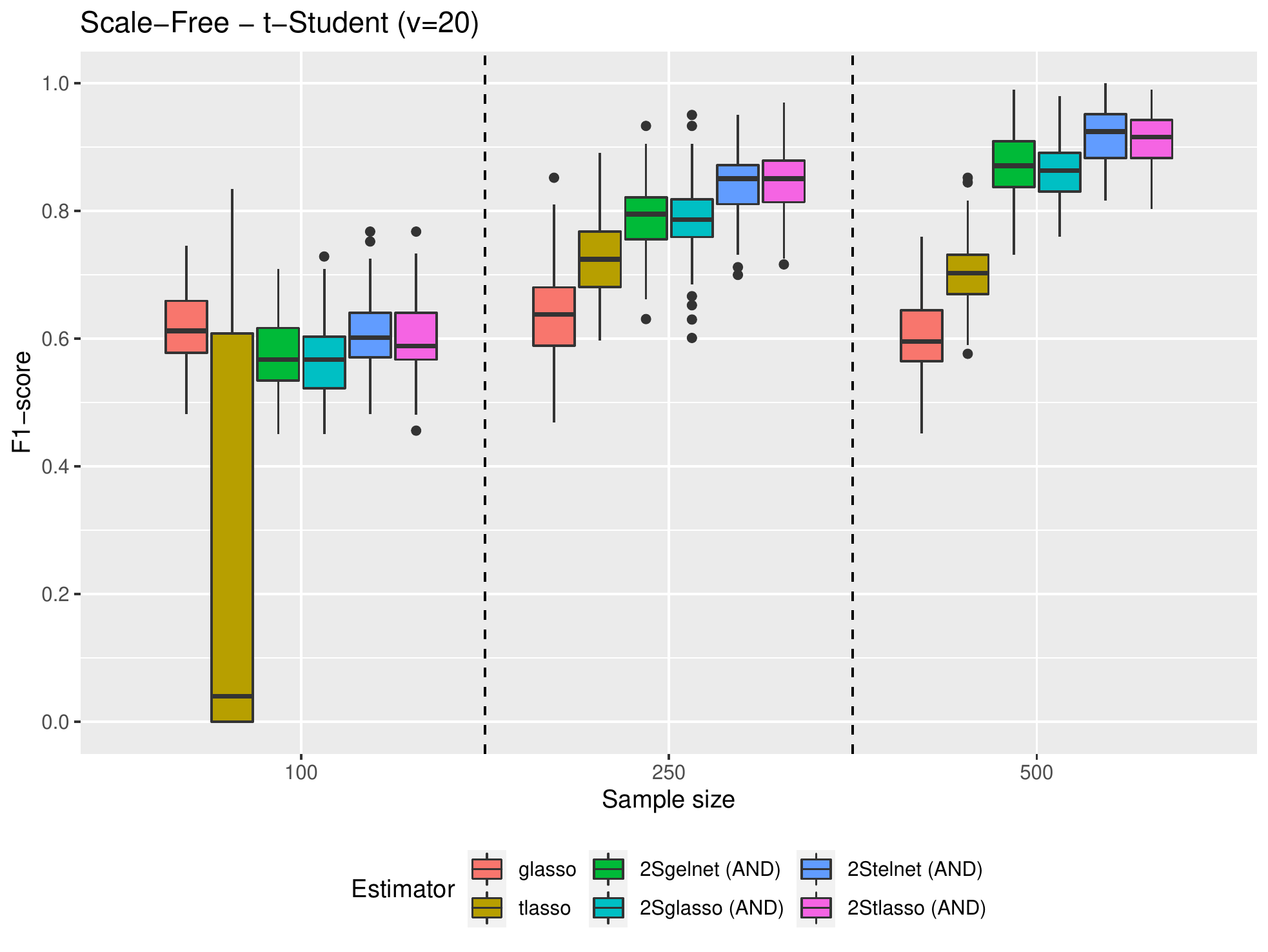}
\end{subfigure}
\begin{subfigure}{0.4\textwidth}
  \includegraphics[width=1\linewidth, angle = 0]{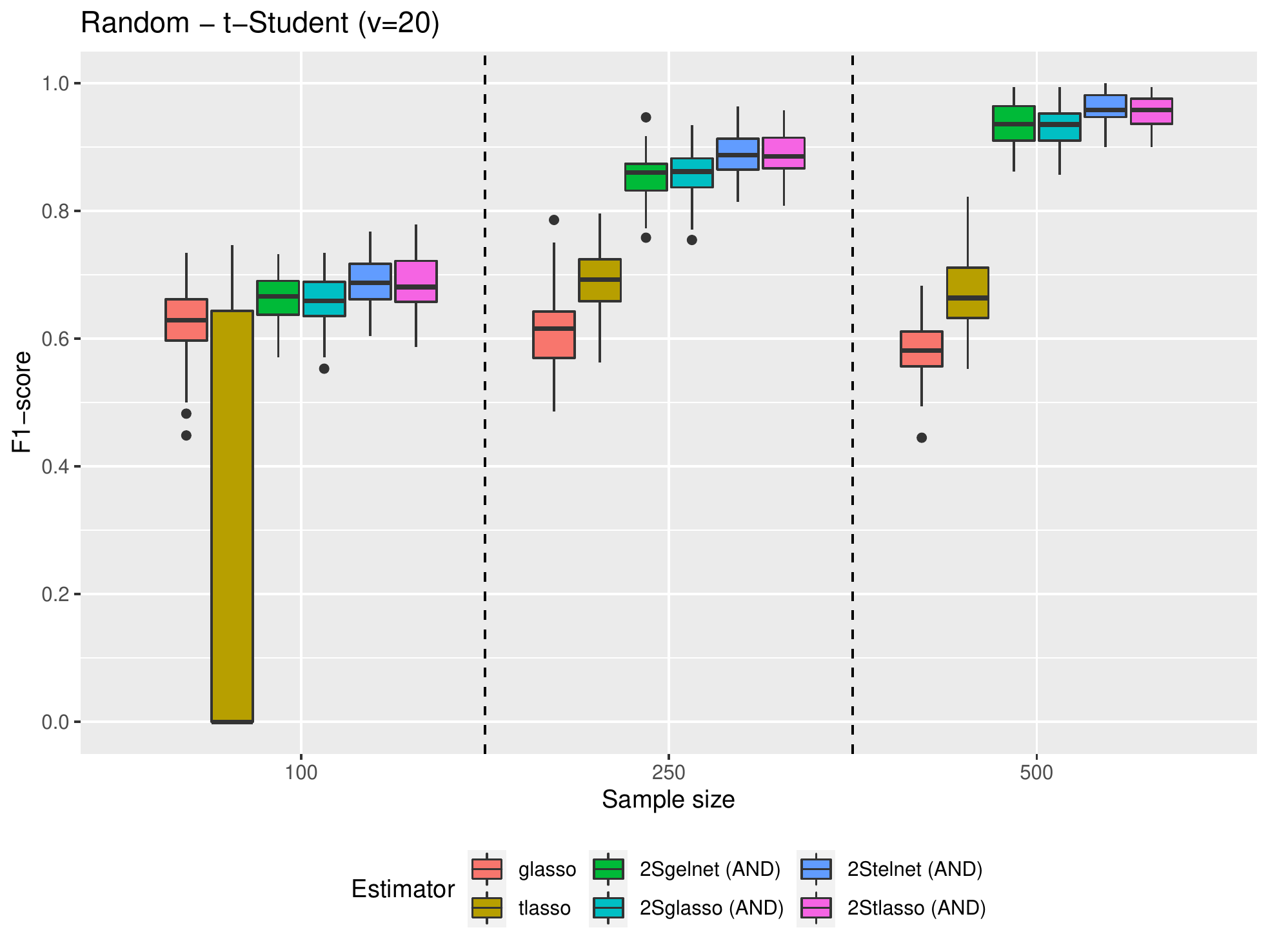}
\end{subfigure}
\newline
\begin{subfigure}{0.4\textwidth}
  \includegraphics[width=1\linewidth, angle = 0]{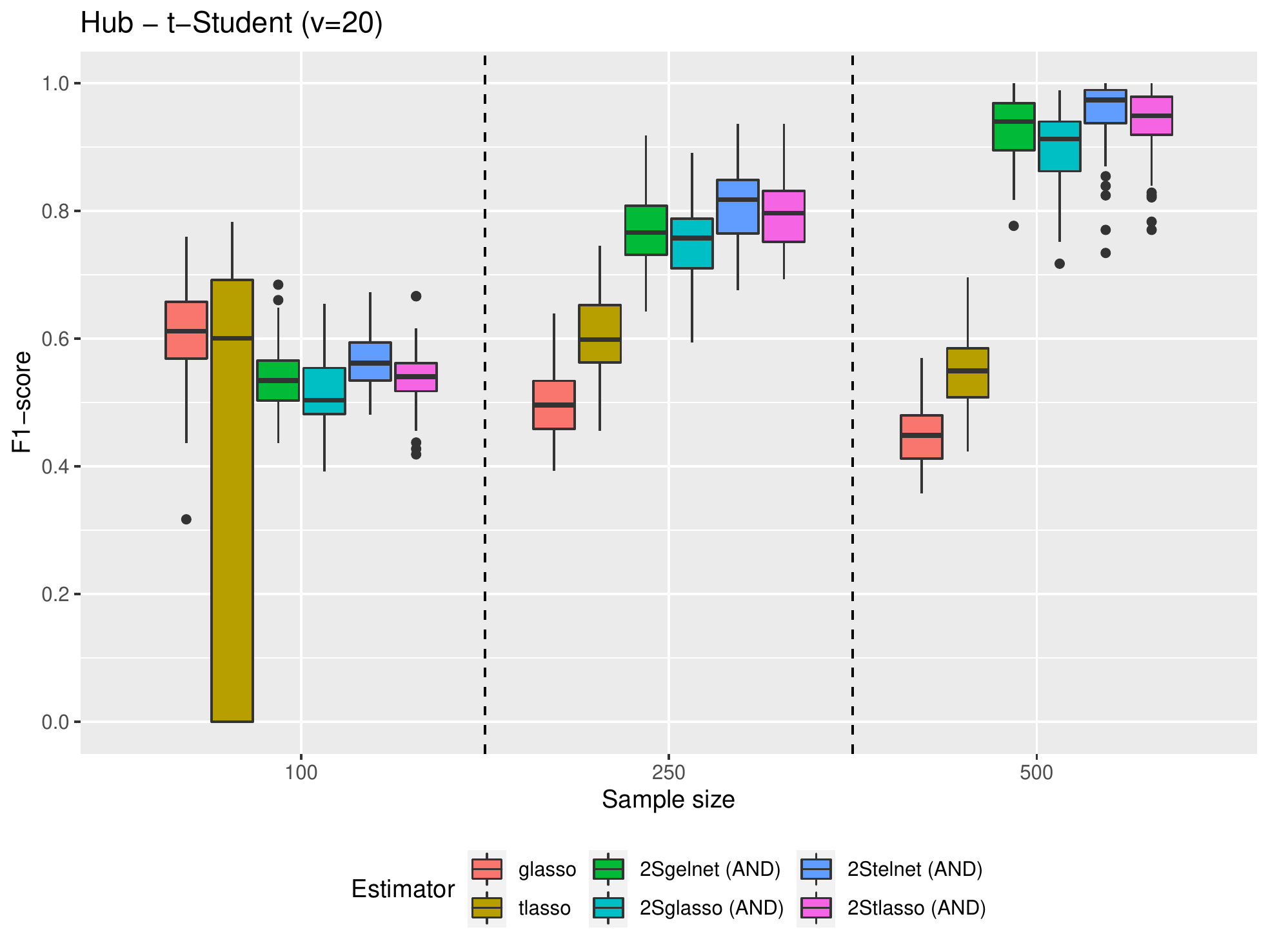}
\end{subfigure}
\begin{subfigure}{0.4\textwidth}
  \includegraphics[width=1\linewidth, angle = 0]{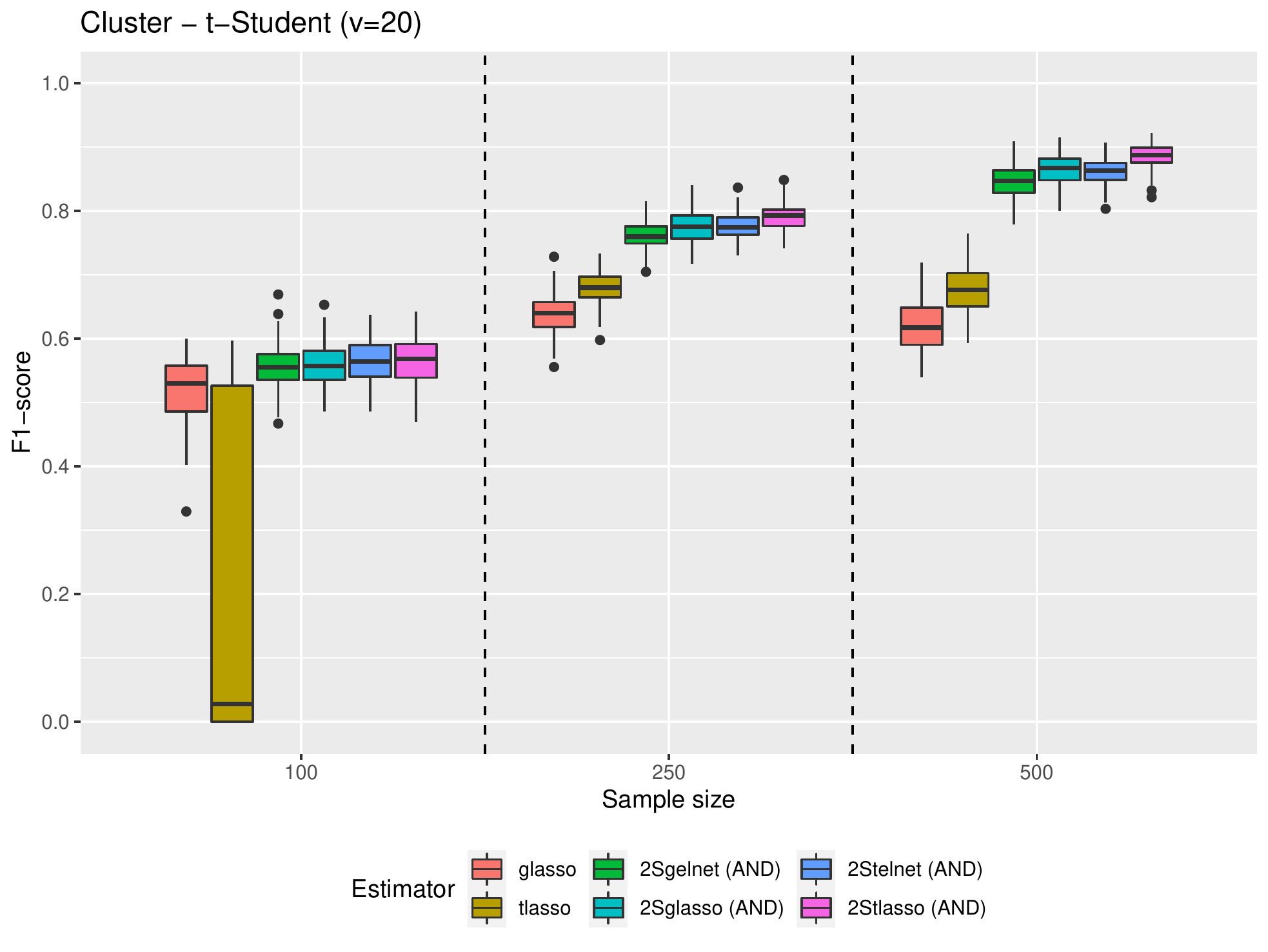}
\end{subfigure}
\newline
\begin{subfigure}{0.4\textwidth}
  \includegraphics[width=1\linewidth, angle = 0]{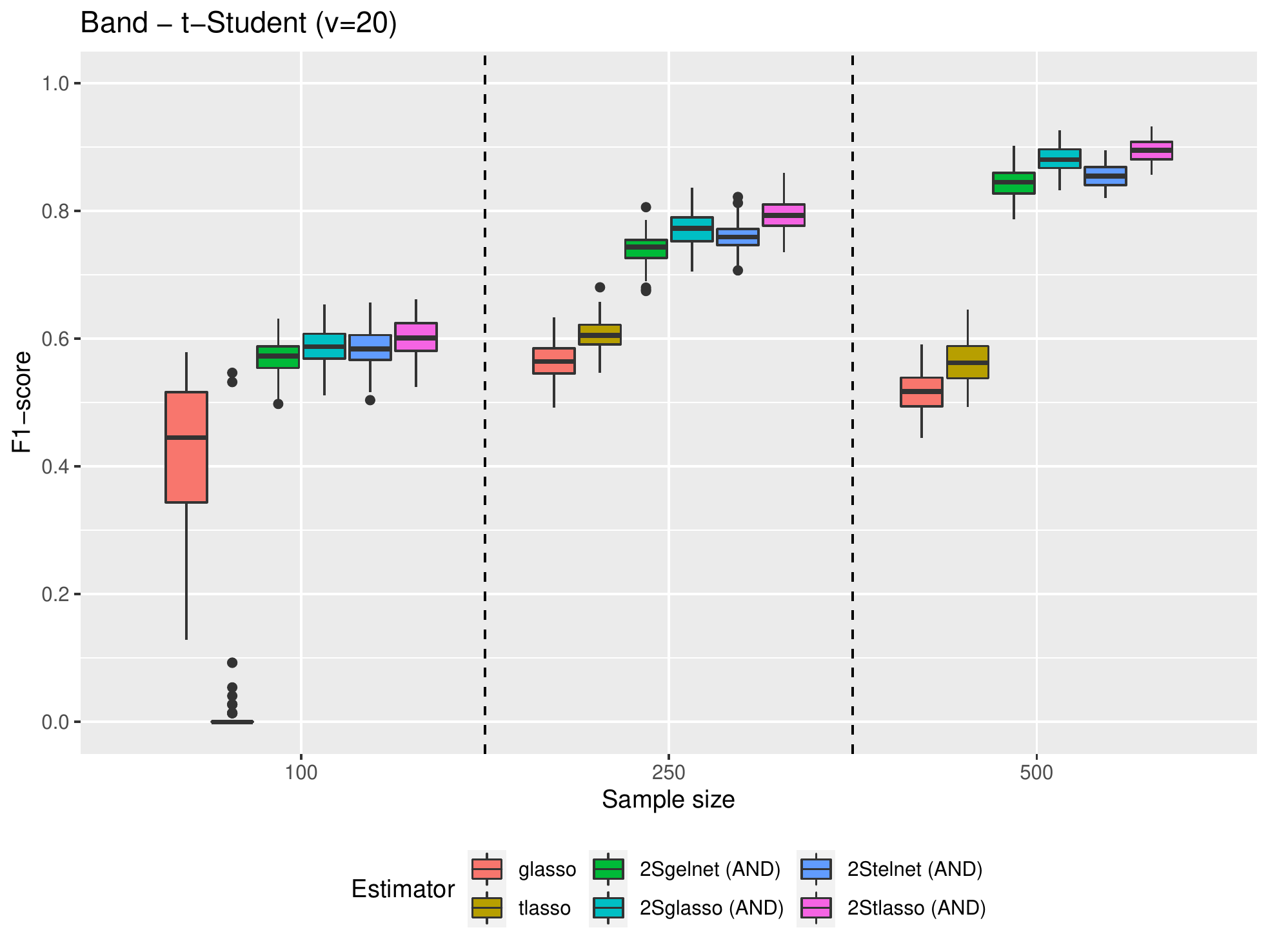}
\end{subfigure}
\begin{subfigure}{0.4\textwidth}
  \includegraphics[width=1\linewidth, angle = 0]{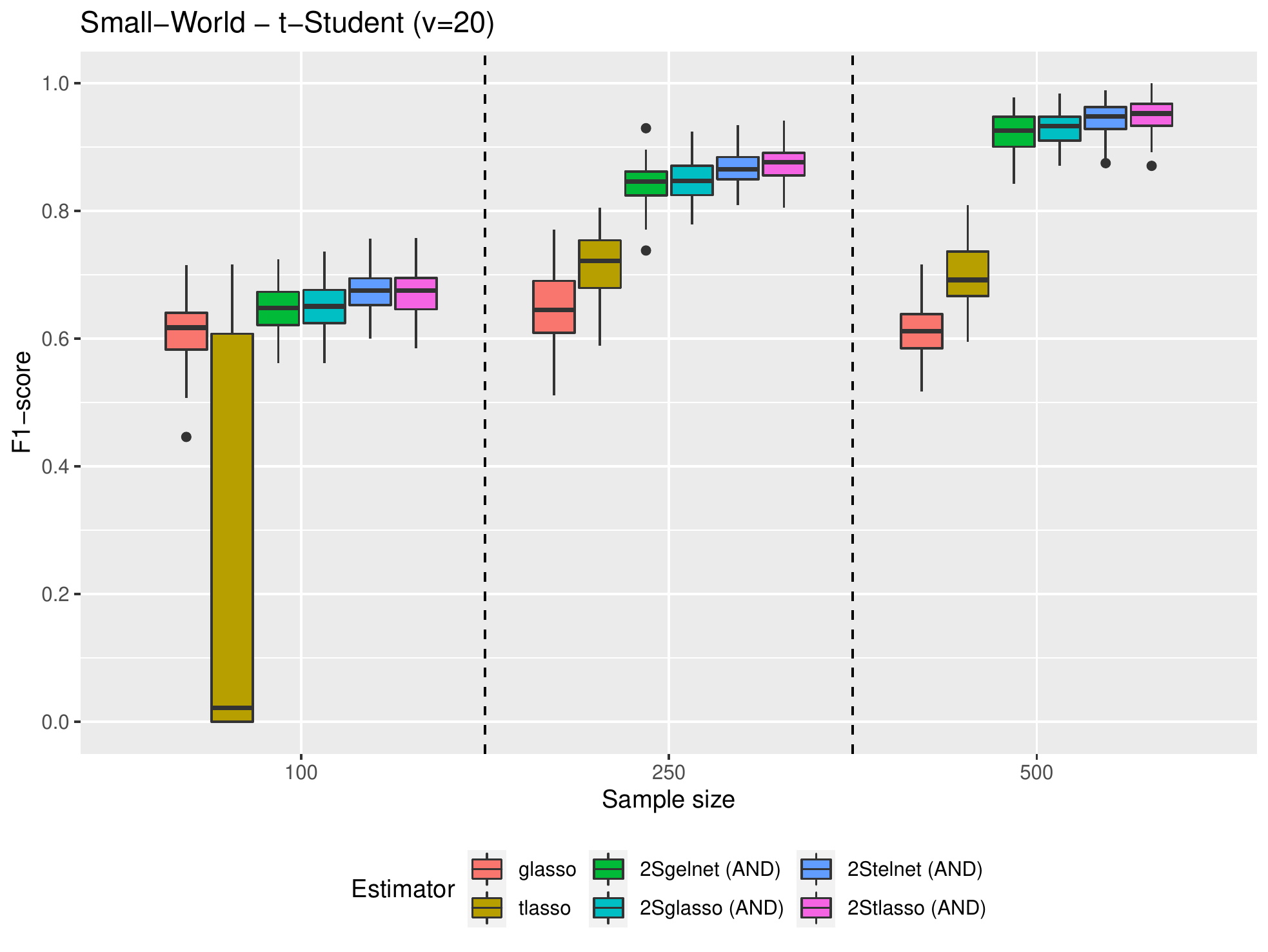}
\end{subfigure}
\newline
\centering
\begin{subfigure}{0.4\textwidth}
  \includegraphics[width=1\linewidth, angle = 0]{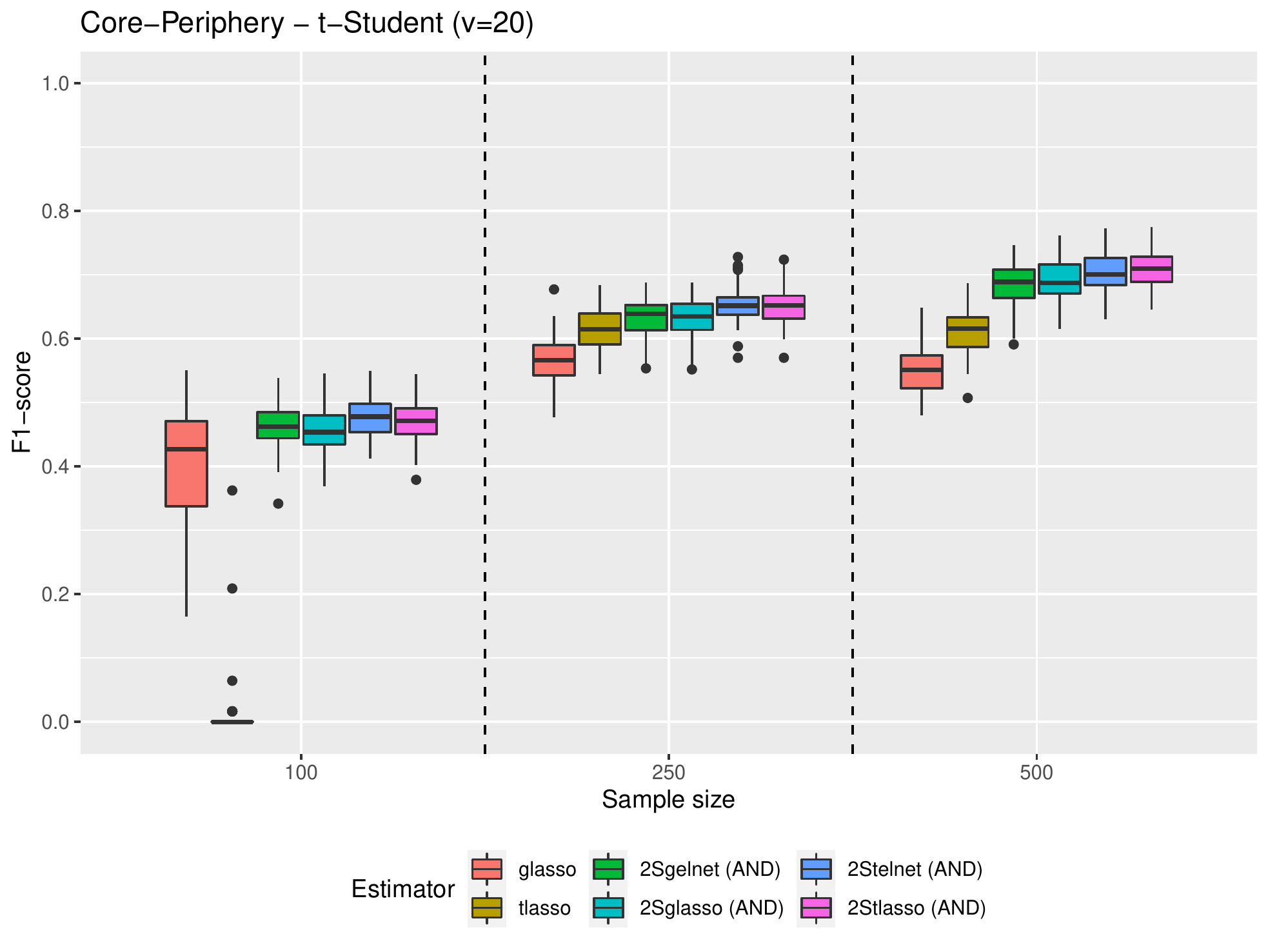}
\end{subfigure}
\caption{F$_1$-score - t-Student distribution with $v=20$}
\label{fig:f1ststd20}
\end{figure}
\clearpage

\begin{figure}[h]
\begin{subfigure}{0.4\textwidth}
  \includegraphics[width=1\linewidth, angle = 0]{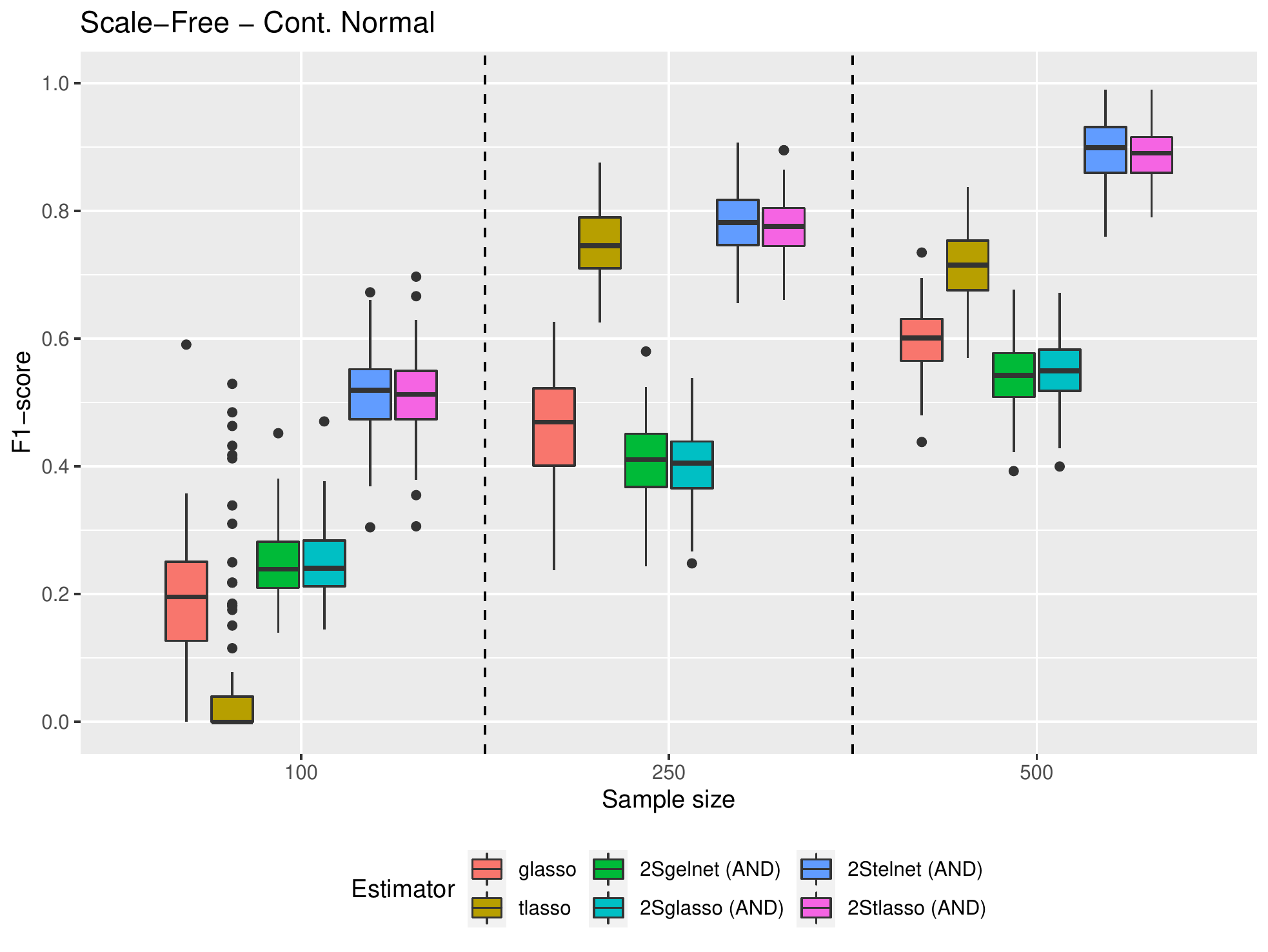}
\end{subfigure}
\begin{subfigure}{0.4\textwidth}
  \includegraphics[width=1\linewidth, angle = 0]{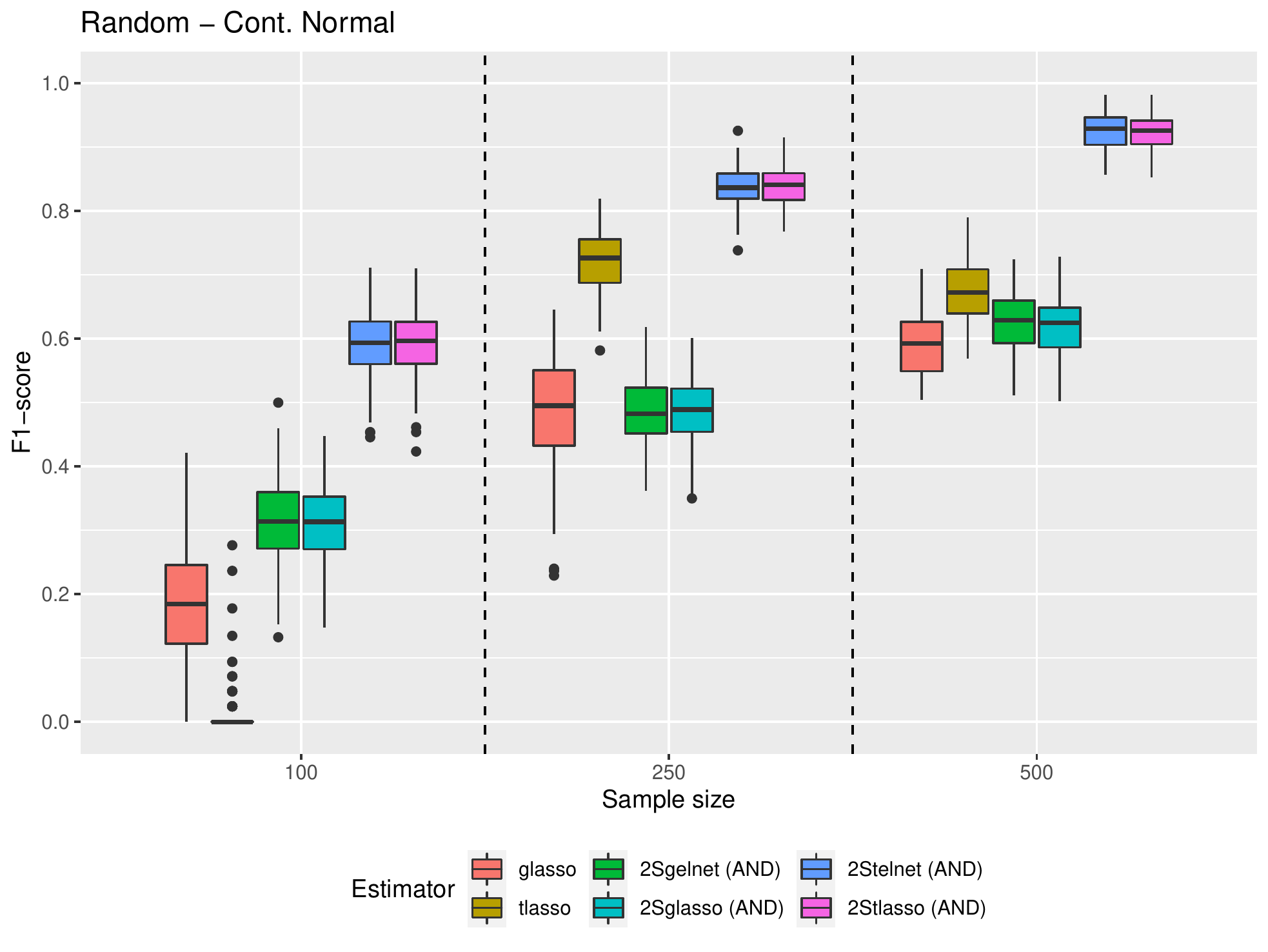}
\end{subfigure}
\newline
\begin{subfigure}{0.4\textwidth}
  \includegraphics[width=1\linewidth, angle = 0]{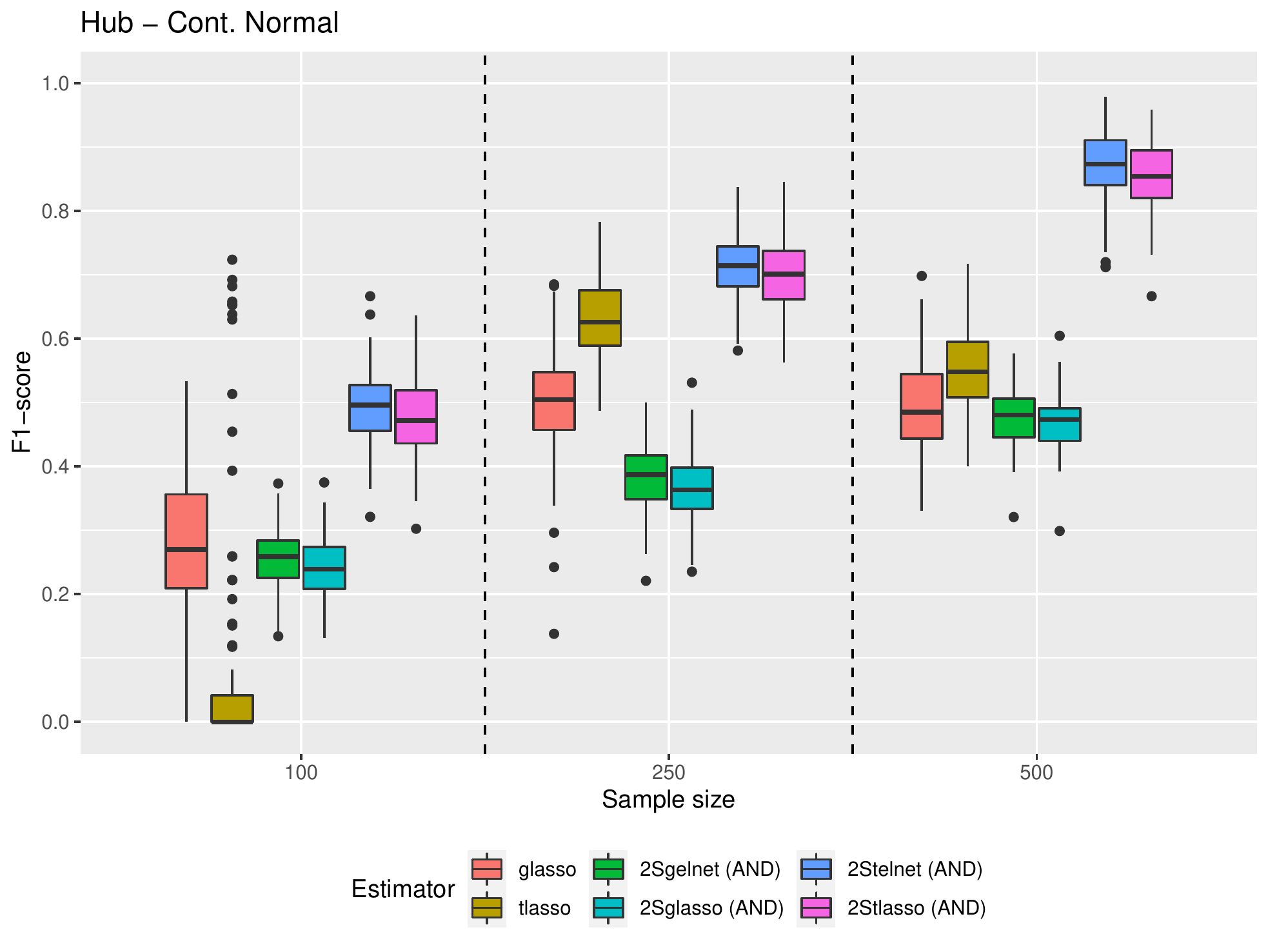}
\end{subfigure}
\begin{subfigure}{0.4\textwidth}
  \includegraphics[width=1\linewidth, angle = 0]{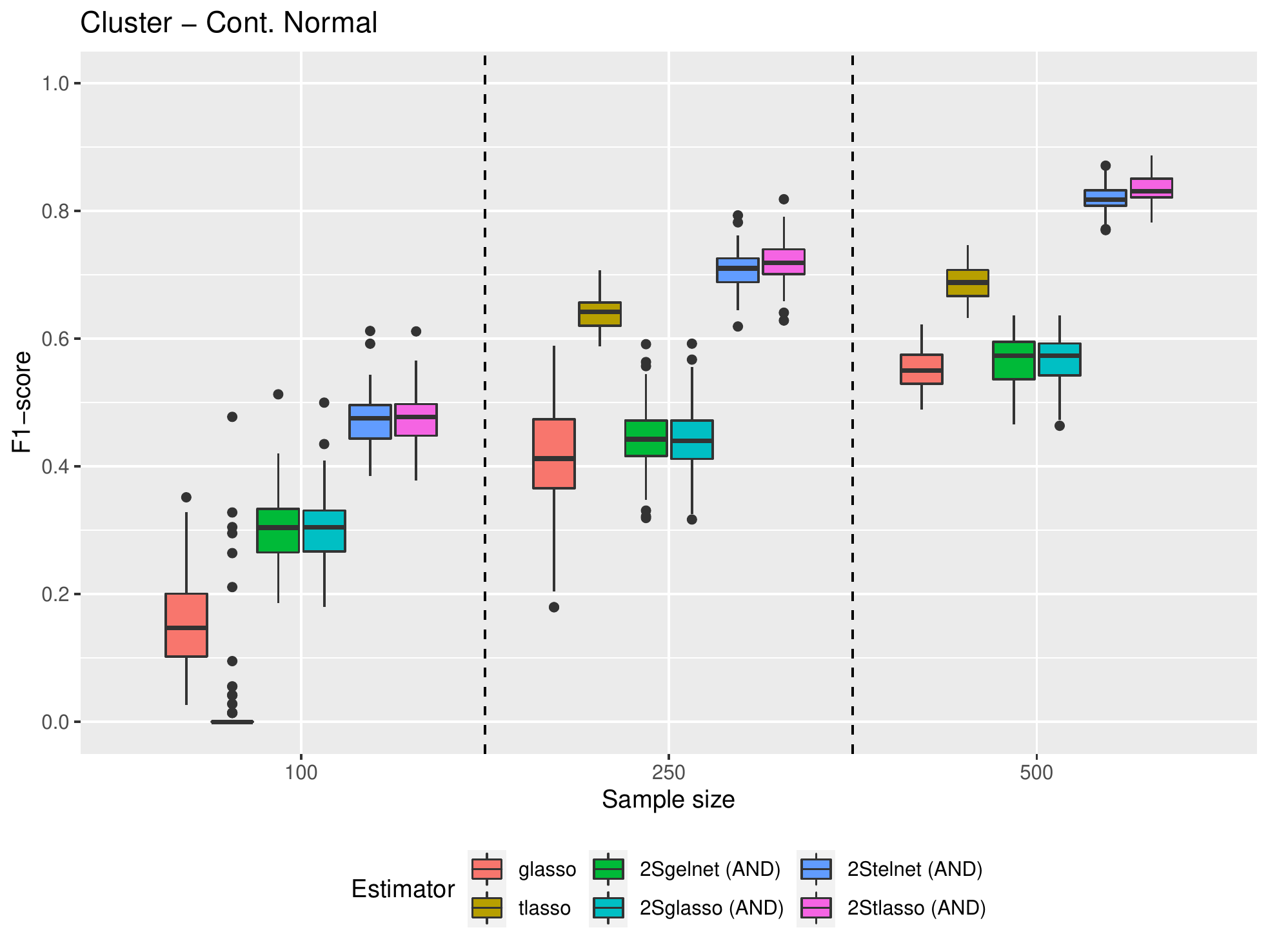}
\end{subfigure}
\newline
\begin{subfigure}{0.4\textwidth}
  \includegraphics[width=1\linewidth, angle = 0]{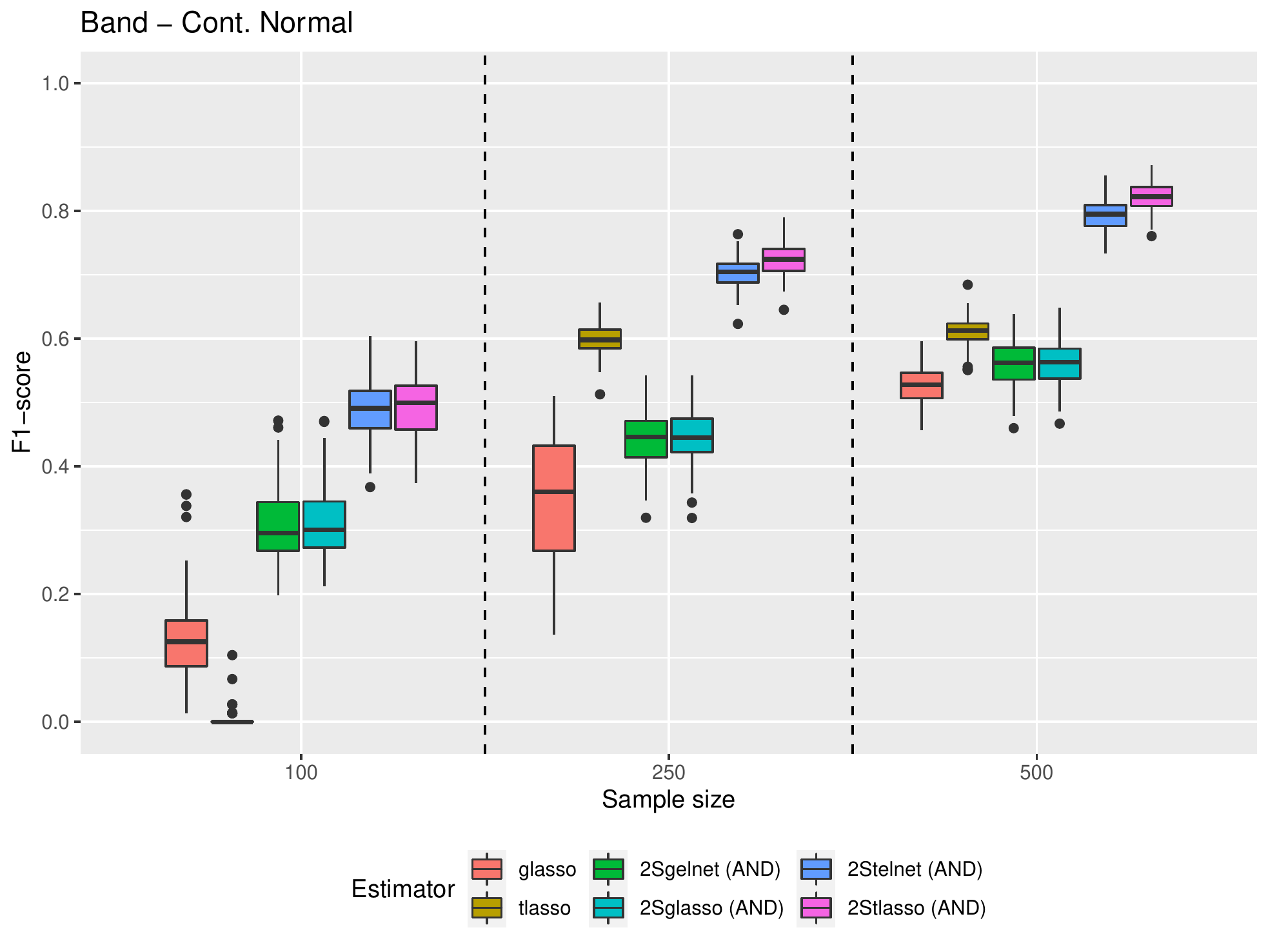}
\end{subfigure}
\begin{subfigure}{0.4\textwidth}
  \includegraphics[width=1\linewidth, angle = 0]{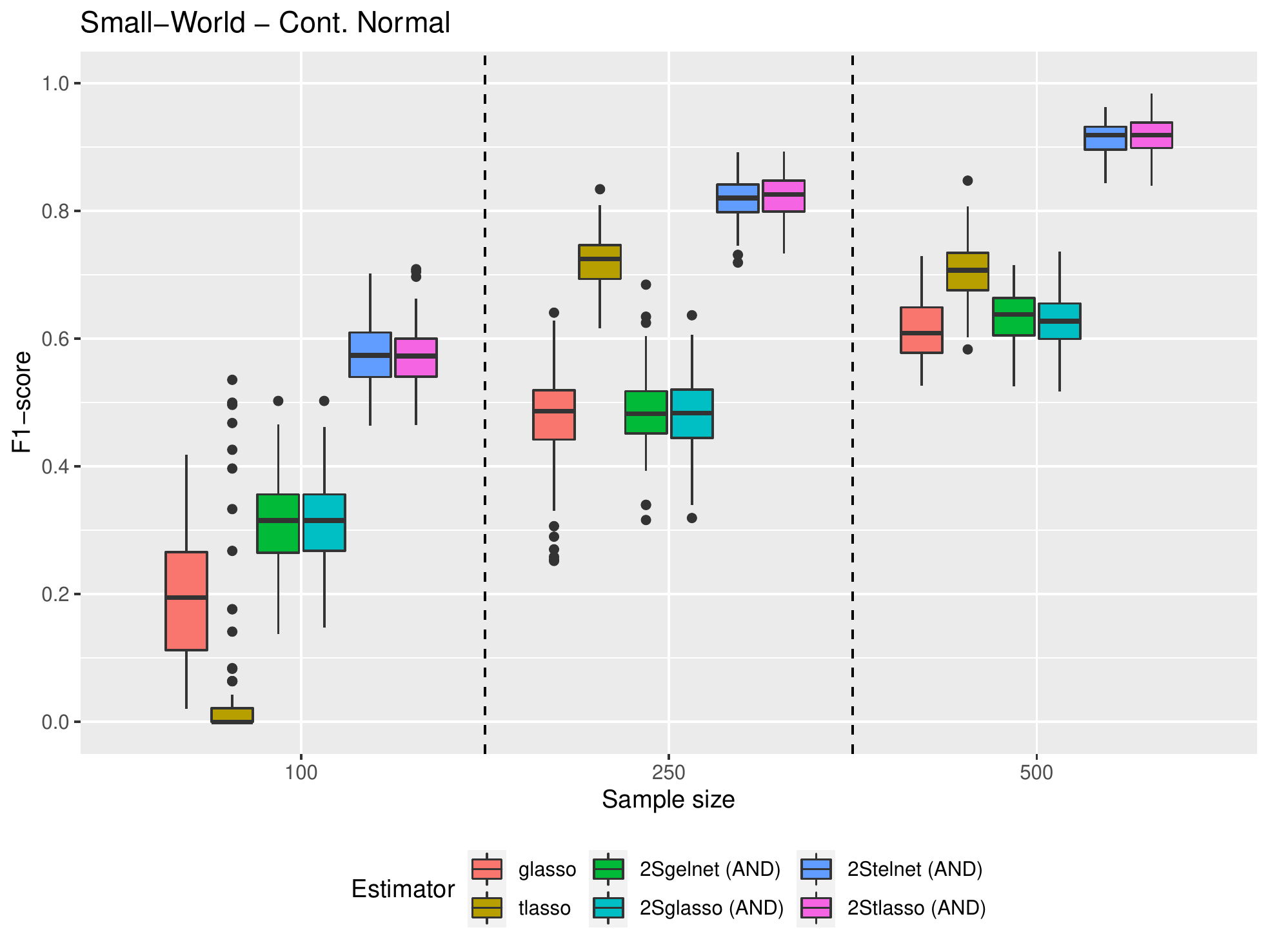}
\end{subfigure}
\newline
\centering
\begin{subfigure}{0.4\textwidth}
  \includegraphics[width=1\linewidth, angle = 0]{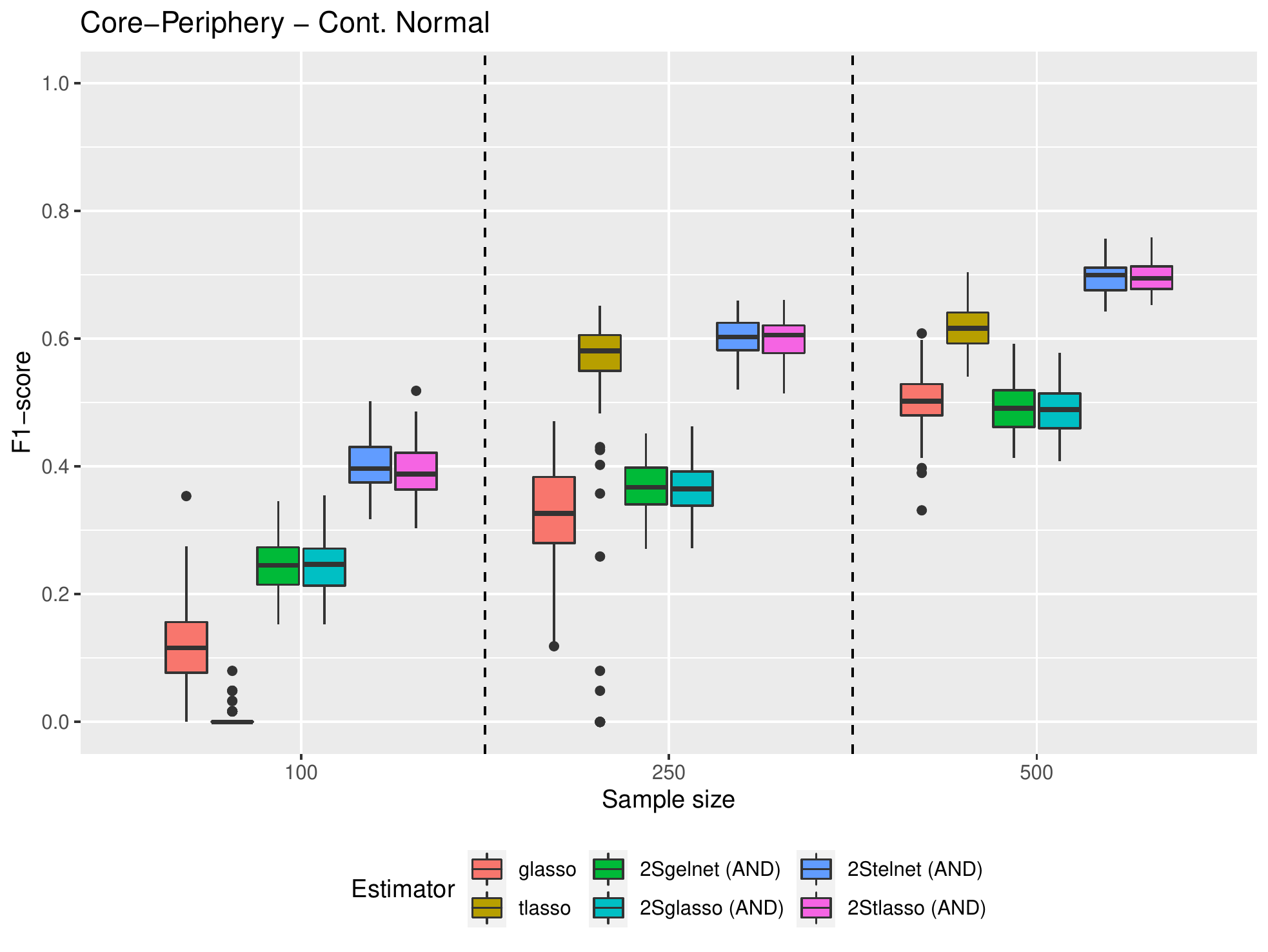}
\end{subfigure}
\caption{F$_1$-score - Contaminated normal distribution}
\label{fig:f1scontnorm}
\end{figure}
\clearpage

\begin{figure}[h]
\begin{subfigure}{0.4\textwidth}
  \includegraphics[width=1\linewidth, angle = 0]{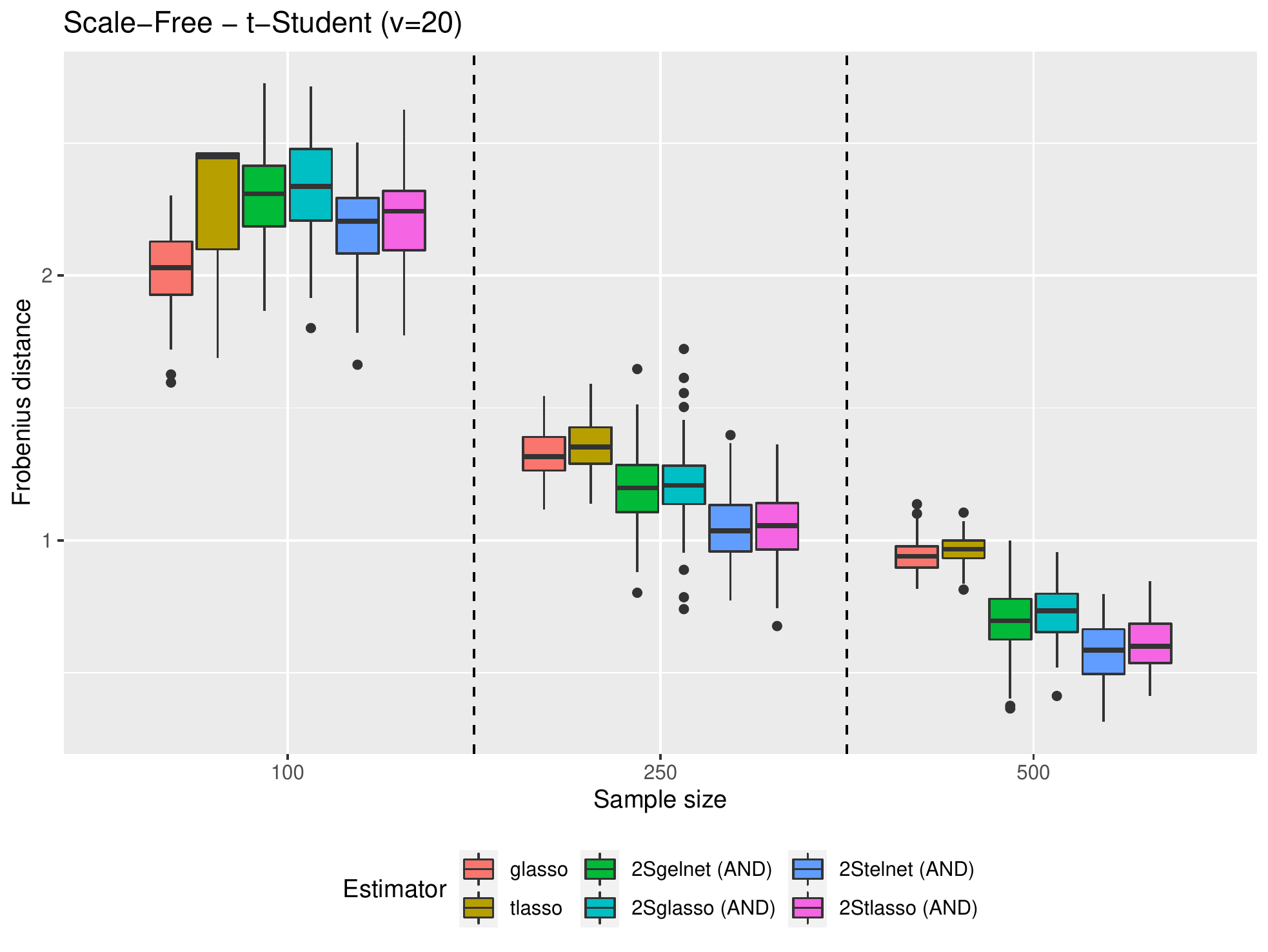}
\end{subfigure}
\begin{subfigure}{0.4\textwidth}
  \includegraphics[width=1\linewidth, angle = 0]{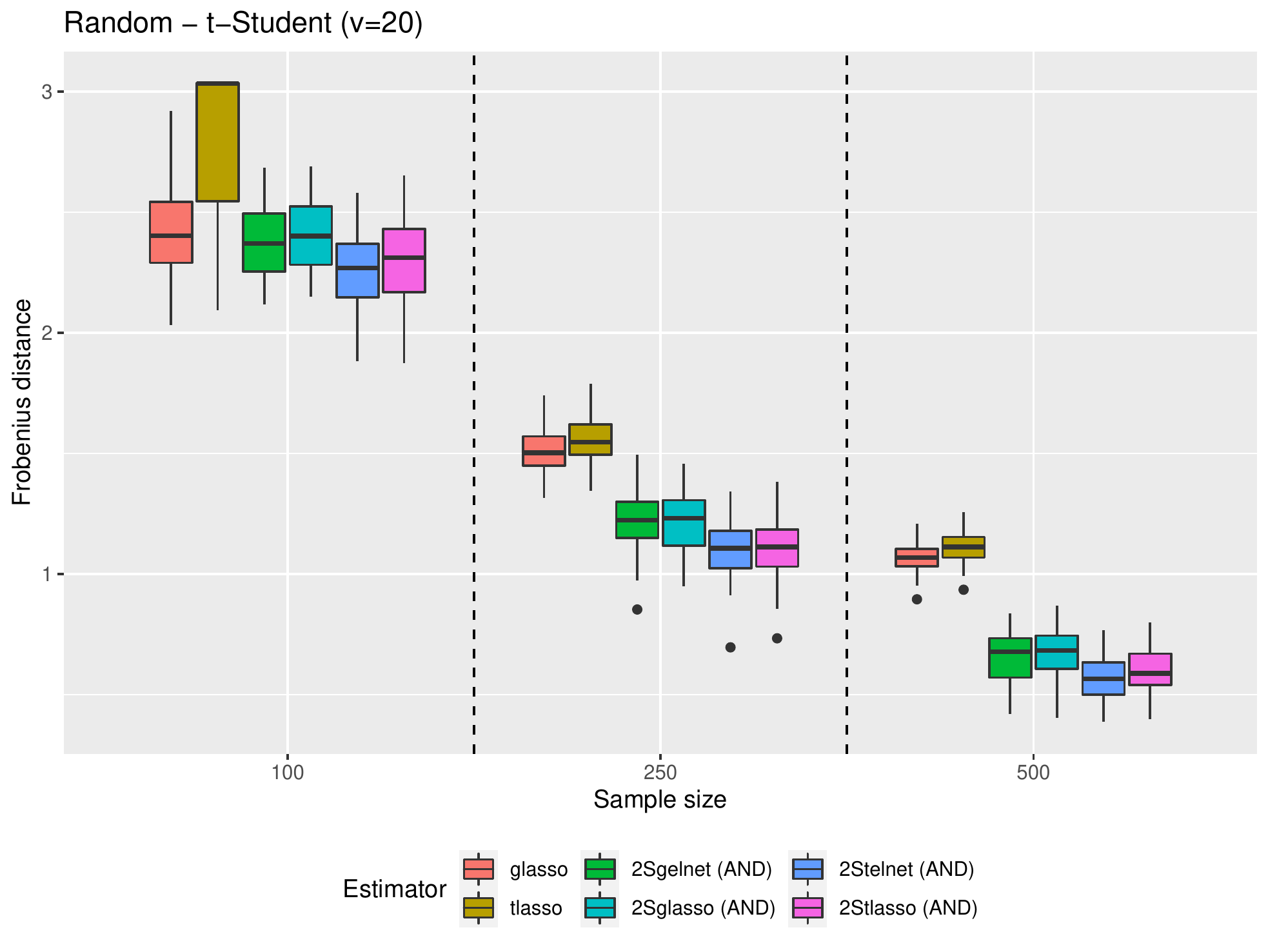}
\end{subfigure}
\newline
\begin{subfigure}{0.4\textwidth}
  \includegraphics[width=1\linewidth, angle = 0]{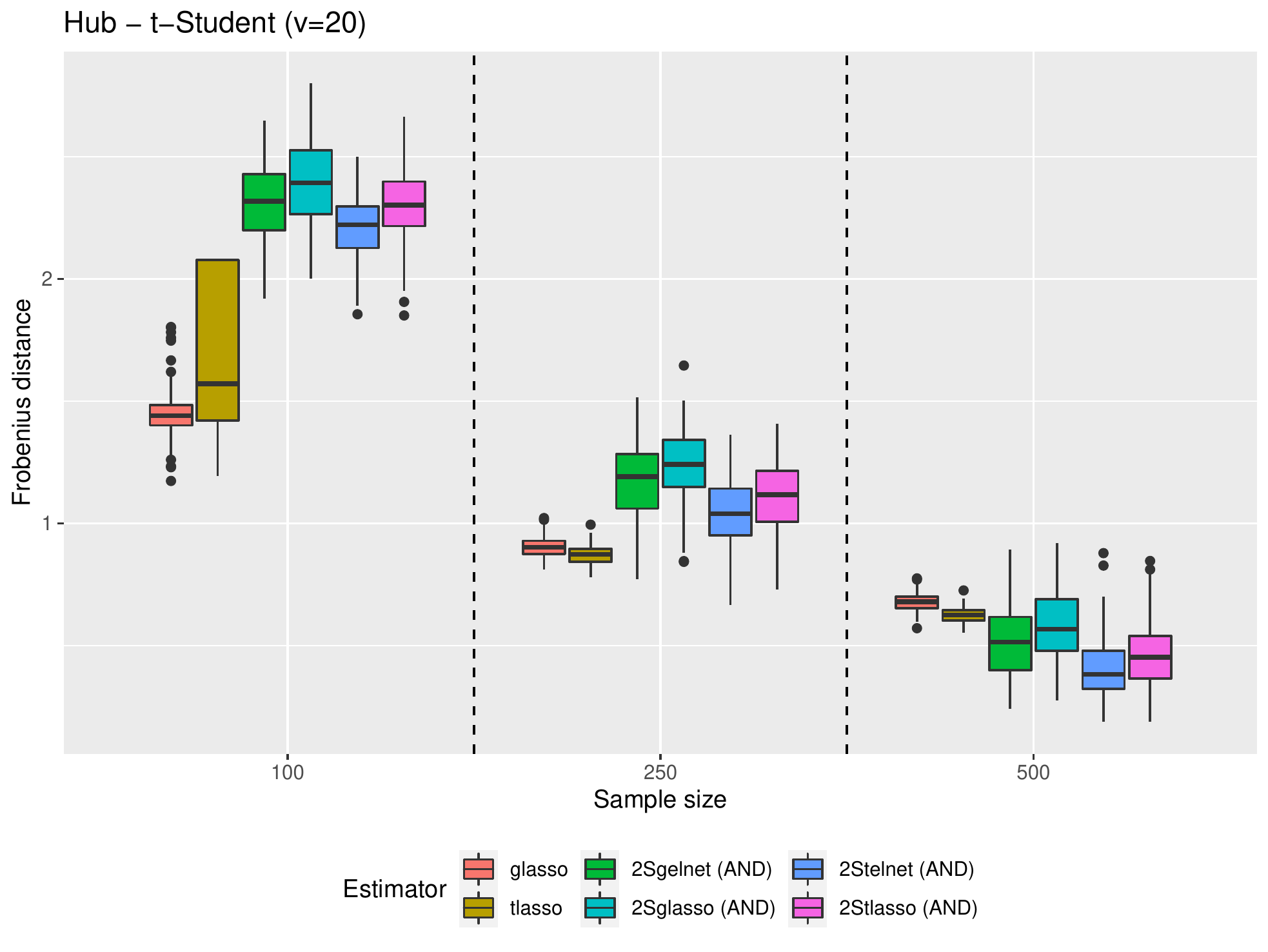}
\end{subfigure}
\begin{subfigure}{0.4\textwidth}
  \includegraphics[width=1\linewidth, angle = 0]{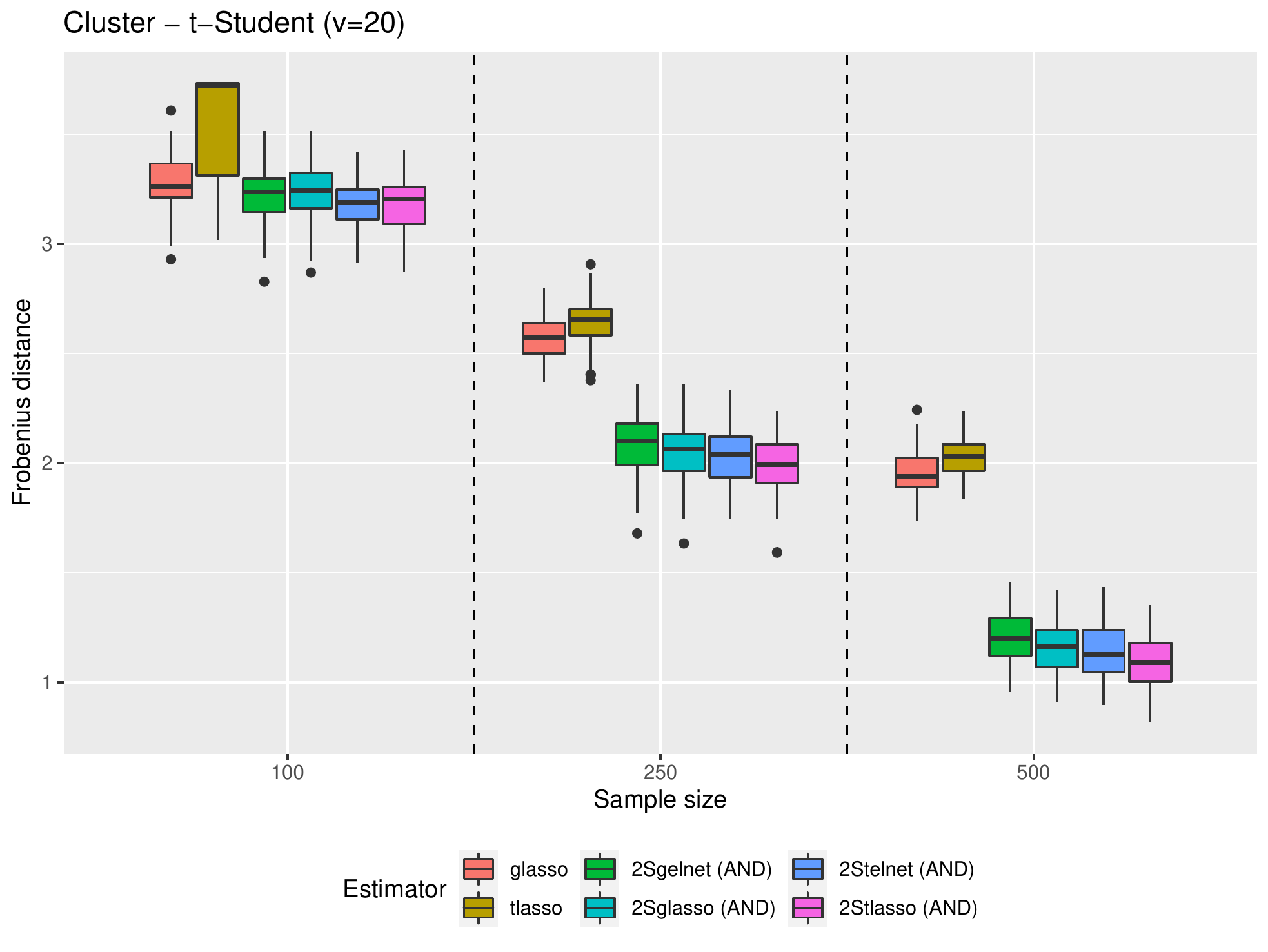}
\end{subfigure}
\newline
\begin{subfigure}{0.4\textwidth}
  \includegraphics[width=1\linewidth, angle = 0]{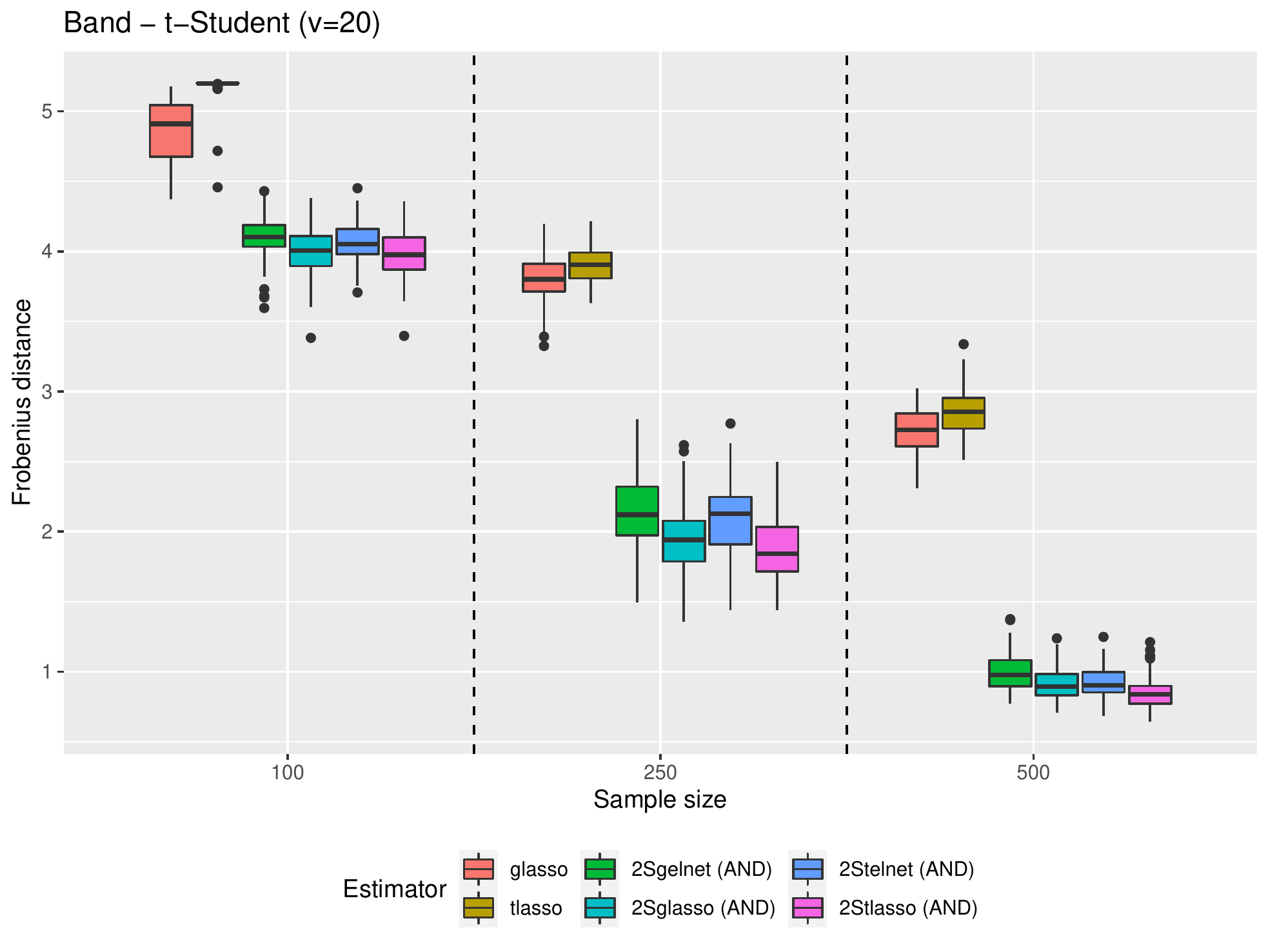}
\end{subfigure}
\begin{subfigure}{0.4\textwidth}
  \includegraphics[width=1\linewidth, angle = 0]{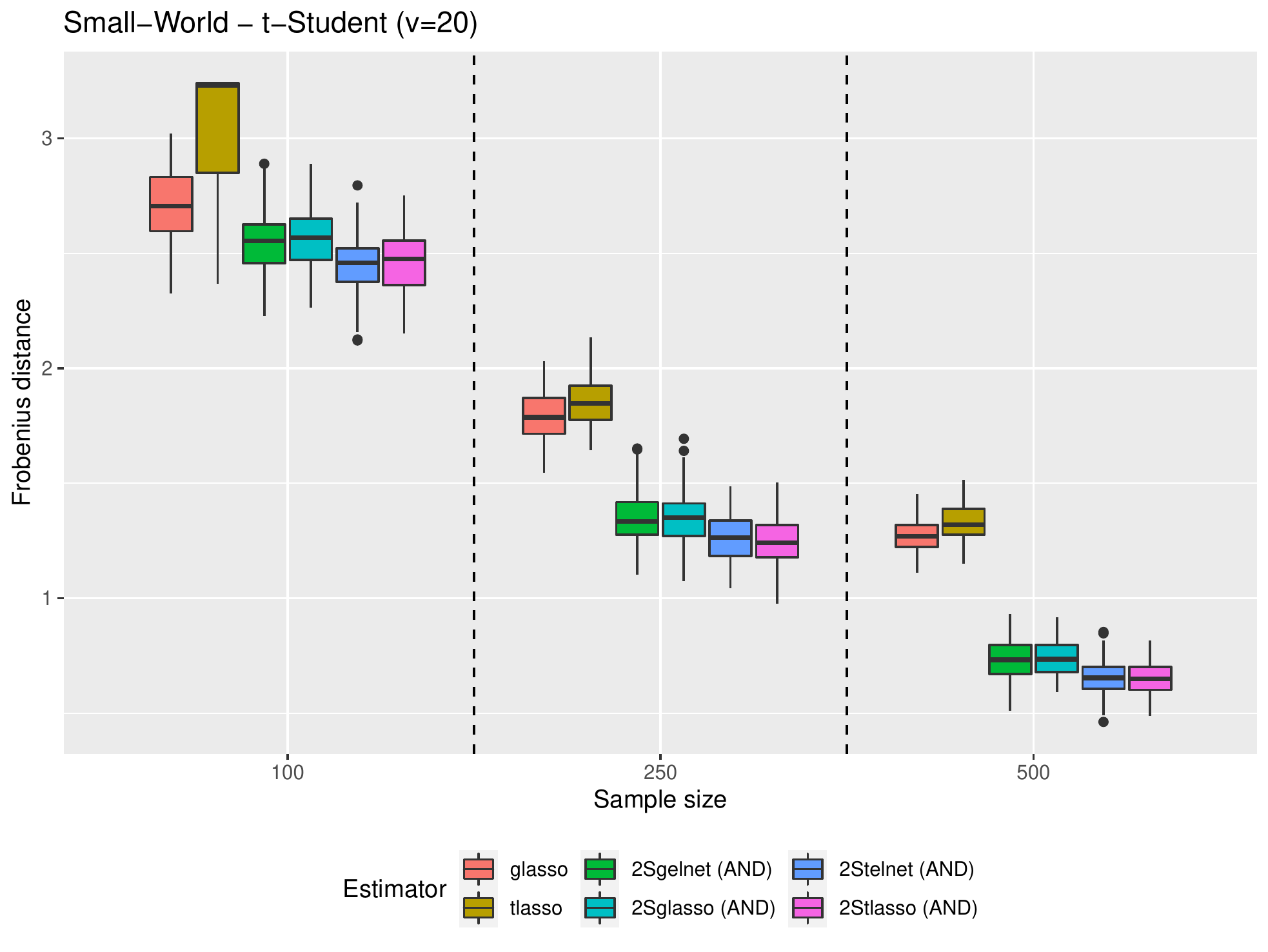}
\end{subfigure}
\newline
\centering
\begin{subfigure}{0.4\textwidth}
  \includegraphics[width=1\linewidth, angle = 0]{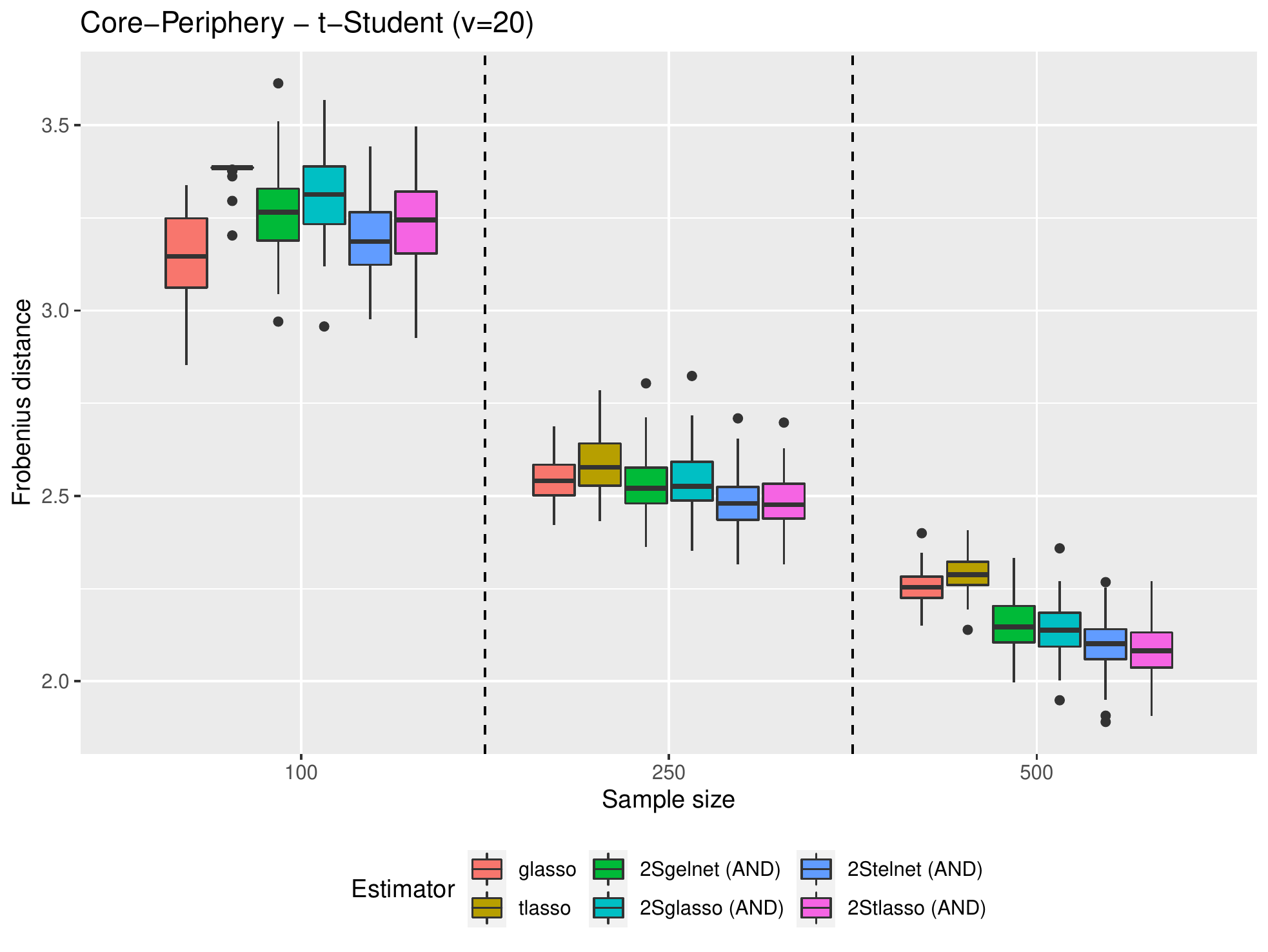}
\end{subfigure}
\caption{Frobenius distance - t-Student distribution with $v=20$}
\label{fig:fdpctstd20}
\end{figure}
\clearpage

\begin{figure}[h]
\begin{subfigure}{0.4\textwidth}
  \includegraphics[width=1\linewidth, angle = 0]{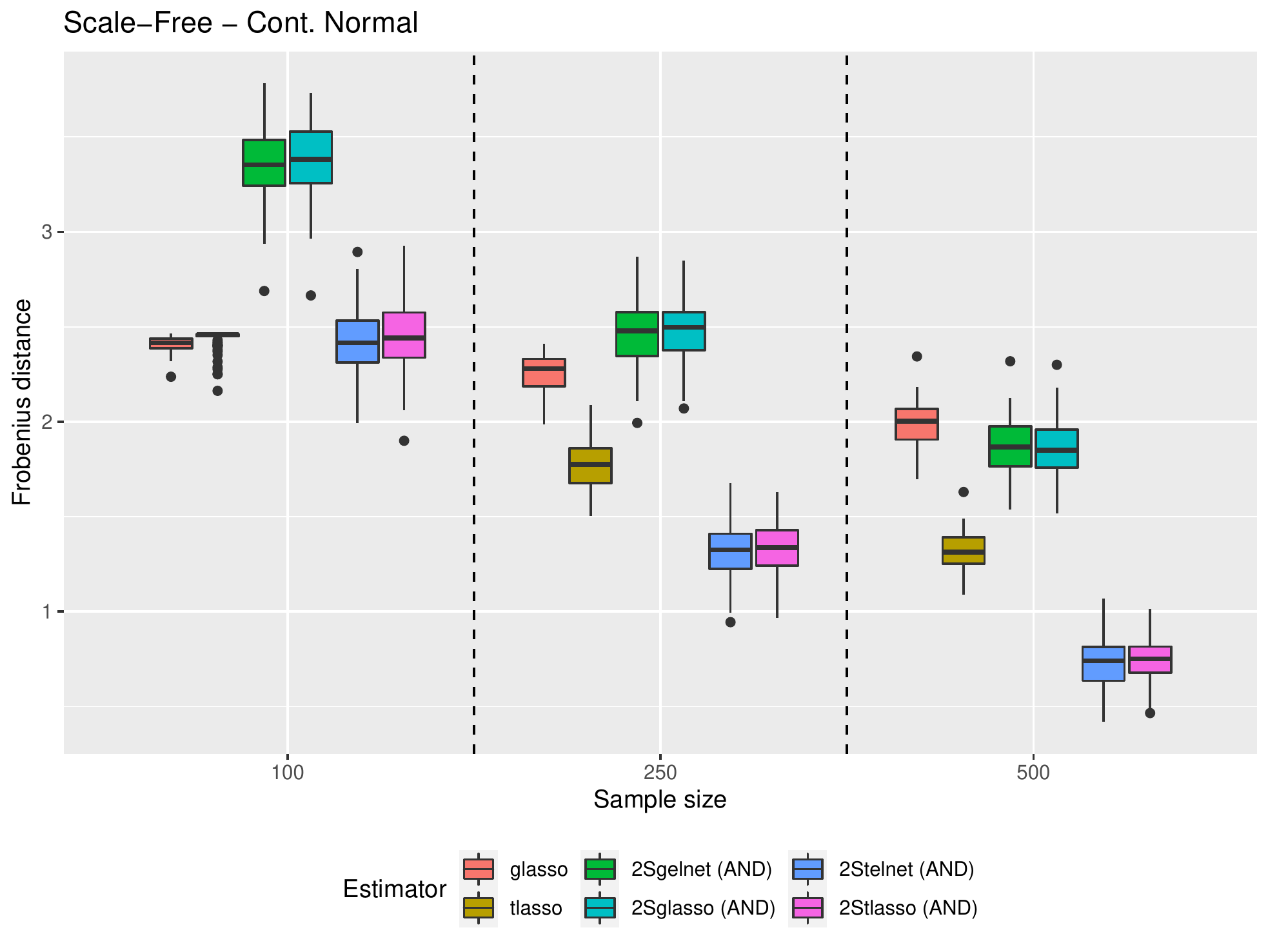}
\end{subfigure}
\begin{subfigure}{0.4\textwidth}
  \includegraphics[width=1\linewidth, angle = 0]{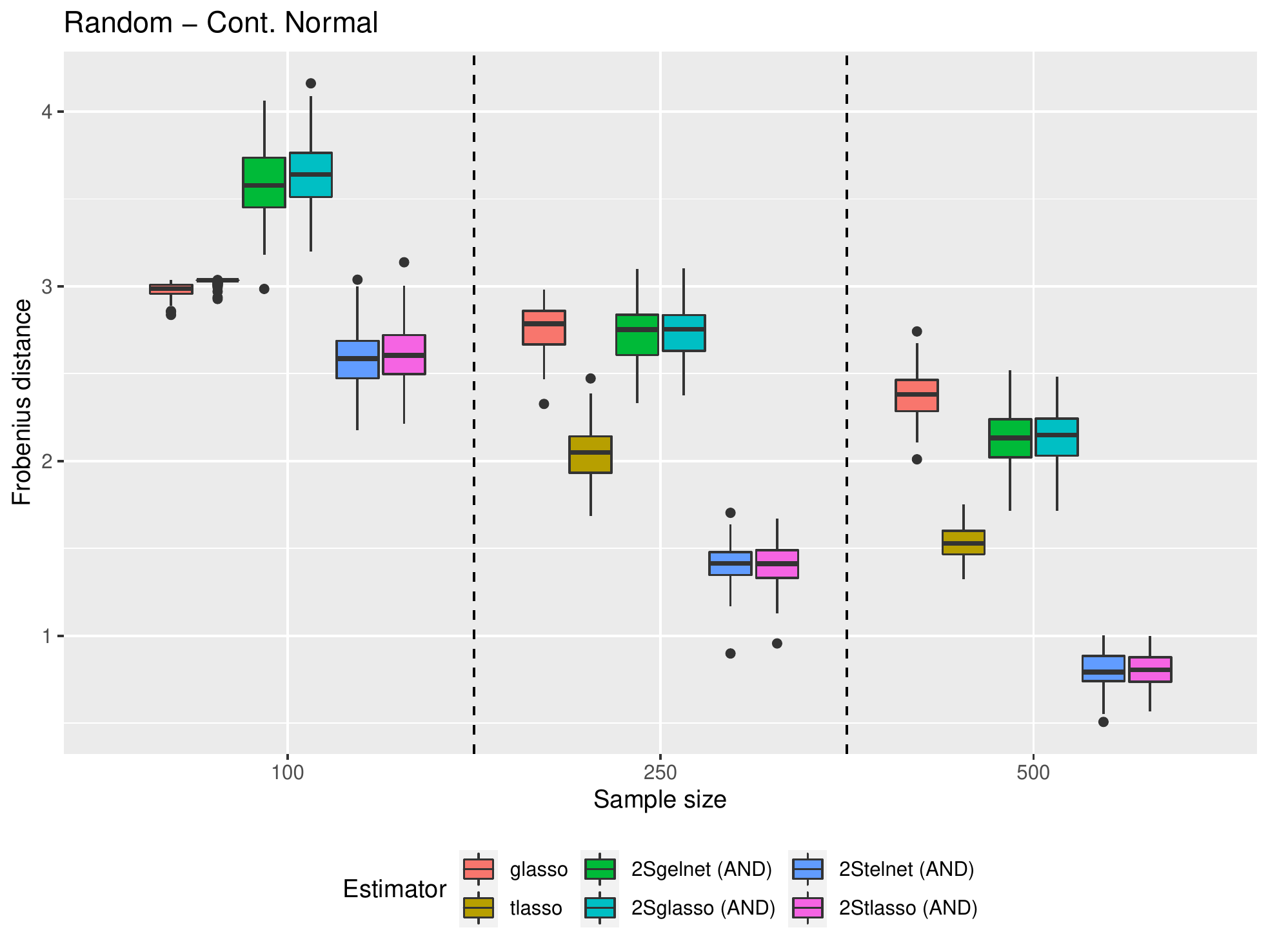}
\end{subfigure}
\newline
\begin{subfigure}{0.4\textwidth}
  \includegraphics[width=1\linewidth, angle = 0]{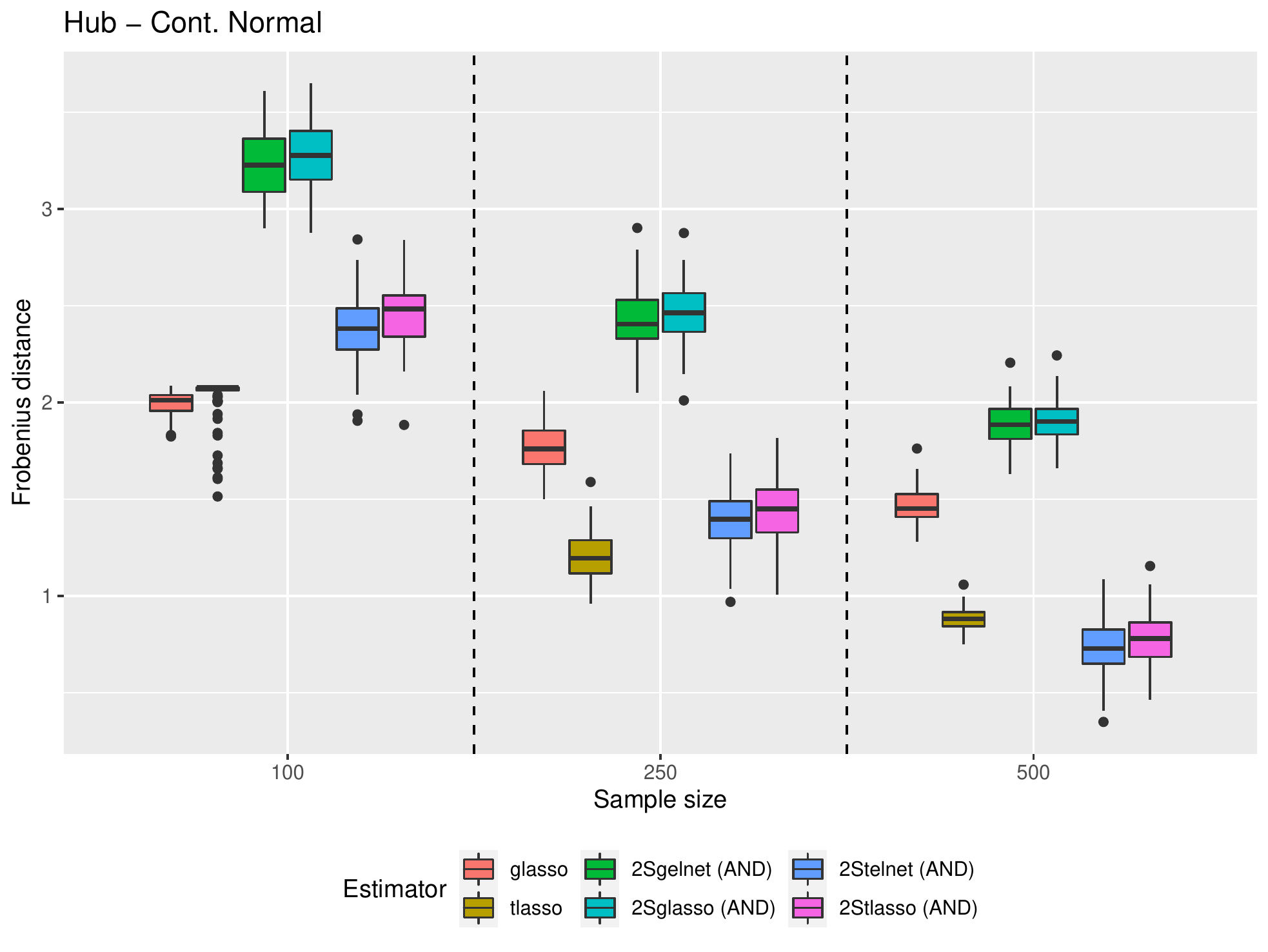}
\end{subfigure}
\begin{subfigure}{0.4\textwidth}
  \includegraphics[width=1\linewidth, angle = 0]{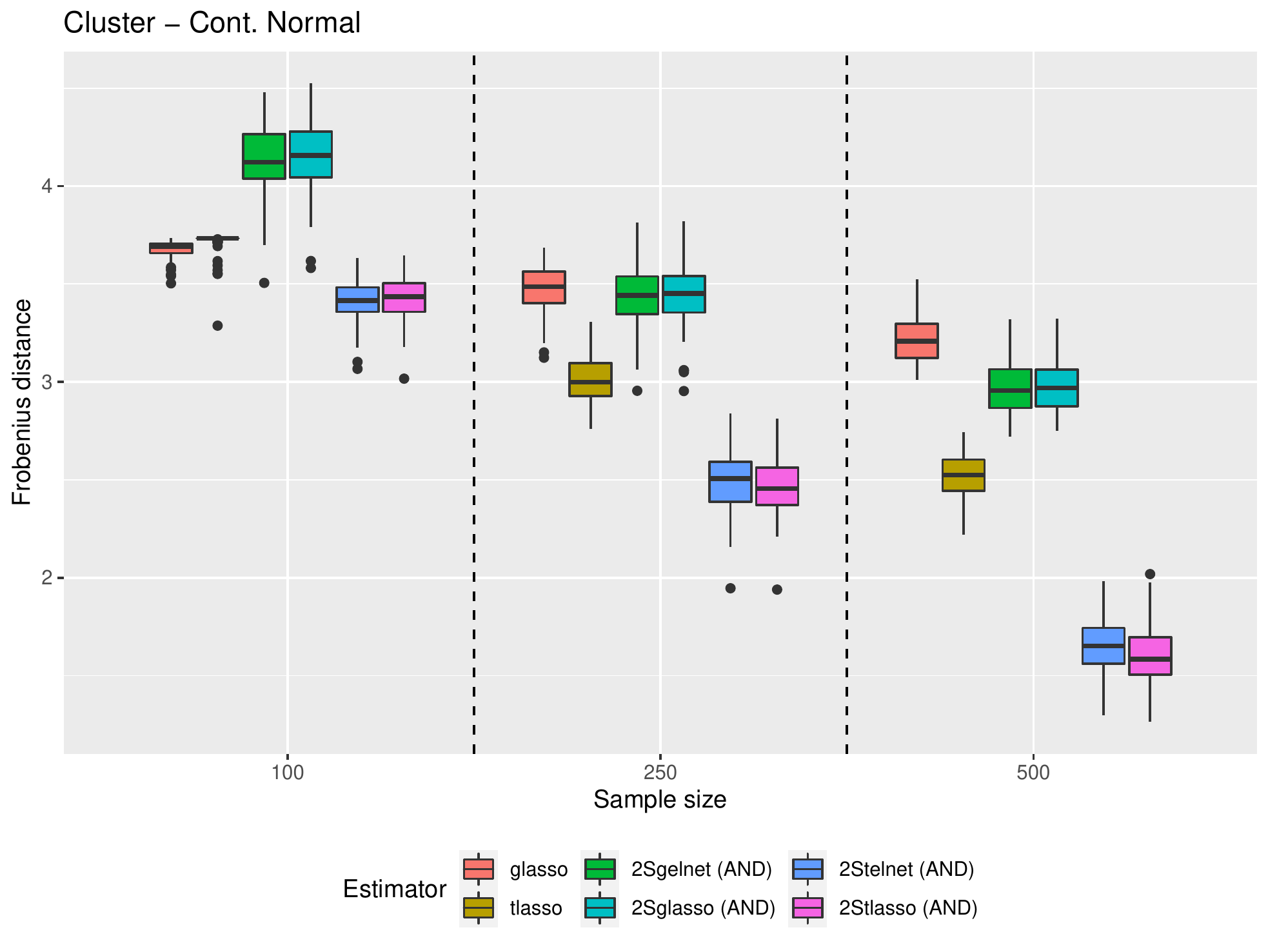}
\end{subfigure}
\newline
\begin{subfigure}{0.4\textwidth}
  \includegraphics[width=1\linewidth, angle = 0]{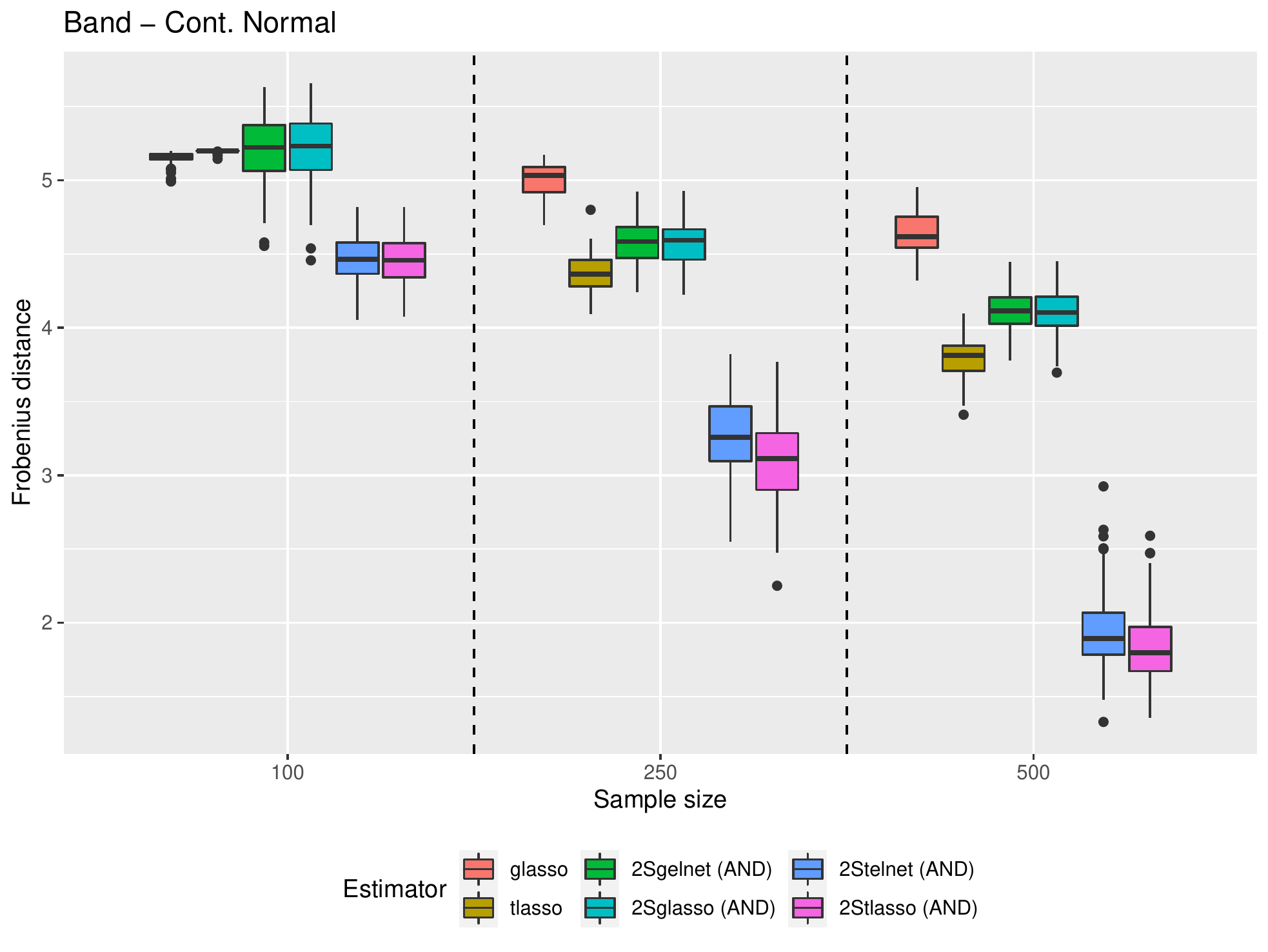}
\end{subfigure}
\begin{subfigure}{0.4\textwidth}
  \includegraphics[width=1\linewidth, angle = 0]{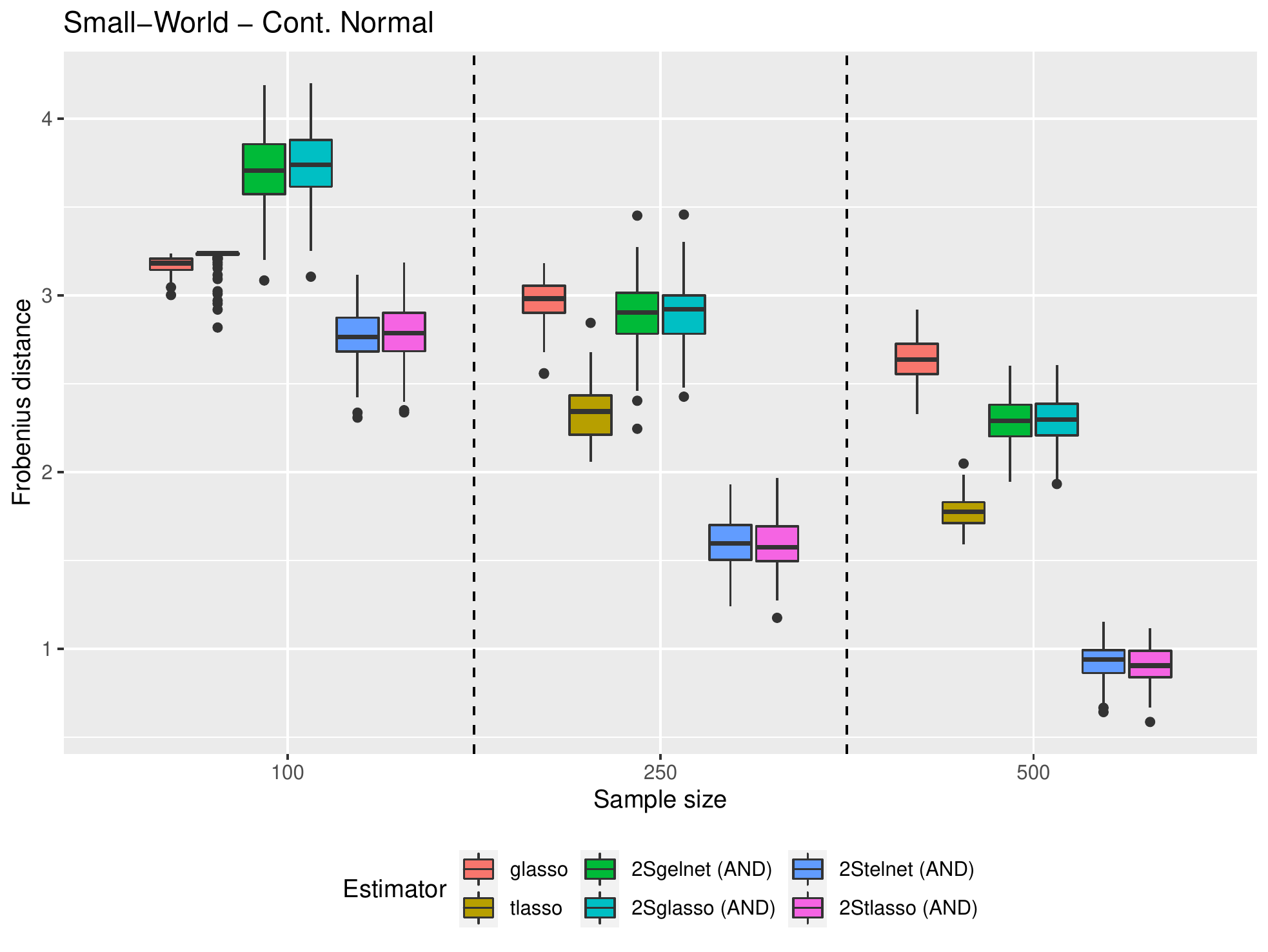}
\end{subfigure}
\newline
\centering
\begin{subfigure}{0.4\textwidth}
  \includegraphics[width=1\linewidth, angle = 0]{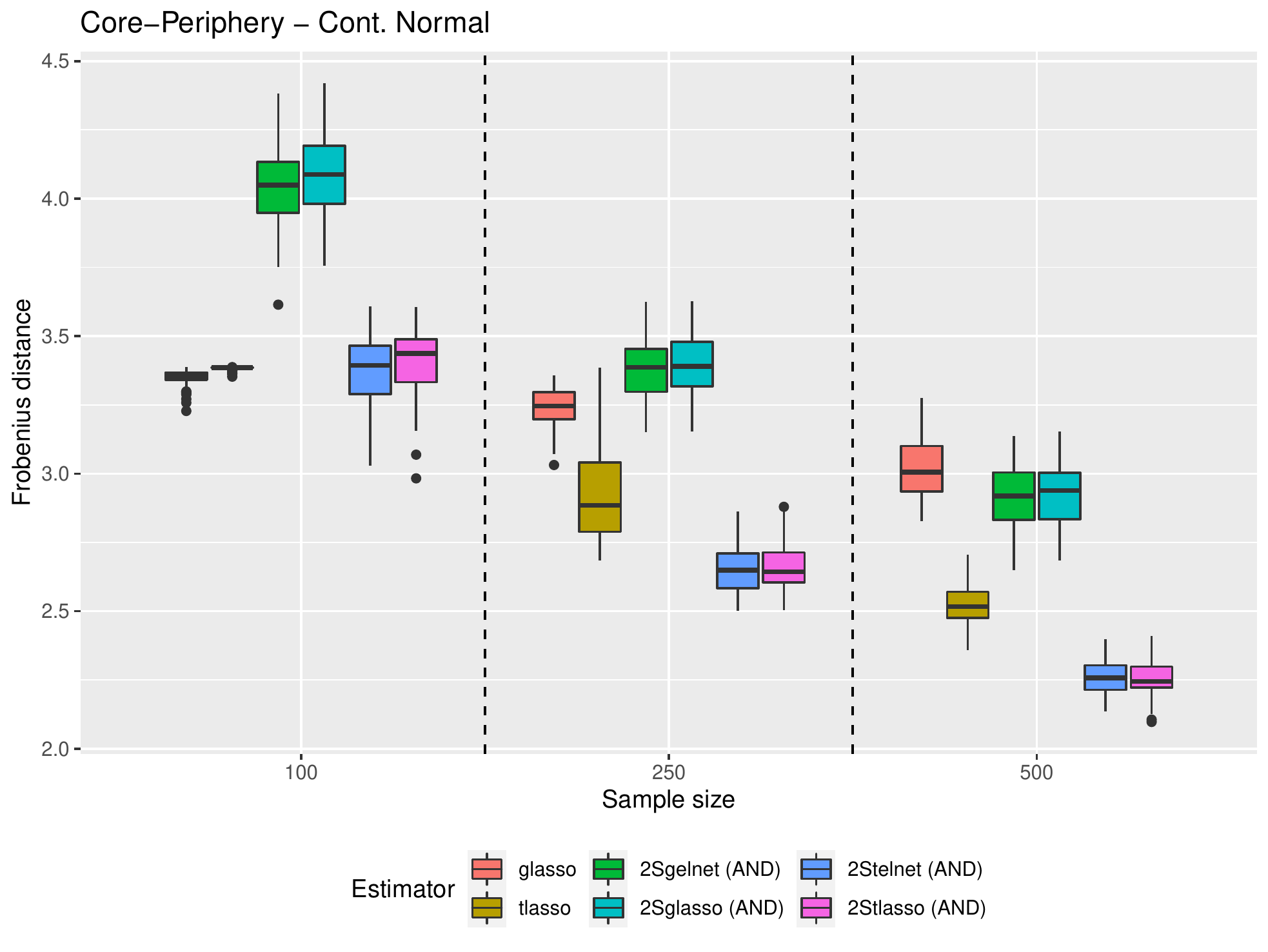}
\end{subfigure}
\caption{Frobenius distance - Contaminated normal distribution}
\label{fig:fdpccontnorm}
\end{figure}
\clearpage
\clearpage
\section{Additional results: European banking network} \label{sec:app3}

\begin{table}[ht]
\centering
\resizebox{0.73\textwidth}{!}{

\begin{tabular}{rllcc}
  \hline
 & Bank & ShortName & Normal & t-Student \\ 
  \hline
1 & HSBC Holdings PLC & HSBC & R & NR \\ 
  2 & BNP Paribas SA & BNP Paribas & R & NR \\ 
  3 & Credit Agricole SA & Cr. Agricole & R & NR \\ 
  4 & Banco Santander SA & B. Santander & R & NR \\ 
  5 & Societe Generale SA & Soc. General & R & NR \\ 
  6 & Barclays PLC & Barclays & R & NR \\ 
  7 & Lloyds Banking Group PLC & Lloyds & R & NR \\ 
  8 & ING Groep NV & ING Group & R & NR \\ 
  9 & UniCredit SpA & Unicredit & R & NR \\ 
  10 & Natwest Group PLC & Natwest & R & NR \\ 
  11 & Intesa Sanpaolo SpA & Intesa SP & R & NR \\ 
  12 & Banco Bilbao Vizcaya Argentaria SA & B. Bilbao VA & R & NR \\ 
  13 & Standard Chartered PLC & Std. Chartered & R & NR \\ 
  14 & Danske Bank A/S & Danske Bank & R & NR \\ 
  15 & Commerzbank AG & Commerzbank & R & NR \\ 
  16 & Svenska Handelsbanken AB & Svenska & R & NR \\ 
  17 & KBC Groep NV & KBC Group & R & NR \\ 
  18 & Skandinaviska Enskilda Banken AB & Skandinaviska & R & NR \\ 
  19 & Dnb ASA & Dnb & R & NR \\ 
  20 & Erste Group Bank AG & Erste Group & R & NR \\ 
  21 & Swedbank AB & Swedenbank & R & NR \\ 
  22 & Banco de Sabadell SA & B. Sabadell & R & NR \\ 
  23 & Raiffeisen Bank International AG & Raiffeisen & R & NR \\ 
  24 & Banca Monte dei Paschi di Siena SpA & Mt. Paschi & R & NR \\ 
  25 & Bank of Ireland Group PLC & B. Ireland & R & NR \\ 
  26 & AIB Group plc & AIB Group & R & NR \\ 
  27 & Jyske Bank A/S & Jyske Bank & R & NR \\ 
  28 & National Bank of Greece SA & N.B. of Greece & R & NR \\ 
  29 & Komercni Banka as & Komercni & R & NR \\ 
  30 & Banco BPM SpA & Banco BPM & R & NR \\ 
  31 & Nordea Bank Abp & Nordea Bank & R & NR \\ 
  32 & Deutsche Bank AG & Deutsche Bank & R & NR \\ 
  33 & UBI Banca SPA & UBI & R & R \\ 
  34 & Credit Suisse Group & Credit Suisse & R & NR \\ 
  35 & UBS Group AG & UBS Group & R & NR \\ 
  36 & Natixis & Natixis & R & NR \\ 
   \hline
\end{tabular}}
\caption{List of banks with KS-tests ($\alpha=0.05$) on residuals (R: rejected - NR: not rejected)}\label{tab:banksnames}
\end{table}

\clearpage

\begin{figure}[ht]
\vspace{-2cm}
\centering
\begin{subfigure}{0.65\textwidth}
  \centering
  \includegraphics[width=1\linewidth, angle = 0]{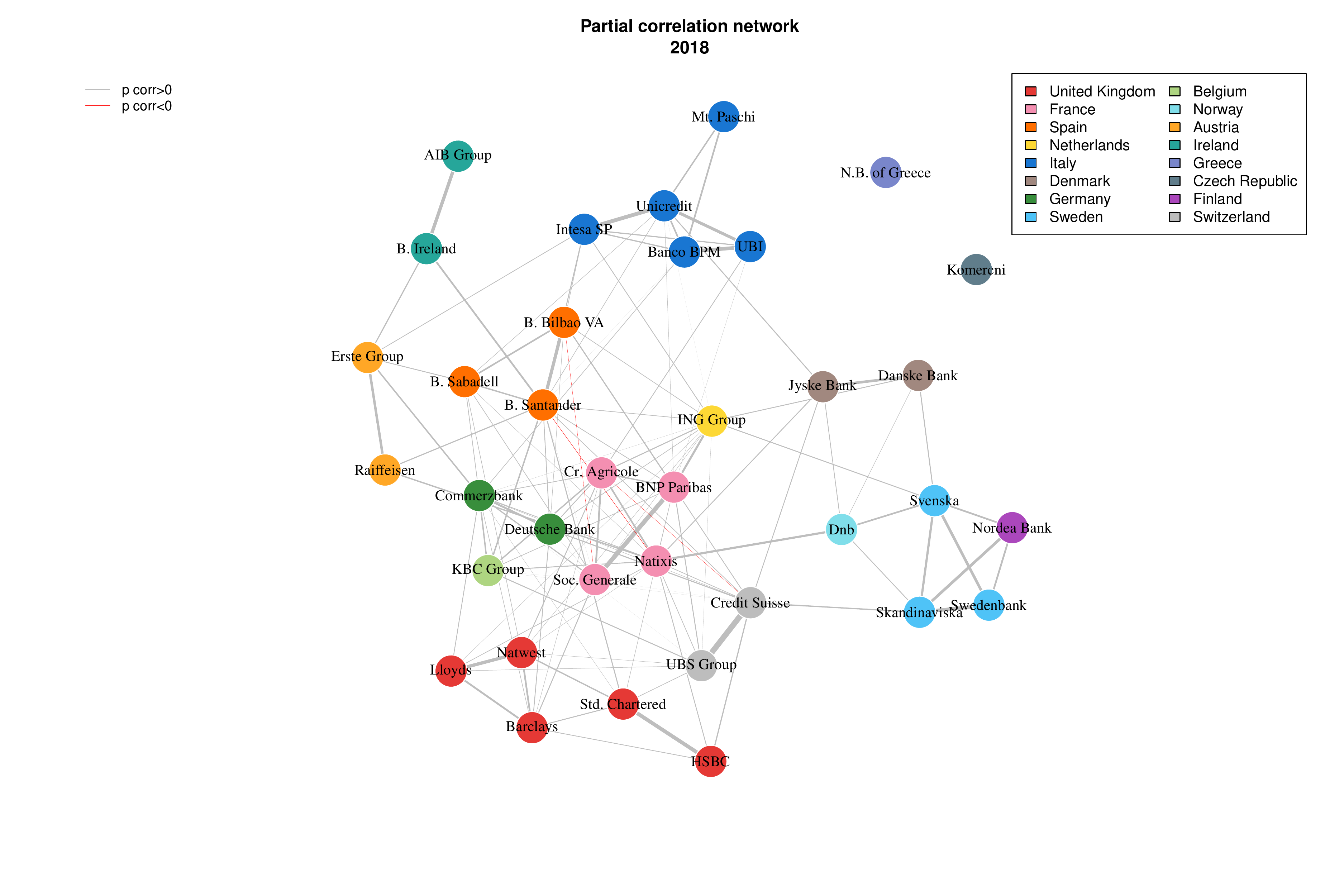}
\end{subfigure}
\newline
\centering
\begin{subfigure}{0.65\textwidth}
  \centering
  \includegraphics[width=1\linewidth, angle = 0]{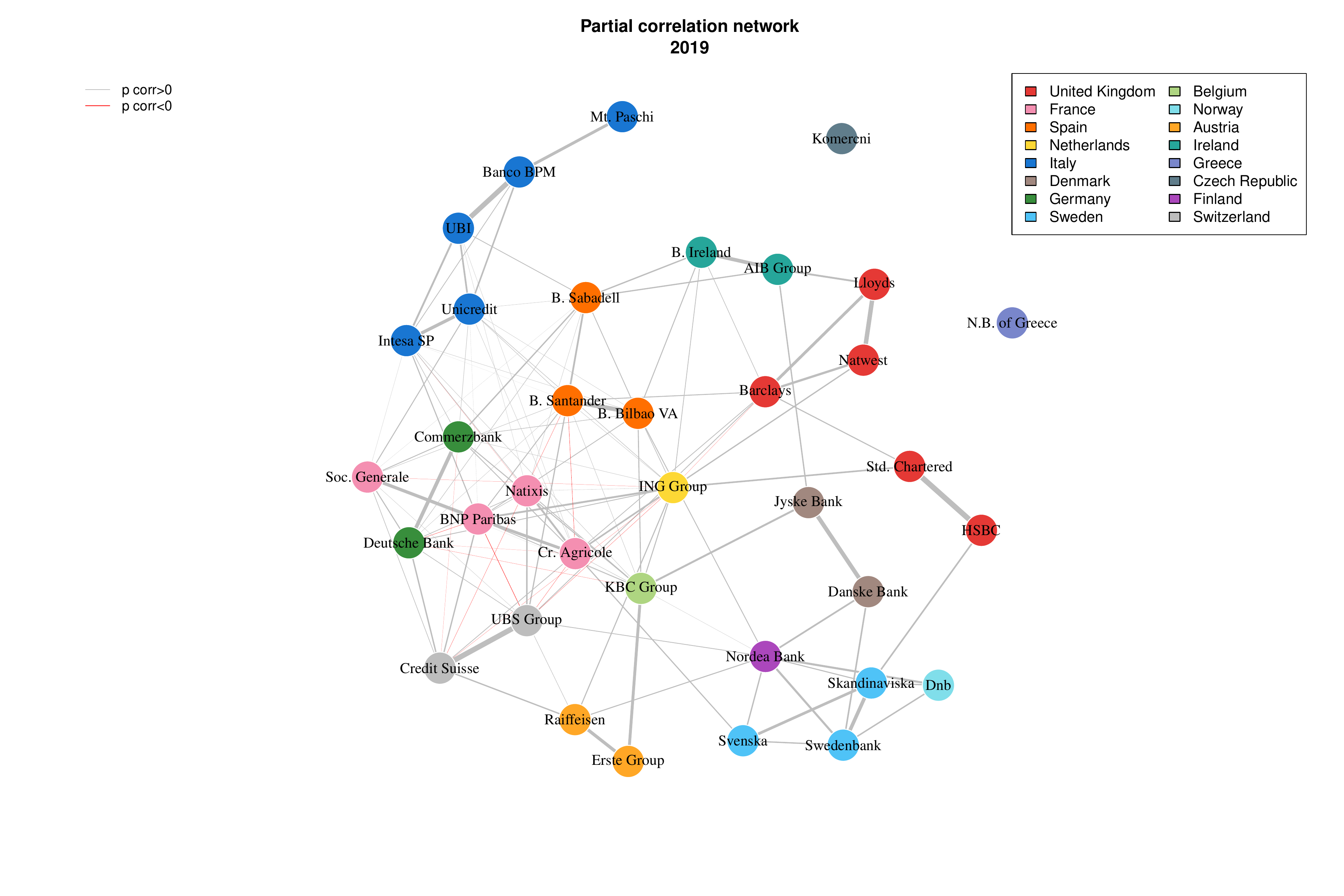}
\end{subfigure}
\newline
\centering
\begin{subfigure}{0.65\textwidth}
  \centering
  \includegraphics[width=1\linewidth, angle = 0]{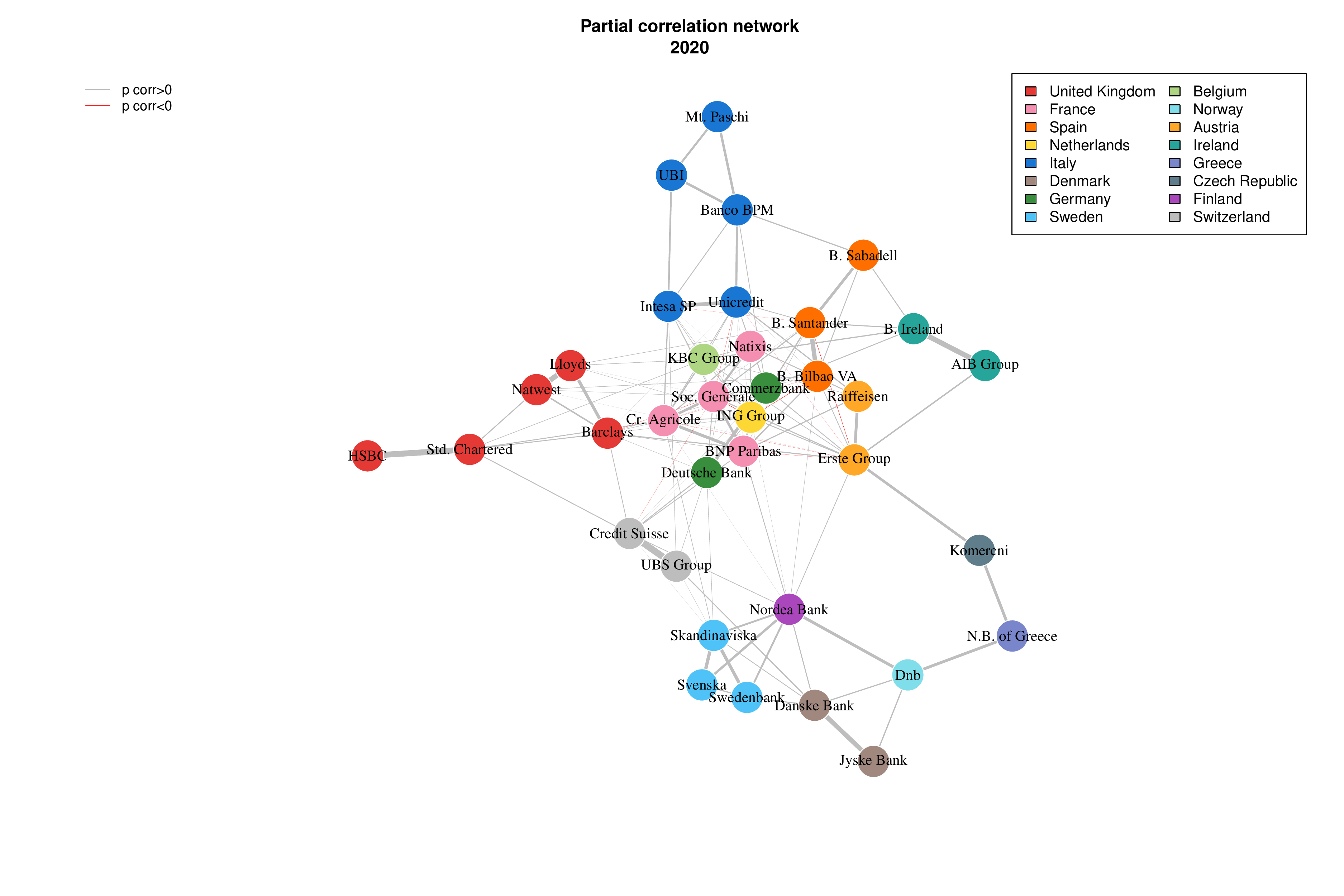}
\end{subfigure}
\caption{Estimated partial correlation networks}
\label{fig:estimatedparcorr}
\end{figure}

\clearpage

\begin{figure}[ht]
\vspace{-2cm}
\centering
\begin{subfigure}{1\textwidth}
  \centering
  \includegraphics[width=1\linewidth, angle = 0 ]{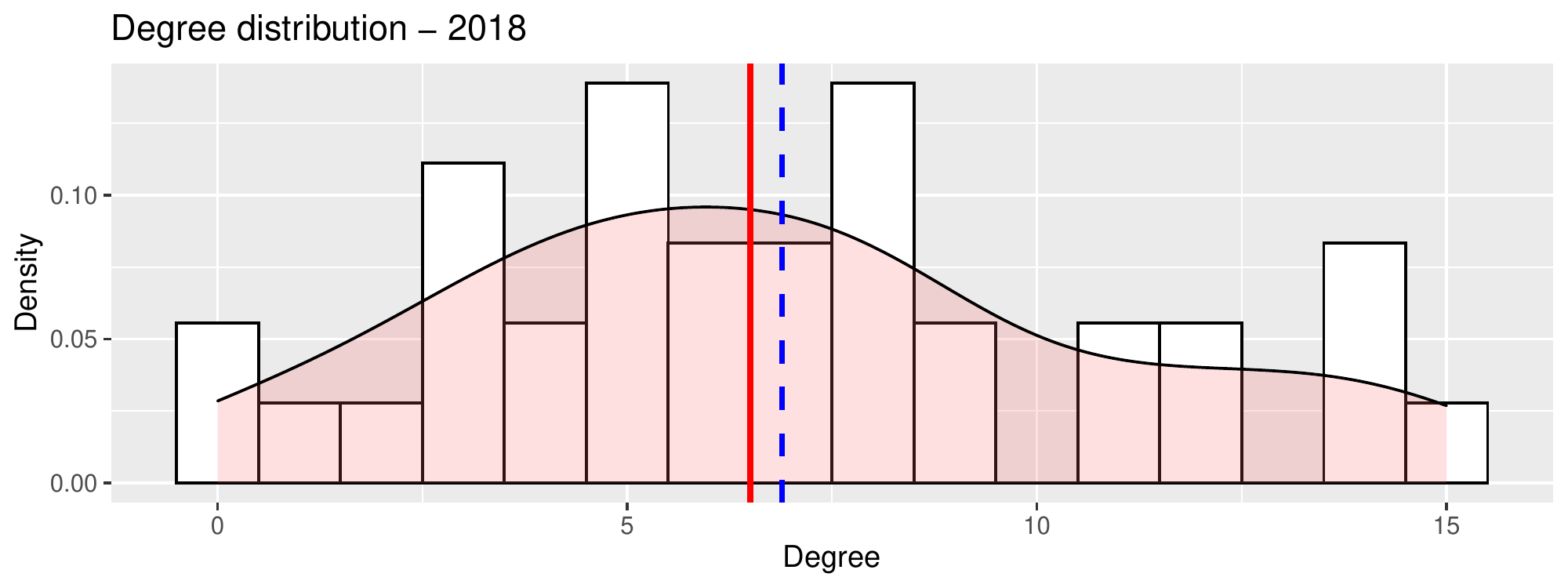}
\end{subfigure}
\newline
\centering
\begin{subfigure}{1\textwidth}
  \centering
  \includegraphics[width=1\linewidth, angle = 0 ]{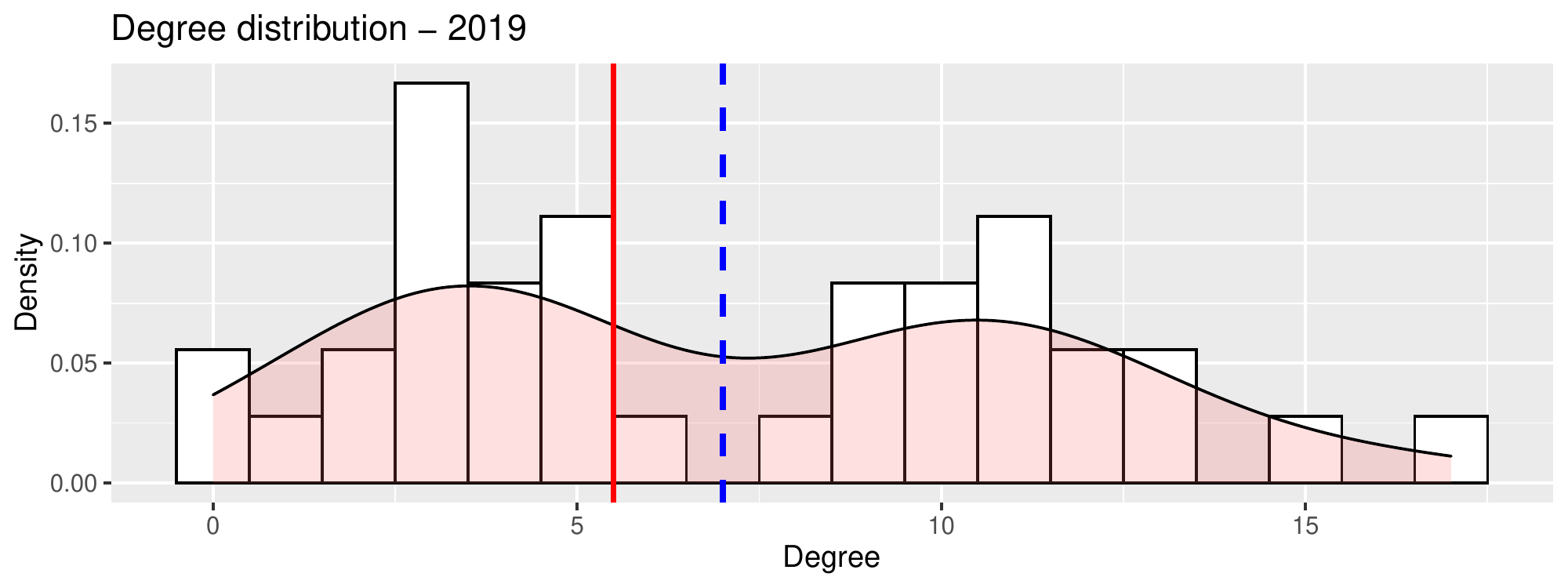}
\end{subfigure}
\newline
\centering
\begin{subfigure}{1\textwidth}
  \centering
  \includegraphics[width=1\linewidth, angle = 0 ]{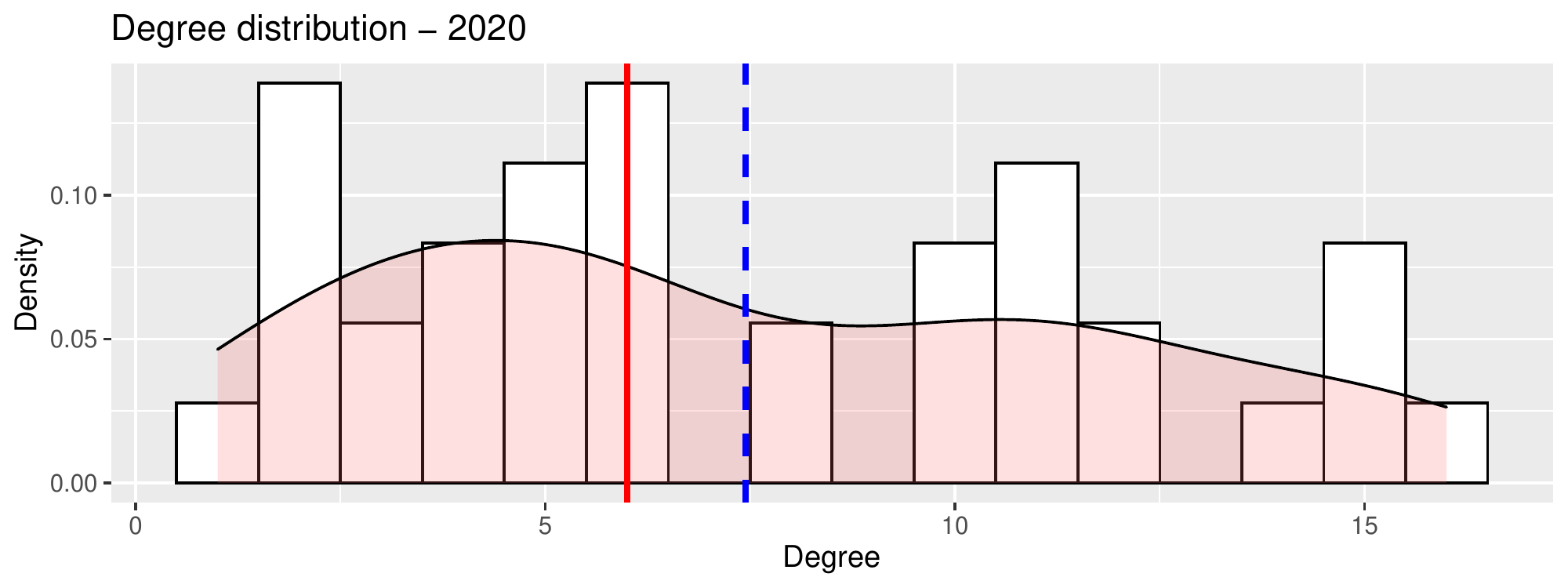}
\end{subfigure}
\caption{Degree distribution (median: red line - mean: blue dashed line)}
\label{fig:degdists}
\end{figure}

\clearpage

\end{document}